\documentclass[a4paper,10pt]{elsarticle}
\usepackage{jcomp}
\usepackage{framed,multirow}
\usepackage[utf8]{inputenc}
\usepackage{bm}
\usepackage{amsmath}
\usepackage{mathrsfs}
\usepackage{graphicx}
\usepackage{epsfig, setspace}
\usepackage{url}
\usepackage{graphicx, transparent, color}
\usepackage{algorithm}
\usepackage{algorithmicx}
\usepackage{etex}
\usepackage[bookmarks=true]{hyperref}
\usepackage{amsmath}
\usepackage{xcolor, soul}
\usepackage{rotating} % Rotating the figures but keeps the page in portrait mode
\usepackage{pdflscape} % Rotates single page into Landscape mode
\usepackage{silence}
\usepackage{ulem,cancel}

\newtheorem{example}{Example}[section]
\usepackage{graphicx}
\usepackage{float}
\usepackage{subfigure}% subcaptions for subfigures
\usepackage{subfigmat}% matrices of similar subfigures,

\usepackage{tikz}
\usepackage{capt-of}
\usepackage{setspace,lipsum}
\usepackage{lineno}
\newcommand{\half}{\frac{1}{2}}
\usepackage{setspace,lipsum}
\usepackage{listings}
\usepackage{xcolor}
\definecolor{aquamarine}{rgb}{0.5, 1.0, 0.83}
\definecolor{OliveGreen}{rgb}{0,0.6,0}

\sethlcolor{aquamarine}

\definecolor{codegreen}{rgb}{0,0.6,0}
\definecolor{codegray}{rgb}{0.5,0.5,0.5}
\definecolor{codepurple}{rgb}{0.58,0,0.82}
\definecolor{backcolour}{rgb}{0.95,0.95,0.92}

\lstdefinestyle{mystyle}{
    backgroundcolor=\color{backcolour},   
    commentstyle=\color{codegreen},
    keywordstyle=\color{magenta},
    numberstyle=\tiny\color{codegray},
    stringstyle=\color{codepurple},
    basicstyle=\ttfamily\footnotesize,
    breakatwhitespace=false,         
    breaklines=true,                 
    captionpos=b,                    
    keepspaces=true,                 
    numbers=left,                    
    numbersep=5pt,                  
    showspaces=false,                
    showstringspaces=false,
    showtabs=false,                  
    tabsize=2
}
\begin{document}

\begin{frontmatter}
\title{Wave or Physics-Appropriate Multidimensional Upwinding Approach for Compressible Multiphase Flows}

\author[AA_address]{Amareshwara Sainadh Chamarthi \cortext[cor1]{Corresponding author. \\ 
E-mail address: sainath@caltech.edu (Amareshwara Sainadh  Ch.).}}
\address[AA_address]{Division of Engineering and Applied Science, California Institute of Technology, Pasadena, CA, USA}
\begin{abstract}
This paper introduces multidimensional algorithms for simulating multiphase flows, leveraging the wave structure of the Euler equations in characteristic space and the physical properties of variables in physical space. The algorithm applies different reconstruction schemes to acoustic, vorticity, and entropy waves in characteristic space to enhance accuracy and minimize numerical artifacts. In characteristic space, upwind schemes are used for acoustic waves, central schemes for vorticity and entropy waves, and Tangent of Hyper-bola for INterface Capturing (THINC) reconstruction for material interfaces and contact discontinuities (a subset of entropy waves). This approach prevents spurious vortices in periodic shear layers, accurately captures vortical structures in gas-gas and gas-liquid interactions, and improves the accuracy of shock-entropy wave interactions. In physical space, phasic densities are computed using THINC in regions of contact discontinuities and material interfaces, while tangential velocities are calculated with central schemes to improve vortical structures. An adaptive reconstruction technique is also introduced to mitigate oscillations near shocks, which arise from primitive variable reconstruction, by combining primitive and characteristic variable reconstructions with the liquid phase being identified using the stiffened gas parameter. The proposed multidimensional upwinding approach outperforms traditional schemes, demonstrating superior accuracy in capturing physical phenomena, reducing numerical artifacts, and better matching experimental results across complex test cases.
\end{abstract}
\end{frontmatter}
\section{Introduction}\label{sec:intro}
The numerical simulation of compressible flows has evolved significantly since Godunov’s method, which solved the Riemann problem at cell interfaces \cite{godunov1959}. While stable and capable of capturing shocks, it suffered from excessive numerical dissipation, spurring the development of higher-order methods. In the 1970s, van Leer introduced upwind schemes with monotonicity preservation, flux limiters, and the Monotonic Upstream-centered Scheme for Conservation Laws (MUSCL) approach, achieving second-order accuracy and robustness near discontinuities \cite{vanleer1979}. Weighted Essentially Non-Oscillatory (WENO) schemes by Liu, Osher, and Chan further advanced accuracy by adaptively combining stencils using smoothness indicators \cite{Liu1994}. Jiang and Shu’s WENO-JS improved accuracy and stability, becoming a standard for complex flows \cite{Jiang1995}. Later adaptive WENO schemes, such as those by Hu and Adams \cite{Hu2010} and Balsara, Garain and Shu \cite{Balsara2016}, dynamically blended central and upwind methods, enhancing efficiency and accuracy for resolving shocks and turbulence. To the author's knowledge, one common theme in all these approaches is that the numerical schemes initially designed for the linear advection equation are often used as building blocks for simulating more complex fluid dynamics equations like the Euler and Navier-Stokes equations. For example, the linear advection equation is typically used to analyze a numerical scheme’s spectral properties (dispersion and dissipation). While such a one-dimensional analysis carried out for linear advection equation is useful—for instance, it has been shown in Ref. \cite{chamarthi2023gradient} that improving the curvature estimates of Monotonicity Preserving (MP) limiter \cite{suresh1997accurate} enhances dispersion properties and improves the results, as corroborated by simulations—it may not be sufficient. The paper aims to develop algorithms directly for the Euler (application to multiphase flows) equations rather than for the linear advection by considering the relevant physics to the extent possible.

In \cite{van2006upwind}, van Leer mentioned that a numerical method should use upwinding for advection and distributed omnidirectionally (presumably central scheme) when representing subsonic acoustic propagation. While, for example, the above-mentioned adaptive central-upwind WENO scheme of Hu and Adams \cite{Hu2010} does try to blend upwind and central schemes, it is still designed based on linear advection, and the scheme does not take into account the contributions of different physical aspects of the variables and constraints as suggested by van Leer. Roe proposed such a fundamental approach called “genuinely multidimensional” upwind schemes \cite{roe1986discrete}. Roe mentioned in the corresponding paper that:

\textit{“The chief difficulty is that there are just three types of elementary waves in one dimension. Only one model can be constructed in one dimension with three parameters: the unknown wave strengths. Matching the model to the spatial gradients of the three data quantities, density, velocity, and pressure, gives three simple linear equations. In two dimensions, the data will allow us to estimate gradients in two directions of four quantities, yielding eight information items. Whatever model we choose must have “eight free” parameters.”}

The algorithms to be proposed in this paper share some similarities with that of Roe \cite{roe1986discrete} and will be explained briefly. The following discussion assumes that the variables are stored at the cell centres, $i$, and are reconstructed to the cell interfaces, ${i+\frac{1}{2}}$, as in a finite volume method \cite{godunov1959}. $u^{L}_{i+\frac{1}{2}}$ represents the left-biased interface value and $u^{R}_{i+\frac{1}{2}}$ represents the right biased value.  Suppose a numerical scheme is designed for the linear advection equation, $\frac{\partial u}{\partial t} + \frac{\partial u}{\partial x} = 0$. In that case, it will have \textit{one free parameter} for interface reconstruction:
\begin{equation}
u_{i+\frac{1}{2}} = \eta  u^{L}_{i+\frac{1}{2}} + \left( 1 - \eta \right) u^{R}_{i+\frac{1}{2}}.
\end{equation}
In a first-order upwind method $u^{L}_{i+\frac{1}{2}}$ = $u_i$ and $u^{R}_{i+\frac{1}{2}}$ =  and $u_{i+1}$.  The parameter $\eta$ can be either one, then the scheme will be purely upwind ($u_i$), or half, then the scheme will be purely central $((u_i+u_{i+1})/2)$. Unlike the linear advection equation, the Euler equations involve multiple variables and are inherently multidimensional. For one-dimensional (1D) Euler equations, there will be three parameters for interface reconstruction for each of the three variables (density ($\rho$), velocity ($u$) and pressure ($p$) - assuming primitive variables are reconstructed). The complexity increases significantly for 2D Euler equations. Eight parameters ($\rho$, $u$, $v$, and $p$—four variables in two directions, $x$ and $y$) must be considered, underscoring the problem’s multidimensional nature, as shown below, and the need for a sophisticated algorithm to handle it.
\begin{equation*}
\begin{aligned}[t]
\rho^{x}_{i+\frac{1}{2}} &= \eta_1 \rho^{x,L}_{i+\frac{1}{2}} + \left( 1 - \eta_1 \right)   \rho^{x, R}_{i+\frac{1}{2}}\\
u^{x}_{i+\frac{1}{2}} &= \eta_2  u^{x, L}_{i+\frac{1}{2}} + \left( 1 - \eta_2 \right) u^{x, R}_{i+\frac{1}{2}}\\
v^{x}_{i+\frac{1}{2}} &= \eta_3 v^{x, L}_{i+\frac{1}{2}} +   \left( 1 - \eta_3 \right) v^{x, R}_{i+\frac{1}{2}}\\
p^{x}_{i+\frac{1}{2}} &= \eta_4  p^{x, L}_{i+\frac{1}{2}} + \left( 1 - \eta_4 \right) p^{x, R}_{i+\frac{1}{2}}\\
\end{aligned}
\qquad 
\begin{aligned}[t]
\rho^{y}_{i+\frac{1}{2}} &=  \eta_5 \rho^{y, L}_{i+\frac{1}{2}} +  \left( 1 - \eta_5 \right)  \rho^{y, R}_{i+\frac{1}{2}}\\
u^{y}_{i+\frac{1}{2}} &=   \eta_6 u^{y, L}_{i+\frac{1}{2}} + \left( 1 - \eta_6 \right) u^{y, R}_{i+\frac{1}{2}}\\
v^{y}_{i+\frac{1}{2}} &=  \eta_7 v^{y, L}_{i+\frac{1}{2}} +  \left( 1 - \eta_7 \right)  v^{y, R}_{i+\frac{1}{2}}\\
p^{y}_{i+\frac{1}{2}} &=  \eta_8 p^{y, L}_{i+\frac{1}{2}} +  \left( 1 - \eta_8 \right)  p^{y, R}_{i+\frac{1}{2}}\\
\end{aligned}
\end{equation*}

If a numerical method is \textit{designed} for the linear advection equation, then, for example, a fifth-order upwind scheme is considered; it would remain a fifth-order scheme for all the variables in all directions when applied for the simulations of Euler equations. The scheme thus designed will not become a sixth-order scheme and compute the tangential velocities in Euler simulations in each direction using a central scheme. In this paper, the value of $\eta$ is either 0.5 or 1 and is different for different variables in various directions. Suppose reconstruction is carried out in characteristic space; the upwind scheme computes the acoustic waves, which means $\eta$ is one. The vorticity waves (and entropy waves in some cases) are computed by the central scheme, meaning $\eta$ is 0.5, implying eight different values of $\eta$. While adaptive central-upwind schemes exist in the literature, as mentioned earlier, they are not \textit{always} upwinded for some variables and computed using a central scheme for certain variables. The eight free parameters discussed here may or may not be the same as those envisioned by Roe \cite{roe1986discrete}, but the ideas are similar, and the current approach can be called “multidimensional upwinding” - as it has many similarities.

Apart from the reconstruction procedure, physical variables are also subjected to various physical constraints depending on the flow situation. For example, across a contact discontinuity, density is discontinuous, whereas velocity and pressure remain continuous \cite{hirschvol2,abgrall1996prevent,abgrall2001computations,chargy1990comparisons}. These constraints necessitate using an upwind scheme with a nonlinear discontinuity-capturing mechanism for density, while velocity and pressure can be computed using a central scheme. A scheme designed solely for the linear advection equation may not account for such physical requirements and constraints. Similarly, tangential velocities are continuous across material interfaces (in viscous flows \cite{batchelor1967introduction} or due to the presence of artificial viscosity \cite{meng2018numerical})—$u-$velocity is continuous in the y-direction, and $v-$velocity is continuous in the x-direction—highlighting the need for schemes that address these \textit{multidimensional} aspects. Across a shockwave, the tangential velocity is continuous, and density, pressure, and normal velocities are discontinuous. Another aspect is the choice of variables for reconstruction, primitive or characteristic variables. While a discretization using primitive variables for reconstruction is robust for large-density regions, it will lead to oscillations near shockwaves. Characteristic variable reconstruction prevents oscillations near shockwaves \cite{van2006upwind,coralic2014finite} but is not robust near regions with the gas-liquid interface (significant jump in density) \cite{bryngelson2021mfc}. A numerical method can improve the numerical simulations if it accounts for these constraints and considers the flow’s multidimensional aspects. Furthermore, neither van Leer \cite{van2006upwind} nor Roe \cite{roe1986discrete} mentioned in their papers (again, to the authors’ knowledge) which variables or waves should use upwind and central schemes. The present paper introduces an algorithm that accounts for these constraints and multidimensional aspects, targeting multiphase flows as the primary application. The paper has the following objectives:

\begin{itemize}
	\item Use primitive or characteristic variables appropriately for multiphase simulations to achieve as much as possible oscillation-free results.
	\item Compute vorticity waves (tangential velocities in physical space) using a central scheme and analyze the effects and advantages of this approach, including a qualitative agreement with the experimental results.
	\item Ensure that only the variable density (phasic densities) is modified in regions of material interfaces and contact discontinuities. Such an algorithm should selectively modify the entropy wave in characteristic space. When primitive variables are reconstructed, shock regions may also be improved since density is discontinuous across them. Not all regions of the entropy waves may be contact discontinuities, and those regions should not be altered by an interface-capturing method. It is possible to compute certain regions of the entropy waves using a central scheme.
\end{itemize}

Finally, Hoffman, Chamarthi, and Frankel \cite{hoffmann2024centralized} employed an upwind scheme for acoustic waves and a central scheme for the remaining waves in characteristic space similar to the current work. \textbf{In the literature, central schemes are typically applied to all waves or variables \cite{van2022immersed,hendrickson2018improved,kuya2018kinetic,chamarthi2023role}. These non-dissipative schemes are well-suited for turbulent flow simulations and were target applications in \cite{hoffmann2024centralized}. However, it was observed that employing central schemes for acoustic waves led to oscillations due to the Ducros sensor and eventual simulation failure. Therefore, using upwind schemes for acoustic waves in \cite{hoffmann2024centralized} was driven by stability concerns rather than deliberate choice. In contrast, the current approach deliberately uses upwind schemes for the acoustic waves. As demonstrated in the periodic shear layer test case, later in results, even in scenarios without discontinuities, relying solely on central schemes can result in numerical instabilities and oscillations.}  Ref. \cite{hoffmann2024centralized} aimed to predict the transition in hypersonic single-species flows, which required a low dispersion-dissipation method, whereas the current work’s objective is to simulate multiphase flows. Ref. \cite{hoffmann2024centralized} used the Gradient-based reconstruction \cite{chamarthi2023gradient,chamarthi2023implicit,chamarthi2023wave,chamarthi2023efficient} due to its superior spectral properties. Gradient-based reconstruction was found to be not robust for multiphase flows due to its long stencil. Similarly, the Ducros sensor, used for shock capturing and centralization for multicomponent flows in \cite{sainadh2024consistent}, lacked robustness in multiphase flow simulations, necessitating a simpler alternative. Additionally, multiphase flows involve material interfaces that must be captured with minimal dissipation, which requires sophisticated interface-capturing techniques like THINC \cite{xiao2011revisit,shyue2014eulerian,deng2018high,deng2018limiter}. Furthermore, the current paper addresses the complexities of multiphase flows, utilizing standard (well-known and most used) fifth- or sixth-order reconstruction schemes that demonstrate the \textit{benefits} of a “multidimensional-upwind” scheme, which very high-order methods could obscure.

The rest of the manuscript is organized as follows: Section \ref{sec:eqns-gov} presents the governing equations. Section \ref{sec:num} presents the numerical methods, including the novel wave-appropriate multidimensional upwind schemes. Section \ref{results} consists of the proposed algorithm's numerical results, and the conclusions are presented in Section \ref{clusions}.

\section{Governing equations}\label{sec:eqns-gov}
The quasi-conservative five equation model for compressible inviscid multiphase flows \cite{allaire2002five} is as follows:

{\begin{equation}\label{5eqn-base}
\frac{\partial \mathbf{Q}}{\partial t}+\frac{\partial \mathbf{F}}{\partial x}+\frac{\partial \mathbf{G}}{\partial y}=\mathbf{S},
\end{equation}
where the state vector ($\mathbf{Q}$), convective flux vectors ($\mathbf{F}$ and $\mathbf{G}$) and source term, $\mathbf{S}$, are given by:
\begin{equation}
\mathbf{Q}=\left[\begin{array}{c}
\alpha_{1} \rho_{1} \\
\alpha_{2} \rho_{2} \\
\rho u \\
\rho v \\
E \\
\alpha_{1}
\end{array}\right], \quad \mathbf{F}=\left[\begin{array}{c}
\alpha_{1} \rho_{1} u \\
\alpha_{2} \rho_{2} u \\
\rho u^{2}+p \\
\rho v u \\
(E+p) u \\
\alpha_{1} u
\end{array}\right], \quad \mathbf{G}=\left[\begin{array}{c}
\alpha_{1} \rho_{1} v \\
\alpha_{2} \rho_{2} v \\
\rho u v \\
\rho v^{2}+p \\
(E+p) v \\
\alpha_{1} v
\end{array}\right],  \quad \mathbf{S}=\left[\begin{array}{c}
0 \\
0 \\
0 \\
0 \\
0 \\
\alpha_{1} \nabla \cdot \mathbf{u}
\end{array}\right],
\end{equation}
where $\rho_1$ and $\rho_2$ correspond to the densities of fluids $1$ and $2$, $\alpha_{1}$ and $\alpha_{2}$ are the volume fractions of the fluids $1$ and $2$, $\rho$, $u$,$v$, $p$ and $E$ are the density, $x-$ and $y-$ velocity components, pressure, total energy per unit volume of the mixture, respectively.  Under the isobaric assumption, the following equation is used to close the system
\begin{equation}\label{eqn:pressure-new}
p = (\gamma -1) (E - \rho \frac{(u^2+v^2)}{2})-\gamma \pi_{\infty},
\end{equation}
where $\gamma$ and $\pi_{\infty}$ are fitting parameters, and $c=\sqrt{\gamma\left(p+\pi_{\infty}\right) / \rho}$ is the sound speed. The five-equation model requires a set of mixture rules for various properties of the fluids. The mixture rules for the volume fractions of the two fluids $\alpha_1$ and $\alpha_2$, as well as the density and mixture rules for the fitting parameters, are given by:
\begin{equation}
\alpha_{2}=1-\alpha_{1},
\end{equation}
\begin{equation}
\rho=\rho_{1} \alpha_{1}+\rho_{2} \alpha_{2},
\end{equation}
\begin{equation}
\frac{1}{\gamma-1}=\frac{\alpha_{1}}{\gamma_{1}-1}+\frac{\alpha_{2}}{{\gamma}_{2}-1},
\end{equation}
 
\begin{equation}
\frac{{\gamma} {\pi_{\infty}}}{{\gamma}-1} =\frac{\alpha_1 \gamma_1 \pi_{\infty,1}}{\gamma_1-1}+\frac{\alpha_2 \gamma_2 \pi_{\infty,2}}{\gamma_2-1}.
\end{equation}
The primitive variable vector for the multicomponent flows for the five-equation model in the two-dimensional scenario is $\mathbf{U}$ = $( \alpha_{1}\rho_{1}, \alpha_{2}\rho_{2}, u, v, p,\alpha_1)^T$. The following section presents the numerical discretization procedure of these equations.

\section{Numerical discretization}\label{sec:num}
The five-equation model described earlier (Equations (\ref{5eqn-base})) is discretized using a finite volume method on a uniform Cartesian grid with cell sizes $\Delta x$ and $\Delta y$ in the $x$- and $y$-directions, respectively. The conservative variables, $\mathbf{Q}$, are stored at the center of each cell $I_{i, j}$, where the indices $i$ and $j$ denote the $i$-th cell in the $x$-direction and the $j$-th cell in the $y$-direction. The resulting semi-discrete form of the equations for the cell $\mathbf{Q}_{i, j}$ is expressed as:\\
\begin{equation}
\begin{aligned}
\frac{d \mathbf{Q}_{i, j}}{d t} & =-\left(\frac{\mathbf {\hat{F}^c}_{i+1 / 2 ,j}-\mathbf {\hat{F}^c}_{i-1 / 2,j}}{\Delta x}\right)-\left(\frac{\mathbf {\hat{G}^c}_{i,j+1 / 2}-\mathbf {\hat{G}^c}_{i,j-1 / 2}}{\Delta y}\right)+\mathbf{S}_{i, j}, \\
& =R\left(\mathbf{Q}_{i, j}\right).
\end{aligned}
\end{equation}
Here, $\mathbf{\hat{F}^c}$ and $\mathbf{\hat{G}^c}$ are the numerical approximations of the convective fluxes in the $x$- and $y$-directions, respectively, at the cell interfaces $i \pm \frac{1}{2}$ and $j \pm \frac{1}{2}$. The term $R\left(\mathbf{Q}_{i, j}\right)$ represents the residual function and $\mathbf{S}_{i, j}$ is the source term. The resulting system of ordinary differential equations is integrated in time using a high-order strong stability-preserving (SSP) Runge-Kutta scheme \cite{Jiang1995}, as follows:
\begin{eqnarray}\label{rk}
%\begin{array}{l}
\mathbf{ Q}_{i, j}^{(1)}&=&\mathbf{ Q}_{i, j}^{n}+\Delta t \ \mathbf{Res}\left(\mathbf{ Q}_{i, j}^{n}\right), \\
\mathbf{ Q}_{i, j}^{(2)}&=&\frac{3}{4} \mathbf{ Q}_{i, j}^{n}+\frac{1}{4} \mathbf{ Q}_{i, j}^{(1)}+\frac{1}{4} \Delta t \ \mathbf{Res}\left(\mathbf{ Q}_{i, j}^{(1)}\right), \\
\mathbf{ Q}_{i, j}^{n+1}&=&\frac{1}{3} \mathbf{ Q}_{i, j}^{n}+\frac{2}{3} \mathbf{ Q}_{i, j}^{(2)}+\frac{2}{3} \Delta t \ \mathbf{Res}\left(\mathbf{ Q}_{i, j}^{(2)}\right).
%\end{array}
\end{eqnarray}
In these equations, the superscripts ${(1)}$ and ${(2)}$ denote intermediate steps, while ${n}$ and ${n+1}$ represent the current and next time steps, respectively. The time step, $\Delta t$, is calculated as:
\begin{equation}
\Delta t= \text{CFL} \cdot \left(\min _{i, j}\left(\frac{\Delta\textcolor{black}{x}}{\left|u_{i, j}\right|+c_{i, j}}, \frac{\Delta \textcolor{black}{y}}{\left|v_{i, j}\right|+c_{i, j}}\right)\right),
\end{equation}
where $c$ is the speed of sound. The computation of the numerical approximations for the convective fluxes, $\mathbf{\hat{F}^c}$ and $\mathbf{\hat{G}^c}$, is discussed in the following sections.

\subsection{Numerical Discretization}\label{sec:inviscid flux discretization}

The numerical approximation of the convective fluxes is achieved through two main steps. The first step is the reconstruction phase, where the solution vector, originally defined at the cell centre, is interpolated to the cell interfaces. The second step is the evolution phase, which employs an approximate Riemann solver to compute the fluxes at the interfaces. This step considers the propagation of ``waves'' and their directions. The convective flux at a given interface is represented as:
\begin{equation}
	\mathbf {\hat{F}^c}_{i+\frac{1}{2}} = F_{i+\frac{1}{2}}^{Riemann} \left(\mathbf{U}_{i+\frac{1}{2}}^L,\mathbf{U}_{i+\frac{1}{2}}^R\right),
\end{equation}
where, $\mathbf{U}$ represents the primitive variable vector, while the superscripts $L$ and $R$ indicate the reconstructed solution vectors on the left and right sides of the interface, respectively. For evaluating the convective fluxes, this study primarily employs the Harten-Lax-van Leer-Contact (HLLC) approximate Riemann solver \cite{coralic2014finite, toro2009riemann, johnsen2006implementation}. The following subsections introduce three reconstruction techniques for computing the interface values ($L$ and $R$) and describe the procedures for novel wave-adaptive, multidimensional upwinding algorithms.  

\subsubsection{Monotonicity Preserving Scheme (Upwind and Central)}\label{mp5-sec}
The first reconstruction method considered is the Monotonicity-Preserving (MP) approach introduced by Suresh and Huynh \cite{suresh1997accurate}. This technique leverages a geometry-based framework to effectively capture shocks. Its primary advantage is the ability to resolve discontinuities accurately while preserving extrema and maintaining high-order accuracy in smooth regions. The MP reconstruction process involves two steps. In the first step, the left and right biased linear fifth-order polynomials are constructed at the cell interface values, $\mathbf{U}_{i+\frac{1}{2}}^{L, Linear}$ and $\mathbf{U}_{i+\frac{1}{2}}^{R, Linear}$, using the cell averages from the cells $i-2, i-1, i, i+1, i+2, i+3$. The expressions for these polynomials are as follows:
\begin{subequations}
    \begin{align}
 \mathbf{U}_{i+\frac{1}{2}}^{L, Linear}&=\frac{1}{30} \mathbf{U}_{i-2}-\frac{13}{60} \mathbf{U}_{i-1}+\frac{47}{60} \mathbf{U}_{i+0}+\frac{9}{20} \mathbf{U}_{i+1}-\frac{1}{20} \mathbf{U}_{i+2},
 \\[15pt]
\mathbf{U}_{i+\frac{1}{2}}^{R, Linear}&=\frac{1}{30} \mathbf{U}_{i+3}-\frac{13}{60} \mathbf{U}_{i+2}+\frac{47}{60} \mathbf{U}_{i+1}+\frac{9}{20} \mathbf{U}_{i+0}-\frac{1}{20} \mathbf{U}_{i-1}.
    \end{align}
    \label{eqn:unlimitedCharacteristicInterpolation-mp5}
\end{subequations}
In the original paper of Suresh and Huynh \cite{suresh1997accurate}, only upwind polynomials are considered. Here, the following central sixth-order polynomial is also considered:
\begin{subequations}
    \begin{align}
 \mathbf{U}^{C,Linear}_{i+\frac{1}{2}}&=\frac{1}{2}\left( \mathbf{U}_{i+1 / 2}^{L, Linear} +  \mathbf{U}_{i+1 / 2}^{R, Linear}\right)=\frac{1}{60}\left( \mathbf{U}_{i-2}-8  \mathbf{U}_{i-1}+37  \mathbf{U}_i+37  \mathbf{U}_{i+1}-8  \mathbf{U}_{i+2}+ \mathbf{U}_{i+3}\right).
    \end{align}
    \label{eqn:central-mp6}
\end{subequations}
After forming the upwind or central linear schemes, denoted as $\mathbf{U}^{Linear}_{i+\frac{1}{2}}$- representing both upwind and central schemes, using the above-mentioned reconstructions, the following condition, MP criteria, is checked to determine the necessity of applying the MP limiter:
\begin{equation} \label{eqn:mp5Condition}
    \left( {\mathbf{U}}^{Linear}_{i+1/2} - {{\mathbf{U}}}_i \right) \left( {\mathbf{U}}^{Linear}_{i+1/2} - {\mathbf{U}}^{MP} \right) \leq 10^{-40},
\end{equation}
where ${\mathbf{U}^{MP}}$ is given by the following equation:
\begin{equation} \label{eqn:alpha_mp}
\begin{aligned}
 &{\mathbf{U}}^{M P} ={\mathbf{U}}_{j}+\operatorname{minmod}\left[{\mathbf{U}}_{i+1}-{\mathbf{U}}_{i}, 4\left({\mathbf{U}}_{i}-{\mathbf{U}}_{i-1}\right)\right]\,\,,\\
\text{and,} &\operatorname{minmod}(a,b) = \frac{1}{2} \left(\operatorname{sign}(a)+\operatorname{sign}(b)\right)\min(|a|,|b|)\,\,,
\end{aligned}
\end{equation}
If the linear scheme fails to satisfy the Equation (\ref{eqn:mp5Condition}) the procedure of the MP limiter described through the following set of equations is applied:
\begin{equation}\label{eqn:mp-procedure}
\begin{aligned}
\mathbf{U}_{i+1 / 2}^{\text {Nonlinear }} &=\mathbf{U}_{i+1 / 2}^{\text {Linear }}+\operatorname{minmod}\left(\mathbf{U}_{i+1 / 2}^{\min }-\mathbf{U}_{i+1 / 2}^{\text {Linear }}, \mathbf{U}_{i+1 / 2}^{\max }-\mathbf{U}_{i+1 / 2}^{\text {Linear }}\right), \\
\mathbf{U}_{i+1 / 2}^{M P} &=\mathbf{U}_{i}+\operatorname{minmod}\left[\mathbf{{U}}_{i+1}-\mathbf{ {U}}_{i}, 4\left(\mathbf{ {U}}_{j}-\mathbf{ {U}}_{i-1}\right)\right], \\
\mathbf{U}_{i+1 / 2}^{\min } &=\max \left[\min \left(\mathbf{ {U}}_{i}, \mathbf{ {U}}_{i+1}, \mathbf{U}_{i+1 / 2}^{M D}\right), \min \left(\mathbf{ {U}}_{i}, \mathbf{U}_{i+1 / 2}^{U L}, \mathbf{U}_{i+1 / 2}^{L C}\right)\right], \\
\mathbf{U}_{i+1 / 2}^{\max } &=\min \left[\max \left(\mathbf{ {U}}_{i}, \mathbf{ {U}}_{i+1}, \mathbf{U}_{i+1 / 2}^{M D}\right), \max \left(\mathbf{ {U}}_{i}, \mathbf{U}_{i+1 / 2}^{U L}, \mathbf{U}_{i+1 / 2}^{L C}\right)\right], \\
\mathbf{U}_{i+1 / 2}^{M D} &=\frac{1}{2}\left(\mathbf{ {U}}_{i}+\mathbf{ {U}}_{i+1}\right)-\frac{1}{2} d_{i+1 / 2}^{M}, \\
\mathbf{U}_{i+1 / 2}^{U L} &=\mathbf{{U}}_{i}+4\left(\mathbf{ {U}}_{i}-\mathbf{ {U}}_{i-1}\right), \\
\mathbf{U}_{i+1 / 2}^{L C} &=\frac{1}{2}\left(3 \mathbf{ {U}}_{i}-\mathbf{ {U}}_{i-1}\right)+\frac{4}{3} d_{i-1 / 2}^{M}, \\
d_{i+1 / 2}^{M} &=\operatorname{minmod}\left(4 d_{i}-d_{i+1}, 4 d_{i+1}-d, d_{i},d_{i+1}\right), \\
d_{i} &=\mathbf{ {U}}_{i-1}-2 \mathbf{ {U}}_{i}+\mathbf{ {U}}_{i+1},
\end{aligned}
\end{equation}
\textcolor{black} {where,
\begin{equation}\label{minmod}
minmod(a,b) = \half \left ( sign(a)+sign(b) \right ) min(|a|,|b|), 
\end{equation}}
The above-mentioned procedure of the MP limiter is the same as that of Suresh and Huynh \cite{suresh1997accurate}. The MP limiter approach is typically used in conjunction with upwind schemes. However, as mentioned earlier, it is used with the central scheme in this paper (for certain waves in characteristic space or some physical variables depending on direction). It will be shown to perform without any issues.
\subsubsection{MUSCL scheme}\label{muscl-sec}
The second reconstruction scheme is the third-order MUSCL developed by van Leer \cite{van1977towards}. The MUSCL scheme uses a three-point stencil for the reconstruction and is more dissipative than the fifth-order upwind scheme presented earlier. However, the scheme is robust and effective in capturing sharp gradients even if reconstruction is carried out using primitive variables. The expressions for the MUSCL scheme are as follows:
\begin{equation}
\begin{aligned}
& \bm{U}_{i+1 / 2}^{R,MUSCL}=\mathbf{{U}}_{i+1}-\frac{1}{4}\left[(1-{\kappa}) \operatorname{minmod}\left(\Delta_{p}, 2 \Delta_{o}\right) +(1+{\kappa})  \operatorname{minmod}\left(\Delta_{o}, 2 \Delta_{p}\right)\right], \\
& \bm{U}_{i+1 / 2}^{L,MUSCL}=\mathbf{{U}}_{i+0}+\frac{1}{4}\left[(1-{\kappa}) \operatorname{minmod}\left(\Delta_{m}, 2 \Delta_{o}\right) +(1+{\kappa}) \operatorname{minmod}\left(\Delta_{o}, 2 \Delta_{m}\right)\right], \\
\end{aligned}\label{mm-5}
\end{equation}
where $\Delta_{p} = \mathbf{{U}}_{i+2} - \mathbf{{U}}_{i+1}$, $\Delta_{o} = \mathbf{{U}}_{i+1} - \mathbf{{U}}_{i}$, and $\Delta_{m} = \mathbf{{U}}_{i} - \mathbf{{U}}_{i-1}$. The $\operatorname{minmod}$ limiter is the same as that of Equation (\ref{minmod}). The value of $\kappa$= $\frac{1}{3}$ in the above Equation, which will lead to third-order accuracy \cite{van1977towards,van2021towards}.
\subsubsection{Interface Capturing scheme (THINC)}
\noindent The third and final candidate reconstruction function is the THINC scheme ($\mathbf{U}^{T}$), a differentiable and monotone Sigmoid function \cite{xiao2011revisit}. Unlike the reconstruction polynomials described above, the THINC scheme is a non-polynomial function. The explicit formula for the left and right interface for the THINC function are as follows \cite{wakimura2022symmetry}:
\begin{equation}\label{THINC}
\textcolor{black}{\mathbf{U}_{i+1 / 2}^{L, T}=\left\{\begin{array}{l}
\textcolor{black}{\mathbf{{u}_{a}}}+\textcolor{black}{\mathbf{{u}_{d}}}\frac{K_1 + (K_2/K_1)}{1+K_2} \text { if }\left({\mathbf{U}}_{i+1}-{\mathbf{U}}_{i}\right)\left({\mathbf{U}}_{i}-{\mathbf{U}}_{i-1}\right)>0, \\
\mathbf{U}_{i} \text { otherwise }.
\end{array}\right.}
\end{equation}

\begin{equation}
\textcolor{black}{\mathbf{U}_{i-1 / 2}^{R, T}=\left\{\begin{array}{ll}
\textcolor{black}{\mathbf{{u}_{a}}}-\textcolor{black}{\mathbf{{u}_{d}}}\frac{K_1 - (K_2/K_1)}{1-K_2} & \text { if }\left({\mathbf{U}}_{i+1}-{\mathbf{U}}_{i}\right)\left({\mathbf{U}}_{i}-{\mathbf{U}}_{i-1}\right)>0, \\
\mathbf{U}_{i} \text { otherwise },
\end{array}\right.}
\end{equation}
where
\begin{eqnarray*}
&&K_1=\tanh\left(\frac{\beta}{2}\right),\ K_2=\tanh\left(\frac{ \textcolor{black}{\bm{\alpha}_i}\beta}{2}\right),\ \textcolor{black}{\bm{\alpha}_i}=\frac{{\mathbf{U}}_{i}-\textcolor{black}{\mathbf{{u}_{a}}}}{\textcolor{black}{\mathbf{{u}_{d}}}},\textcolor{black}{\mathbf{{u}_{a}}}=\frac{{\mathbf{U}}_{i+1}+{\mathbf{U}}_{i-1}}{2},\ \textcolor{black}{\mathbf{{u}_{d}}}=\frac{{\mathbf{U}}_{i+1}-{\mathbf{U}}_{i-1}}{2}.
\end{eqnarray*}
The performance of the THINC scheme is influenced by the steepness parameter, $\beta$, as highlighted in \cite{deng2018high,deng2019fifth}. This parameter controls the thickness of the jump: a smaller value of $\beta$ results in a smoother profile, while a larger value creates a sharp transition. When $\beta$ is set to 1.8, the reconstruction function approximates a step-like profile, resolving discontinuities within approximately three mesh cells \cite{xiao2011revisit}. In this study, $\beta$ is set to 1.8 for all the test cases. The THINC scheme is applied to regions involving material interfaces and contact discontinuities, particularly in characteristic space, using the sensor described in \cite{sainadh2024consistent}, and is as follows:

\begin{equation}\label{psi-mp}
\begin{aligned}
&\psi_{i}=\frac{2ab + \varepsilon}{\left(a^2+b^2+\varepsilon\right)}, \ \text{where}
&\varepsilon=\frac{0.9 \psi_{c}}{1-0.9 \psi_{c}} \xi, \quad \xi=10^{-2}, \quad \psi_{c}=0.35,
\end{aligned}
\end{equation}

\begin{equation}\label{detector-new}
\begin{aligned}
&a  = \frac{13}{12} \left|s_{i-2} - 2 s_{i-1} + s_{i}\right| + \frac{1}{4} \left|s_{i-2} - 4 s_{i-1} + 3 s_{i}\right|,\\
&b  = \frac{13}{12} \left|s_{i} - 2 s_{i+1} + s_{i+2}\right| + \frac{1}{4} \left|3 s_{i} - 4 s_{i+1} + s_{i+2}\right|, \ \text{where} \ s=\frac{p}{\rho^{\gamma}}, \ \text{and}\ \rho=\rho_{1} \alpha_{1}+\rho_{2} \alpha_{2}.
\end{aligned}
\end{equation}

The following subsections will present two algorithms utilizing the reconstruction procedures described above.

\subsection{Algorithm for adaptive primitive-characteristic variable reconstruction:}

Shock-capturing methods for the Euler equations often rely on characteristic variables to achieve non-oscillatory solutions \cite{van2006upwind, chamarthi2021high}. Directly reconstructing interface values using primitive variables, while robust in regions with large density variations (e.g., gas-liquid interfaces), can introduce minor oscillations near shockwaves, especially in high-resolution schemes. Conversely, characteristic variable reconstruction effectively suppresses oscillations near shockwaves \cite{van2006upwind, coralic2014finite} but lacks robustness in regions with significant density jumps, such as gas-liquid interfaces \cite{bryngelson2021mfc}. 

High-order methods excel in capturing vortical structures but may struggle with robustness near gas-liquid interfaces, whereas lower-order methods perform better. Meanwhile, the THINC scheme provides a more accurate representation of gas-gas and gas-liquid material interfaces. An adaptive algorithm that considers each region of the flows as needed can address these conflicting requirements. The following outlines the adaptive primitive-characteristic variable reconstruction algorithm for gas-liquid flows.
\begin{itemize}
	\item if $\pi_{\infty,{i+\frac{1}{2}}}$ $\ge$ 2 (the stiffened gas parameter), carry out reconstruction of primitive variables using the MUSCL and THINC schemes:
\end{itemize}

\begin{equation}
    \mathbf{U}^{L,R}_{i+\frac{1}{2},b} = 
    \left\{
    \begin{array}{ll}
        \text{if } b = 1,2\text{:} & 
        \begin{cases}
{\mathbf{U}}_{i+\frac{1}{2},b}^{L,R, MUSCL}
            \\[20pt]
{\mathbf{U}}_{i+\frac{1}{2},b}^{L,R, T} & \text{if } \min \left(\psi_{i-1}, \psi_{i}, \psi_{i+1}\right)<\psi_{c},
        \end{cases}\\ 
        \\[10pt]
\text{if } b = 3,4,5\text{:} & 
         \begin{cases}
{\mathbf{U}}_{i+\frac{1}{2},b}^{L,R, MUSCL}
        \end{cases}\\
        \\[10pt]
        \text{if } b = 6\text{:} & 
        \begin{cases}
            \mathbf{U}^{L,R, T}_{i+\frac{1}{2},b} .
		\end{cases}
    \end{array}
    \right.
    \label{primitive-2}
\end{equation}

\begin{itemize}
\item Otherwise, carry out reconstruction of characteristic variables, $\mathbf{W}$, using the fifth-order upwind or sixth-order central or THINC schemes, depending on the characteristic variable. The characteristic variables are obtained by multiplying the left eigenvector with the primitive variable vector. The left eigenvector for the five-equation model is as follows:
\begin{equation}\label{matrix-left}
\mathbf{W}=\mathbf{LU}, \text{where}\
\mathbf{L}=\left[\begin{array}{ccccccc}
0 & 0 & - \frac{n_x c \rho}{2} & - \frac{n_y c \rho}{2} & \frac{1}{2} & 0\\
\\
1 & 0 & 0 & 0 & - \frac{\alpha_{1} \rho_{1}}{c^{2} \rho} & 0\\
\\
0 & 1 & 0 & 0 & - \frac{\alpha_{2} \rho_{2}}{c^{2} \rho} & 0\\
\\
0 & 0 & n_y & n_x & 0 & 0\\
\\
0 & 0 & 0 & 0 & 0 & 1\\
\\
0 & 0 & \frac{n_x c \rho}{2} &  \frac{n_y c \rho}{2} & \frac{1}{2} & 0
\end{array}\right], \text{and} \ \mathbf{U}=\left[\begin{array}{c}
\alpha_1 \rho_1 \\
\\
\alpha_2 \rho_2 \\
\\
u \\
\\
v \\
\\
p \\
\\
\alpha_1
\end{array}\right].
\end{equation}
where $\bm{n}$ = $[n_x \ n_y]^t$ and $[l_x \ l_y]^t$ is a tangent vector (perpendicular to $\bm{n}$) such as $[l_x \ l_y]^t$ = $[-n_y \ n_x]^t$. By taking $\bm{n}$ = $[1, 0]^t$ and $[0, 1]^t$ we obtain the corresponding eigenvectors in $x-$ and $y-$ directions. The rest of the algorithm is as follows (only left interface values are shown here for brevity):
\end{itemize}
\begin{equation}
    \mathbf{W}^{L}_{i+\frac{1}{2},b} = 
    \left\{
    \begin{array}{ll}
         \text{if } b = 1,6\text{:} & \begin{cases}
           \mathbf{W}^{L,Non-Linear}_{i+\frac{1}{2},b} & \text{if } \left( \mathbf{W}^{L,Linear}_{i+\frac{1}{2}} - \mathbf{W}_i \right) \left( \mathbf{W}^{L,Linear}_{i+\frac{1}{2}} - \mathbf{W}^{L,MP}_{i+\frac{1}{2}} \right) \geq 10^{-40},
            \\[20pt]
            \mathbf{W}^{L,Linear}_{i+\frac{1}{2},b} & \text{otherwise}.
        \end{cases}\\
        \\[20pt]
        \text{if } b = 2,3\text{:} & \begin{cases}
            \mathbf{W}^{L,Non-Linear}_{i+\frac{1}{2},b} & \text{if } \left( \mathbf{W}^{L,Linear}_{i+\frac{1}{2}} - \mathbf{W}_i \right) \left( \mathbf{W}^{L,Linear}_{i+\frac{1}{2}} - \mathbf{W}^{L,MP}_{i+\frac{1}{2}} \right) \geq 10^{-40},
            \\[20pt]
            \mathbf{W}^{L,Linear}_{i+\frac{1}{2},b} & \text{otherwise}.
            \\[20pt]
{\bm{W}}_{i+\frac{1}{2},b}^{L, T} & \text{if } \min \left(\psi_{i-1}, \psi_{i}, \psi_{i+1}\right)<\psi_{c},
        \end{cases}\\ 
        \\[20pt]
        \text{if } b = 5\text{:} & \begin{cases}
            \mathbf{W}^{L,T}_{i+\frac{1}{2},b} .
        \end{cases}
    \end{array}
    \right.
    \label{eqn:contact}
\end{equation}
\begin{equation}
    \mathbf{W}^{L}_{i+\frac{1}{2},b} = 
    \left\{
    \begin{array}{ll}
        \text{if } b = 4\text{:} & \begin{cases}
           \mathbf{W}^{C,Non-Linear}_{i+\frac{1}{2},b} & \text{if } \left( \mathbf{W}^{C,Linear}_{i+\frac{1}{2}} - \mathbf{W}_i \right) \left( \mathbf{W}^{C,Linear}_{i+\frac{1}{2}} - \mathbf{W}^{C,MP}_{i+\frac{1}{2}} \right) \geq 10^{-40},
            \\[20pt]
            \mathbf{W}^{C,Linear}_{i+\frac{1}{2},b} & \text{otherwise}.
        \end{cases}\\
    \end{array}
    \right.
    \label{eqn:contact_2}
\end{equation}
After obtaining ${\mathbf{W}}_{i+\frac{1}{2}}^L$ and ${\mathbf{W}}_{i+\frac{1}{2}}^R$ from the above algorithm the primitive variables are then recovered by projecting the characteristic variables back to physical fields: \\
\begin{equation}
\begin{aligned}
	{\mathbf{U}}_{i+\frac{1}{2}}^{L} &= \bm{R}_{\bm{n}_{i+\frac{1}{2}}} {\mathbf{W}}_{i+\frac{1}{2}}^L, \\[10pt]
  {\mathbf{U}}_{i+\frac{1}{2}}^{R} &= \bm{R}_{\bm{n}_{i+\frac{1}{2}}} {\mathbf{W}}_{i+\frac{1}{2}}^R.
\end{aligned}
\end{equation}

The details of each component of the proposed algorithm are explained below:
\begin{itemize} 
	\item \textbf{Reconstruction of Primitive and Characteristic Variables:} In the algorithm, primitive variables are reconstructed in liquid regions, identified where $\pi_{\infty,{i+\frac{1}{2}}} \geq 2$ (calculated as the arithmetic average of $\pi_{\infty}$ at ${i+\frac{1}{2}}$). Typically, $\pi_{\infty,{i+\frac{1}{2}}}$ is significantly large in liquid regions, often exceeding 1000, while it is nearly zero in gaseous regions. This distinction enables reliable identification of liquid regions. Characteristic variables are reconstructed elsewhere, ensuring robustness for flows involving shocks and liquid interfaces while minimizing oscillations near shocks. Although the volume fraction ($\alpha_1$) could be used to identify interfaces, this approach would also reconstruct primitive variables for gas-gas interfaces, which is unnecessary—previous simulations, such as those in Refs. \cite{chamarthi2023gradient,sainadh2024consistent}, successfully reconstructed characteristic variables for gas-gas flows without issues. The challenges arise only near the gas-liquid interfaces due to the significant jump in densities.\\
	\item \textbf{Pressure, Velocity, and Density in Physical Variable Space:} In the physical variable space (Equations \ref{primitive-2}), pressure, velocity, and density (away from contact discontinuities and material interfaces) are computed using the upwind-biased low-order MUSCL scheme and ensures stability and accuracy in regions dominated by liquid.\\
	\item \textbf{Choice of Reconstruction Schemes:} The algorithm employs a low-order scheme (MUSCL/THINC, Equations (\ref{primitive-2})) in liquid regions and a high-order scheme (MP/THINC, Equations (\ref{eqn:contact}) and (\ref{eqn:contact_2})) in gas regions. Liquid regions characterized by high densities benefit from the robustness of the MUSCL/THINC scheme, which outperforms high-order MP/THINC schemes in such cases. While the MUSCL scheme could be applied to gas regions, it is more dissipative than the MP scheme. However, the MP scheme is less robust in liquid regions (in characteristic space) and requires additional positivity-preserving techniques \cite{meng2018numerical, bryngelson2021mfc}, making it unsuitable for such scenarios.\\
	\item \textbf{Reconstruction of Phasic Densities and Volume Fractions:} Phasic densities are reconstructed using the THINC scheme in both physical (${\mathbf{U}}_{1,2}$) and characteristic spaces (entropy waves, ${\mathbf{W}}_{2,3}$) near the material interfaces and contact discontinuities. Similarly, volume fractions are consistently computed using the THINC scheme in physical (${\mathbf{U}}_{6}$) and characteristic spaces (${\mathbf{W}}_{5}$). The physical reason behind such an approach is that across gas-gas and gas-liquid interfaces, only phasic densities and volume fractions are discontinuous, and the rest of the variables are continuous. Contact discontinuities (density jump within a material) are also computed using the THINC scheme as the sensor, given by Equation (\ref{psi-mp}), can detect them. In \cite{harten1989eno}, Harten highlighted that ENO schemes excel in resolving shocks and apply subcell resolution selectively to linearly degenerate characteristic fields to enhance the resolution of contact discontinuities. Similarly, this study focuses on refining the resolution of contact discontinuities and material interfaces, as the polynomial-based schemes already provide effective shock resolution. It is also possible to compute the phasic densities and entropy waves away from the contact discontinuities using a central scheme and the special case will be shown in Example \ref{shock-entropy}.\\
	\item When primitive variables are reconstructed, in some test cases, even the phasic densities are computed using the THINC if the material interface detector, Equation (\ref{psi-mp}), detects the shocks (sometimes it detects and sometimes it does not, but it consistently detects the material interfaces, which is the target). It is physically consistent, as density is discontinuous across shockwaves. Applying THINC to all the variables is inconsistent, as shockwaves and contact discontinuities have different physical characteristics; for example, pressure is continuous across contact discontinuity but discontinuous across a shock. The current approach is in contrast, although physically consistent, with the recently proposed TENO-THINC schemes \cite{takagi2022novel,li2024high} where all the variables are computed using the THINC scheme when any discontinuity is detected. The numerical results section shows the advantages of the current approach and limitations of the TENO-THINC schemes \cite{takagi2022novel,li2024high}. It will be shown that even TENO-THINC schemes will benefit from considering the appropriate physics of the target equations.\\
	\item \textbf{Computation of Acoustic and Vorticity Waves:} Acoustic waves (${\mathbf{W}}_{1,6}$) are computed using a fifth-order upwind scheme, while vorticity waves (${\mathbf{W}}_{4}$), representing tangential velocities in physical space, are computed using a sixth-order central scheme. Tangential velocities are continuous across shockwaves \cite{hirschvol2} and remain continuous across contact discontinuities and material interfaces in the presence of viscosity—whether artificial \cite{meng2018numerical} or physical \cite{batchelor1967introduction}. The tangential velocities in each direction are automatically computed using a central scheme, as evident from the analysis of the matrix $\mathbf{L}$ in Equation (\ref{matrix-left}). The Ducros sensor \cite{hoffmann2024centralized,ducros1999large}, used for shock capturing and centralization of vorticity waves in \cite{sainadh2024consistent}, lacked robustness for multiphase flow simulations, necessitating a simpler alternative.\\
	\item \textbf{Algorithm Stability and Limitations:} A minor inconsistency in the algorithm is that tangential velocities are computed using the upwind MUSCL scheme in the primitive variable space rather than the central scheme used in characteristic space. This choice prioritizes algorithm stability and will be addressed in subsequent primitive variable algorithm approach.
\item The above-discussed algorithm is denoted as \textbf{Wave-MP} in the rest of the paper. MP5 and MUSCL schemes are the standard reconstruction schemes described in sections \ref{mp5-sec} and \ref{muscl-sec}, denoted as such. It is also possible to use the MUSCL scheme instead of the MP scheme in characteristic space, and the scheme is denoted as \textbf{Wave-MUSCL} in the rest of the paper. Lastly, in the characteristic space, the vorticity waves can be computed using the upwind scheme and the corresponding method is denoted as \textbf{MP5-THINC}. The advantages of Wave-MP over Wave-MUSCL and MP5-THINC will be shown through examples.
\end{itemize}

\subsection{Algorithm for primitive variable reconstruction:}

 The adaptive primitive-characteristic variable reconstruction presented above requires three different reconstructions and is slightly complex (but necessary to avoid oscillations near shocks and is also robust). In contrast, primitive variable reconstruction is straightforward and is as follows: 
 
In all directions:
\begin{equation}
    {\bm{U}}^{L}_{i+\frac{1}{2},b} = 
    \left\{
    \begin{array}{ll}
         \text{if } b = 5\text{:} & 
         \begin{cases}
           {\bm{U}}^{L,Non-Linear}_{i+\frac{1}{2},b} & \text{if } \left( {\bm{U}}^{L,Linear}_{i+\frac{1}{2}} - {\bm{U}}_i \right) \left( {\bm{U}}^{L,Linear}_{i+\frac{1}{2}} - {\bm{U}}^{L,MP}_{i+\frac{1}{2}} \right) \geq 10^{-40},
            \\[20pt]
           {\bm{U}}^{L,Linear}_{i+\frac{1}{2},b} & \text{otherwise}.
        \end{cases}\\
        \\[20pt]
        \text{if } b = 1,2\text{:} & 
        \begin{cases}
            {\bm{U}}^{L,Non-Linear}_{i+\frac{1}{2},b} & \text{if } \left( {\bm{U}}^{L,Linear}_{i+\frac{1}{2}} - {\bm{U}}_i \right) \left( {\bm{U}}^{L,Linear}_{i+\frac{1}{2}} - {\bm{U}}^{L,MP}_{i+\frac{1}{2}} \right) \geq 10^{-40},
            \\[20pt]
            {\bm{U}}^{L,Linear}_{i+\frac{1}{2},b} & \text{otherwise}.
            \\[20pt]
			{\bm{U}}_{i+\frac{1}{2},b}^{L, T} & \text{if } \min \left(\psi_{i-1}, \psi_{i}, \psi_{i+1}\right)<\psi_{c}.
        \end{cases}\\ 
        \\[20pt]
        \text{if } b = 6\text{:} & 
        \begin{cases}
            {\bm{U}}^{L,T}_{i+\frac{1}{2},b} .
        \end{cases}
    \end{array}
    \right.
    \label{eqn:prim}
\end{equation}

In $x$-direction:
\begin{equation}\label{eqn:centralScheme_x}
    \mathbf{U}^{L}_{i+\frac{1}{2},b} = 
    \left\{
    \begin{array}{ll}
        \text{if } b = 3\text{:} & \begin{cases}
           {\bm{U}}^{L,Non-Linear}_{i+\frac{1}{2},b} & \text{if } \left( {\bm{U}}^{L,Linear}_{i+\frac{1}{2}} - {\bm{U}}_i \right) \left( {\bm{U}}^{L,Linear}_{i+\frac{1}{2}} - {\bm{U}}^{L,MP}_{i+\frac{1}{2}} \right) \geq 10^{-40},
            \\[20pt]
            {\bm{U}}^{L,Linear}_{i+\frac{1}{2},b} & \text{otherwise}.
        \end{cases}
        \end{array}
    \right.
\end{equation}
\begin{equation}
    \mathbf{U}^{L}_{i+\frac{1}{2},b} = 
    \left\{
    \begin{array}{ll}
        \text{if } b = 4\text{:} & \begin{cases}
           \mathbf{U}^{C,Non-Linear}_{i+\frac{1}{2},b} & \text{if } \left( \mathbf{U}^{C,Linear}_{i+\frac{1}{2}} - \mathbf{U}_i \right) \left( \mathbf{U}^{C,Linear}_{i+\frac{1}{2}} - \mathbf{U}^{C,MP}_{i+\frac{1}{2}} \right) \geq 10^{-40},
            \\[20pt]
            \mathbf{U}^{C,Linear}_{i+\frac{1}{2},b} & \text{otherwise}.
        \end{cases}\\

    \end{array}
    \right.
    \label{eqn:contact_px}
\end{equation}
In $y$-direction:
\begin{equation}\label{eqn:centralScheme_y}
    \mathbf{U}^{L}_{i+\frac{1}{2},b} = 
    \left\{
    \begin{array}{ll}
        \text{if } b = 4\text{:} & \begin{cases}
           {\bm{U}}^{L,Non-Linear}_{i+\frac{1}{2},b} & \text{if } \left( {\bm{U}}^{L,Linear}_{i+\frac{1}{2}} - {\bm{U}}_i \right) \left( {\bm{U}}^{L,Linear}_{i+\frac{1}{2}} - {\bm{U}}^{L,MP}_{i+\frac{1}{2}} \right) \geq 10^{-40},
            \\[20pt]
            {\bm{U}}^{L,Linear}_{i+\frac{1}{2},b} & \text{otherwise}.
        \end{cases}\\        
        \end{array}
    \right.
\end{equation}
\begin{equation}
    \mathbf{U}^{L}_{i+\frac{1}{2},b} = 
    \left\{
    \begin{array}{ll}

        \text{if } b = 3\text{:} & \begin{cases}
           \mathbf{U}^{C,Non-Linear}_{i+\frac{1}{2},b} & \text{if } \left( \mathbf{U}^{C,Linear}_{i+\frac{1}{2}} - \mathbf{U}_i \right) \left( \mathbf{U}^{C,Linear}_{i+\frac{1}{2}} - \mathbf{U}^{C,MP}_{i+\frac{1}{2}} \right) \geq 10^{-40},
            \\[20pt]
            \mathbf{U}^{C,Linear}_{i+\frac{1}{2},b} & \text{otherwise}.
        \end{cases}\\

    \end{array}
    \right.
    \label{eqn:contact_py}
\end{equation}

The details of the primitive variable algorithm are explained below:

\begin{itemize}
	\item MUSCL scheme is not used in this algorithm, and all the variables, except volume fractions, are computed using either the upwind or central schemes (either linear or non-linear) away from the contact discontinuities and material interfaces. In the primitive algorithm, $\alpha_1 \rho_1$ and $\alpha_2 \rho_2$ (${\mathbf{U}}_{1,2}$)  are reconstructed with the THINC scheme if the discontinuity detector detects a discontinuity in physical space. Volume fractions are computed with the THINC, similar to the adaptive primitive-characteristic variable reconstruction algorithm.
	\item While the tangential velocities in each direction are automatically computed using a central scheme in characteristic space, because of the eigenvector matrix, they must be computed individually according to the direction. In the $x-$ direction, $v$ ($\mathbf{U_4}$) is reconstructed using a central scheme if the MP sensor criterion is satisfied. Likewise, in the $y-$ direction, $ u$ ($\mathbf{U_3}$) is reconstructed using a central scheme if the MP sensor criterion is satisfied. 
\item Unlike in the adaptive primitive-characteristic approach, the tangential velocities are computed using a central scheme even across the gas-liquid interface if the MP criterion is met. They are continuous according to the \cite{batchelor1967introduction}. It is challenging to compute them \textit{always} using a central scheme when there are shockwaves in the simulations. The current approach uses a dimension-by-dimension approach, and the shockwaves are not always aligned with the grid. Even though tangential velocities are continuous, normal velocities are discontinuous across a shockwave. If a shockwave is at an angle with the grid, there will be oscillations if a limiter is not applied. 
\item Finally, as explained in the introduction, there can be many free parameters ($\eta$, different for each variable in each direction) if one can go beyond the linear advection equation. Therefore, the variables can be treated with different reconstruction techniques (upwind or central) depending on the direction. It can be exemplified through the primitive variable algorithm reconstruction. Considering only velocities, where the superscripts denoted the directions, the following equations can be written:
\begin{equation*}
\begin{aligned}[t]
u^{x}_{i+\frac{1}{2}} &= \eta_1  u^{x, L}_{i+\frac{1}{2}} + \left( 1 - \eta_1 \right) u^{x, R}_{i+\frac{1}{2}},\\
v^{x}_{i+\frac{1}{2}} &= \eta_2 v^{x, L}_{i+\frac{1}{2}} +   \left( 1 - \eta_2 \right) v^{x, R}_{i+\frac{1}{2}},\\
\end{aligned}
\qquad 
\begin{aligned}[t]
u^{y}_{j+\frac{1}{2}} &=   \eta_3 u^{y, L}_{j+\frac{1}{2}} + \left( 1 - \eta_3 \right) u^{y, R}_{j+\frac{1}{2}},\\
v^{y}_{j+\frac{1}{2}} &=  \eta_4 v^{y, L}_{j+\frac{1}{2}} +  \left( 1 - \eta_4 \right)  v^{y, R}_{j+\frac{1}{2}}.
\end{aligned}
\end{equation*} 
A scheme based on linear advection, such as the MP5 scheme, inherently maintains fifth-order accuracy with an upwind bias determined by a single parameter, $\eta$. In the proposed algorithm, however, the reconstruction approach adapts to the physics of the flow. For instance, $u^{x}$ can employ an upwind reconstruction, while $u^{y}$ can utilize a central reconstruction. $\eta_1$ can be 0.5 and $\eta_3$ can be one, likewise for $v$ velocities. This flexibility may also ensure consistency with the underlying physics. By incorporating this adaptability, considering various physical characteristics of the variables, and drawing parallels to Roe's multi-dimensional upwinding approach \cite{roe1986discrete}, the methodology is aptly named the wave-appropriate (or physics-consistent) multi-dimensional upwinding approach. The primitive variable algorithm is denoted as \textbf{Wave-MP (Prim)} in the rest of the paper. The following section tests the proposed algorithms (Wave-MP and Wave-MP (Prim)) for benchmark cases and experimental results to assess their performance and showcase their advantages.
\end{itemize}

\newpage

\section{Results and discussion}\label{results}

\begin{example}\label{ex:liquid}{Gas-liquid Riemann problem}
\end{example}
The focus of this test is the gas-liquid Riemann problem, which involves a shock tube scenario. In this test case, the left state features highly compressed air, while the right state is water at atmospheric pressure. The non-dimensional initial conditions for this problem are as follows:
\begin{equation}
\left(\rho_l \alpha_l, \rho_g \alpha_g, u, p\right)= \begin{cases}\left(0,1.241, 0,2.753\right) & -1<x<0 \\ \left(0.991, 0,0,3.059 \times 10^{-4}\right) & 0 \leq x \leq 1,\end{cases}
\end{equation}
with fluid properties $\gamma_l=5.5, \pi_{\infty, l}=1.505, \gamma_g=1.4, \pi_{\infty, g}=0$, as in \cite{garrick2017interface}. The simulation is conducted on a grid of 200 points with a constant CFL number of 0.4 until a final time of 0.2.  Figure \ref{fig_zhang} shows the results obtained by the adaptive primitive-characteristic variable reconstruction schemes, and Figure \ref{fig_zhang_p} shows the results obtained by the primitive variable algorithm, respectively.
\begin{figure}[H]
\centering
\subfigure[Density]{\includegraphics[width=0.43\textwidth]{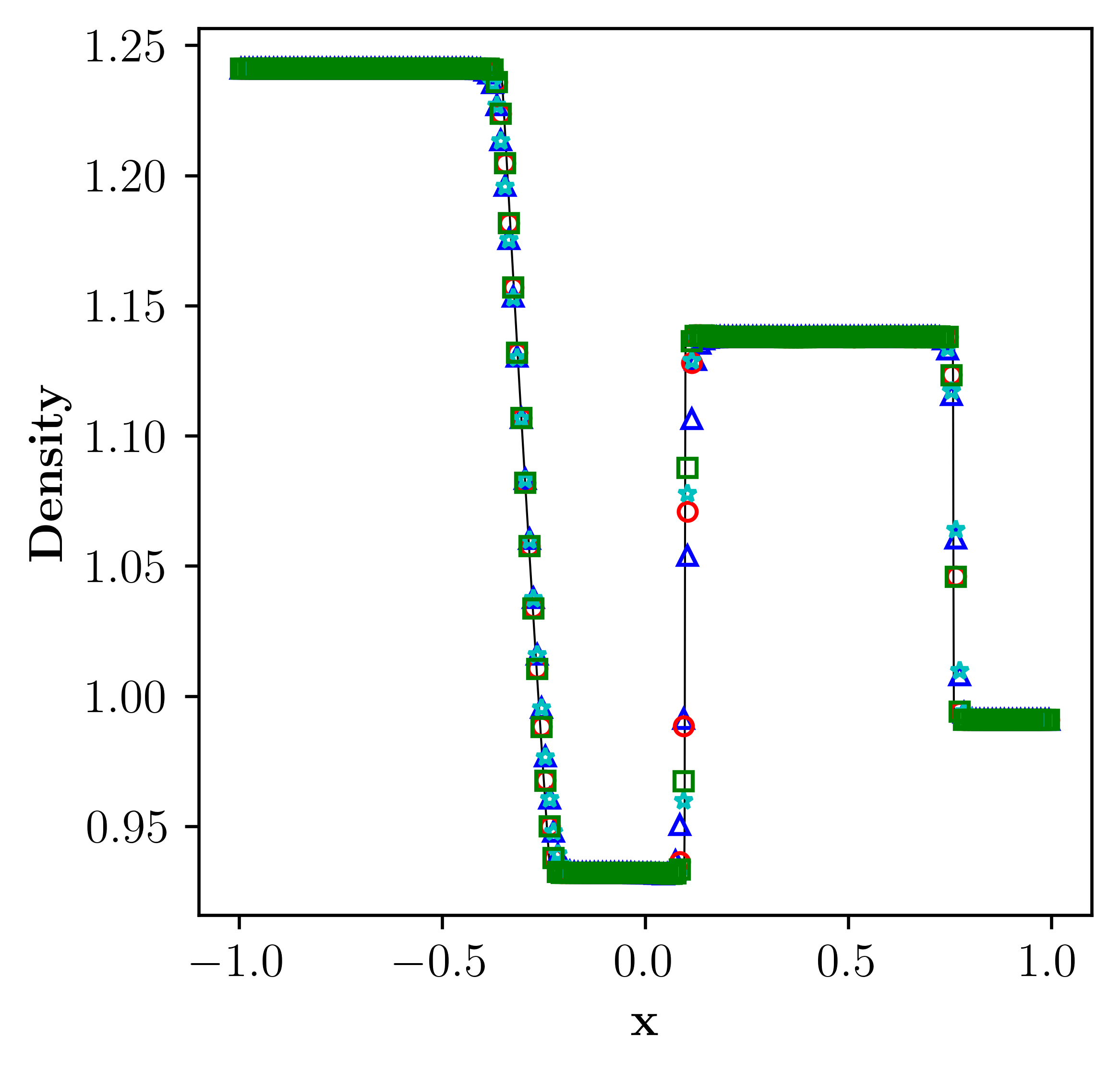}
\label{fig:multi_stiff-den}}
\subfigure[Local density]{\includegraphics[width=0.415\textwidth]{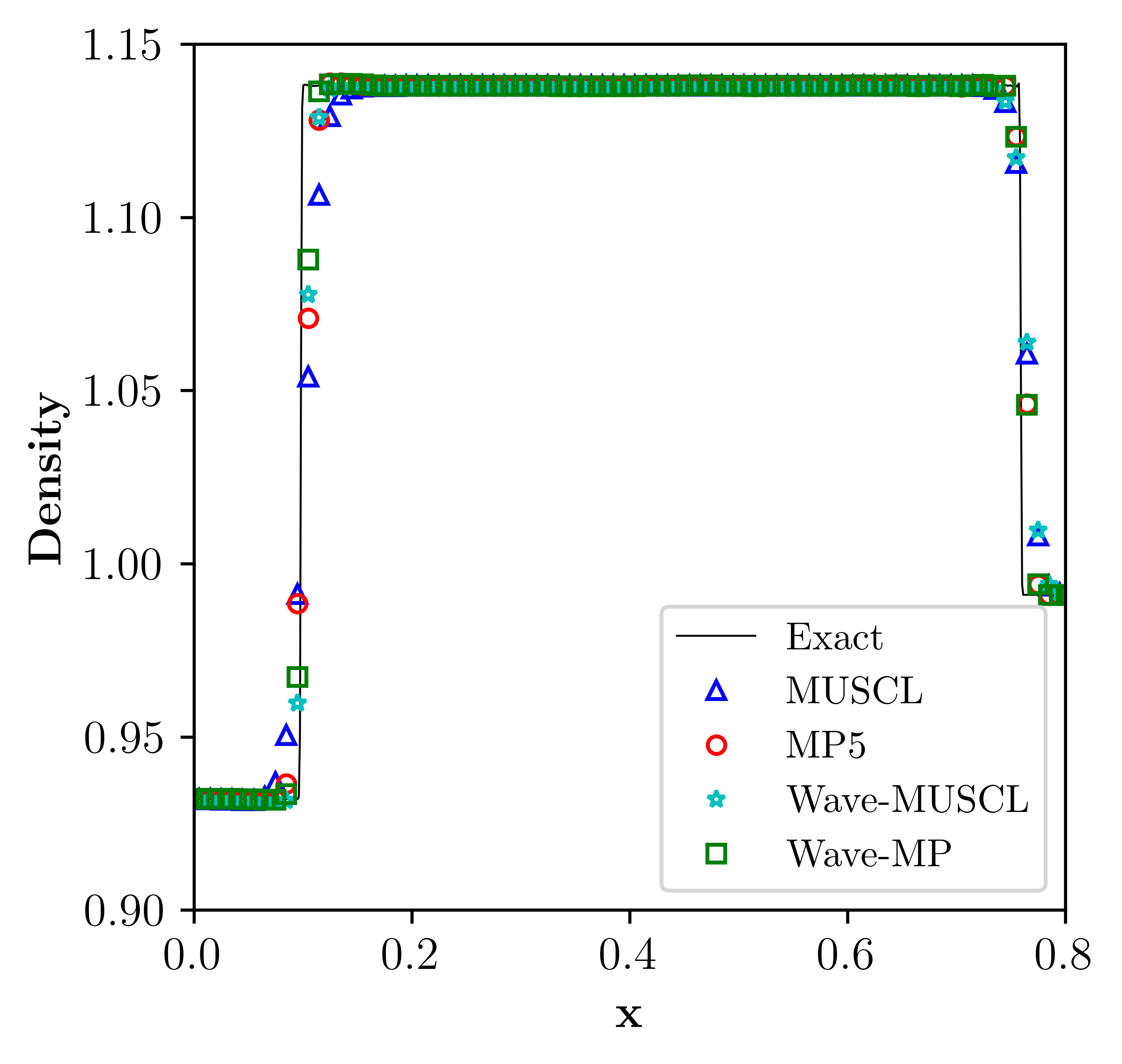}
\label{fig:multi_stiff-pres}}
\subfigure[Pressure]{\includegraphics[width=0.43\textwidth]{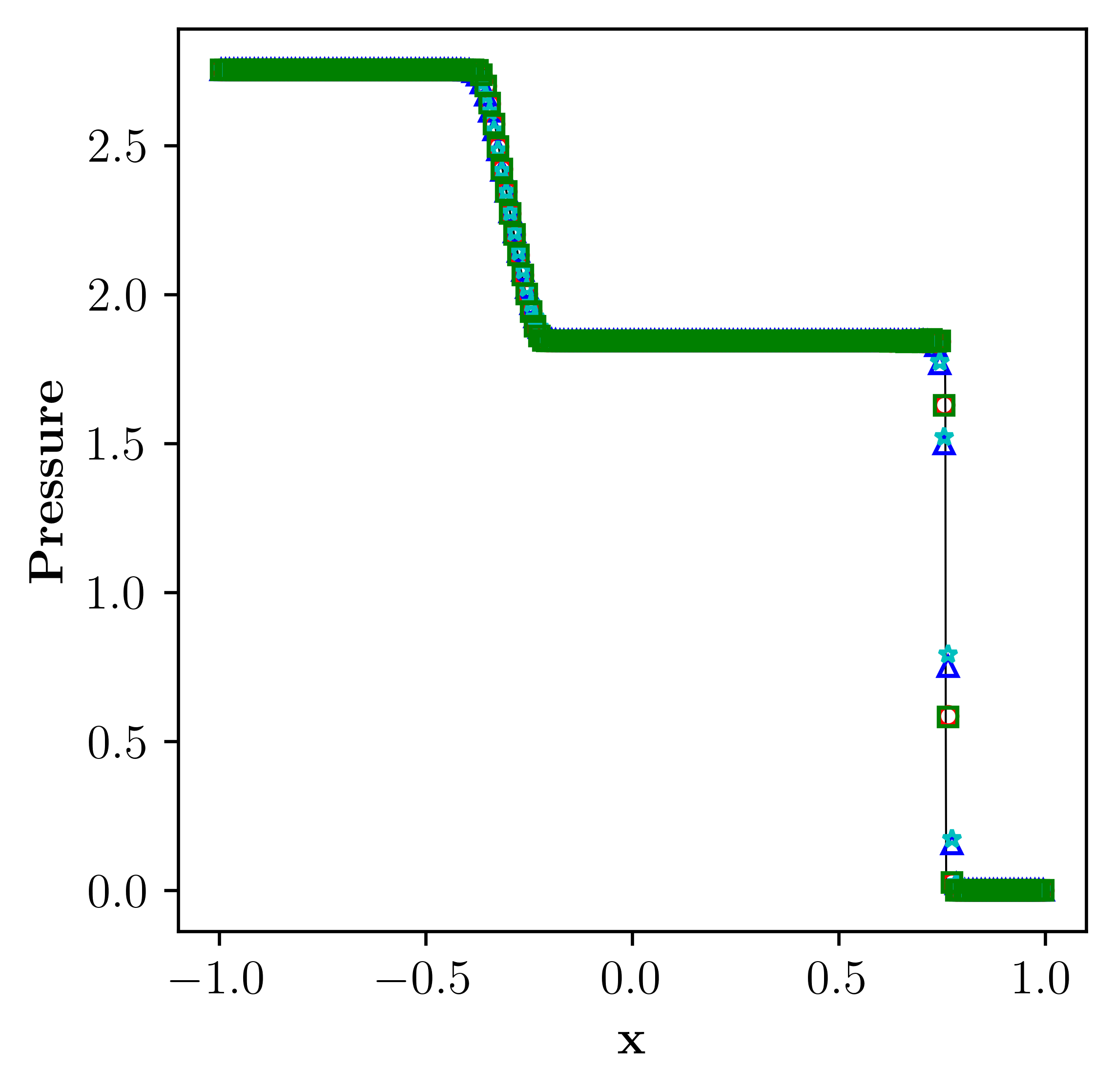}
\label{fig:multi_stiff-alpha}}
\subfigure[Volume fraction, $\alpha_1$]{\includegraphics[width=0.43\textwidth]{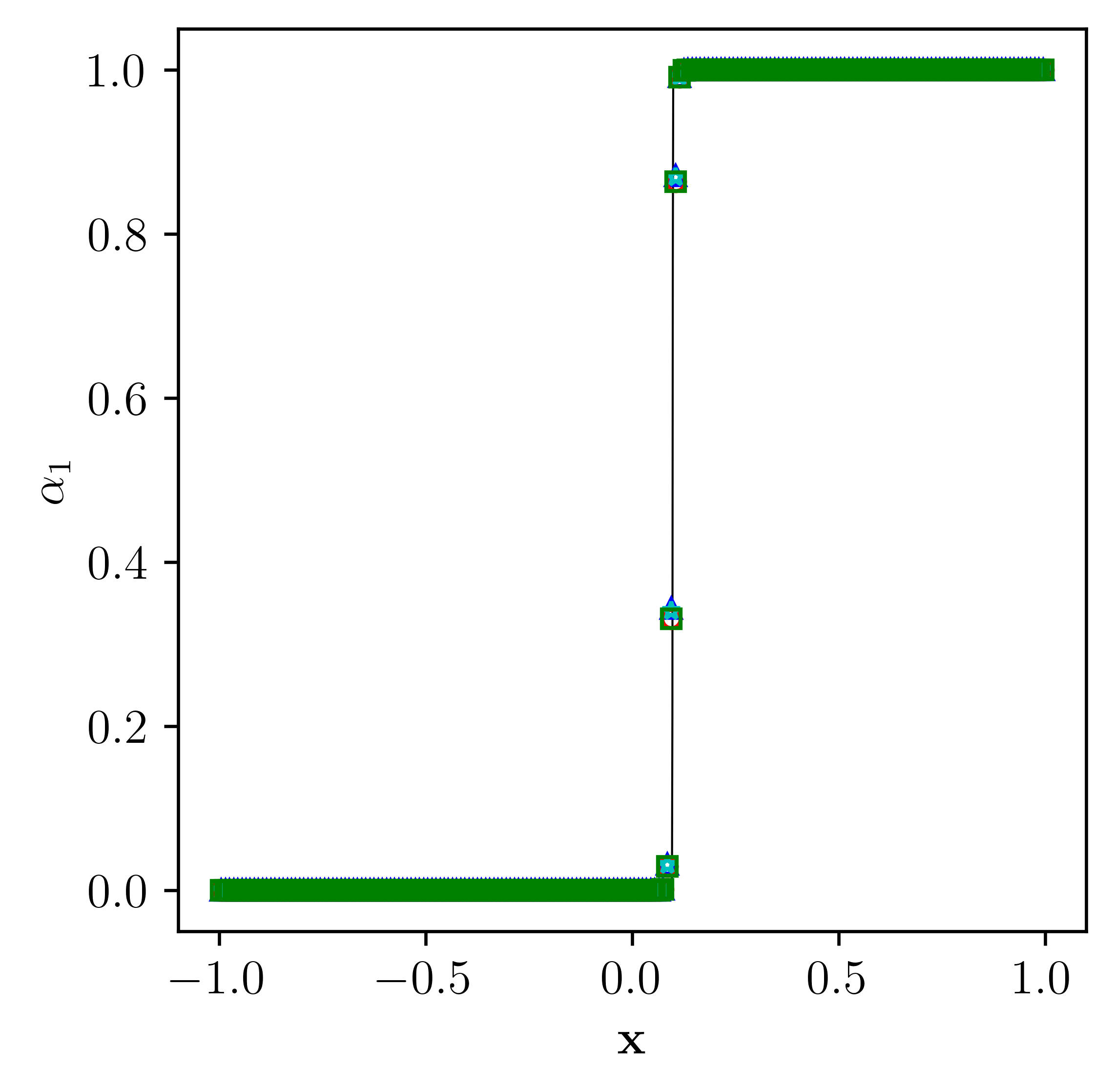}
\label{fig:multi_sod-ent}}
\caption{Numerical solution for shock interface interaction problem in Example \ref{ex:liquid}  on a grid size of $N=200$. Solid line: reference solution; blue triangles: MUSCL; red circles: MP5; cyan stars: Wave-MUSCL and green squares: Wave-MP.}
\label{fig_zhang}
\end{figure}

\begin{figure}[H]
\centering
\subfigure[Density (Prim)]{\includegraphics[width=0.43\textwidth]{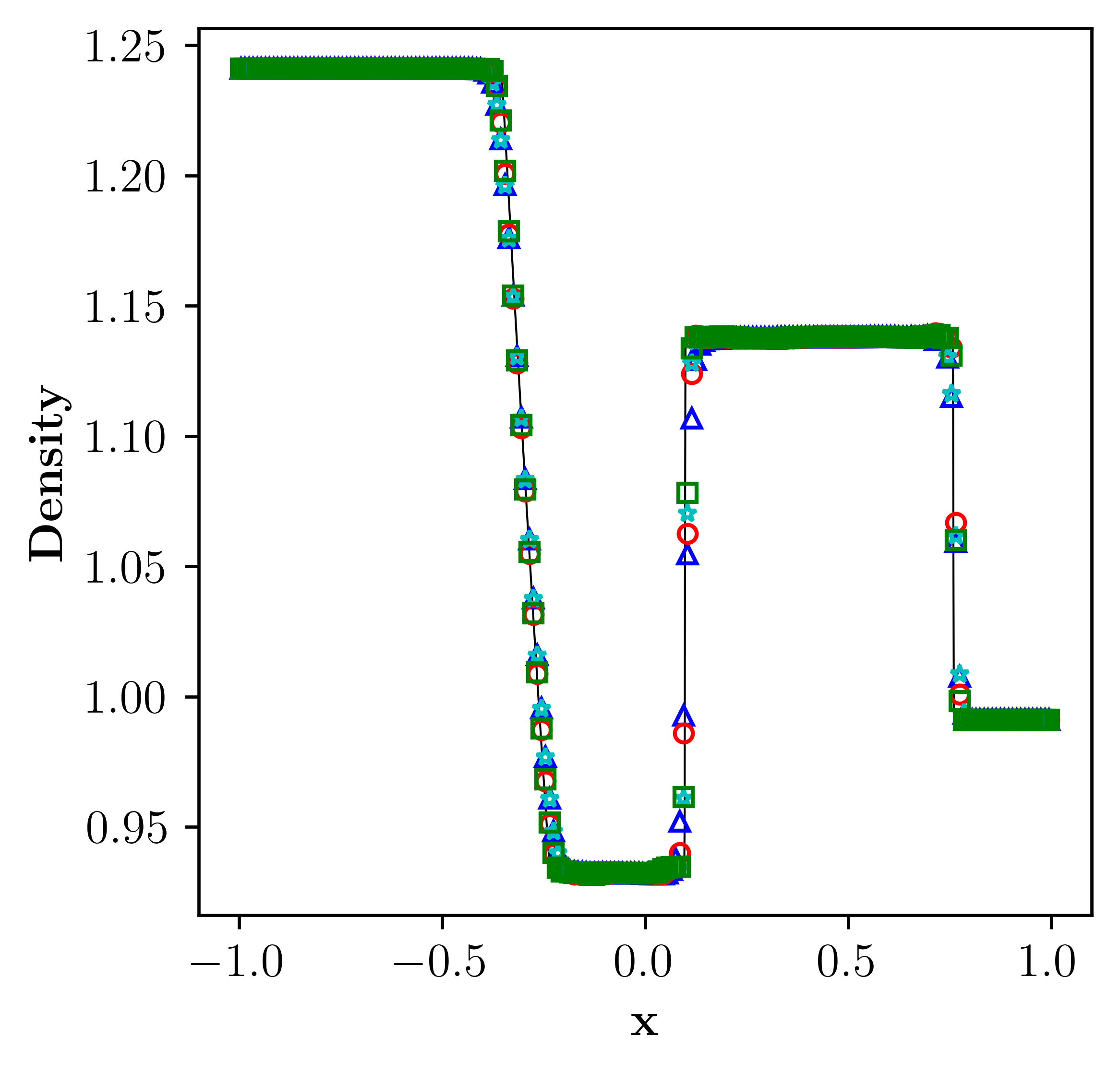}
\label{fig:multi_stiff-den_p}}
\subfigure[Local density (Prim)]{\includegraphics[width=0.415\textwidth]{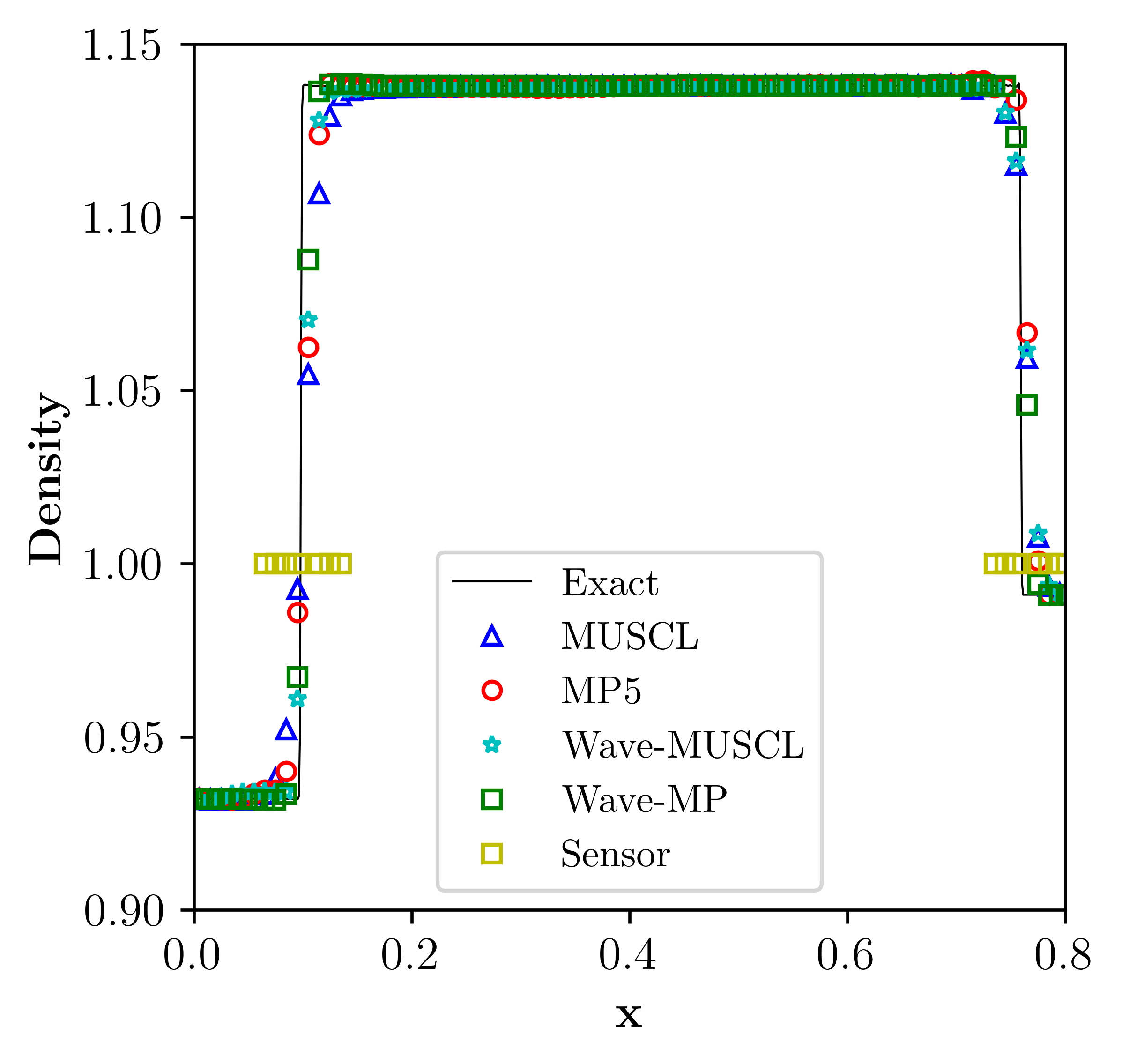}
\label{fig:multi_stiff-pres_p}}
\caption{Numerical solution for shock interface interaction problem in Example \ref{ex:liquid}  on a grid size of $N=200$ using \textbf{primitive variables}. Solid line: reference solution; blue triangles: MUSCL; red circles: MP5; cyan stars: Wave-MUSCL and green squares: Wave-MP.}
\label{fig_zhang_p}
\end{figure}

%The simulation is carried out on a grid of 200 points with a constant CFL of 0.4 until a final time of 0.2. The proposed method is free of oscillations in density and volume fractions, as demonstrated in Figure \ref{fig_zhang}. The Wave-MUSCL scheme is less dissipative than the MUSCL scheme because it captures the material interface with fewer points using THINC near material interface, and the sensor accurately identifies the contact discontinuity. The Wave-MP scheme, being similar to that of Wave-MUSCL near the material interface, on the other hand, is less dissipative (two red circles) for the shockwave than the Wave-MUSCL scheme (four cyan stars) (x$\approx$0.75), which shows the advantages of the adaptive approach (Wave-MUSCL for liquids and Wave-MP for gases). Figure \ref{fig_zhang_p} shows the results computed by the primitive variable algorithm, and even here, the material interface is captured within a few points as opposed to the schemes that do not use THINC. Unlike the adaptive primitive-characteristic variable reconstruction approach, the direct primitive variable reconstruction algorithm has small oscillations. These oscillations will be much more pronounced for the multidimensional cases.

The proposed method is free of oscillations in density and volume fractions, as Figure \ref{fig_zhang} demonstrates. Compared to the MUSCL scheme, the Wave-MUSCL scheme is less dissipative, capturing the material interface with fewer points due to the use of THINC near the interface and an accurate sensor for identifying contact discontinuities. Similarly, the Wave-MP scheme performs identically to the Wave-MUSCL scheme near the material interface. However, it exhibits lower dissipation for the shockwave (evidenced by two red circles at x$\approx$0.75, compared to four cyan stars for Wave-MUSCL). This highlights the advantages of the adaptive approach, which employs Wave-MUSCL for liquids and Wave-MP for gases. Figure \ref{fig_zhang_p} presents the results obtained using the primitive variable algorithm, which also captures the material interface within a few points, unlike schemes that do not utilize THINC. However, the direct primitive variable reconstruction algorithm introduces small oscillations, unlike the adaptive primitive-characteristic variable reconstruction approach. These oscillations are expected to become more pronounced in multidimensional cases.

\begin{example}\label{ex:isol}{Material interface advection}
\end{example}
This one-dimensional, two-species problem involves the advection of an isolated material interface, as described in \cite{deng2018high}. It simulates the transport of a water block through air at a constant velocity. The initial conditions for this test case are as follows:
\begin{equation}
\begin{aligned}
&\begin{array}{llllcc}
\hline x(\mathrm{~m}) & \rho_1 \alpha_1\left(\mathrm{kgm}^{-3}\right) & \rho_2 \alpha_2\left(\mathrm{kgm}^{-3}\right) & u\left(\mathrm{~ms}^{-1}\right) & p(\mathrm{~Pa}) & \alpha_1 \\
\hline 0.25 \leq x<0.75 & 1000.0 & 1.0 \times 10^{-8} & 100.0 & 101325.0 & 1.0-10^{-8} \\
\text { Otherwise } & 1.0 \times 10^{-8} & 1.2 & 100.0 & 101325.0 & 1.0 \times 10^{-8} \\
\hline
\end{array}
\end{aligned}
\end{equation}
with fluid properties $\gamma_l=4.4, \pi_{\infty, l}=6 \times 10^{8}, \gamma_g=1.4, \pi_{\infty, g}=0$. The simulation was performed on a computational domain extending from $x = 0$ to $x = 1$, with $N = 200$ uniformly distributed grid points, as in \cite{deng2018high}, and the final simulation time is $t = 0.1$. Periodic boundary conditions were applied at both ends of the domain.

Figure \ref{fig_iso-ml-p} compares the exact solution with the numerical solutions obtained by various schemes. All the numerical schemes successfully captured the material interface without introducing spurious oscillations. The Wave-MUSCL scheme achieved this using fewer points than the MUSCL scheme, demonstrating the sensor's ability to detect the material interface reliably. For this test case, there is no discernible difference between the Wave-MUSCL and Wave-MP schemes, as previously discussed, as the Wave-MUSCL approach is activated in liquid regions.

The MP5 scheme, being high-order, captured the interface with fewer points than the MUSCL scheme but exhibited more dissipation compared to schemes employing THINC for interface resolution. As shown in Figure \ref{fig:iso-met_p}, pressure remained constant without oscillations. Additionally, the results obtained using the primitive variable reconstruction, shown in Figure \ref{fig:iso-met2_prim}, closely matched those of the adaptive primitive-characteristic variable reconstruction approach.

\begin{figure}[H]
\centering
\subfigure[Density profile, Example \ref{ex:isol}]{\includegraphics[width=0.45\textwidth]{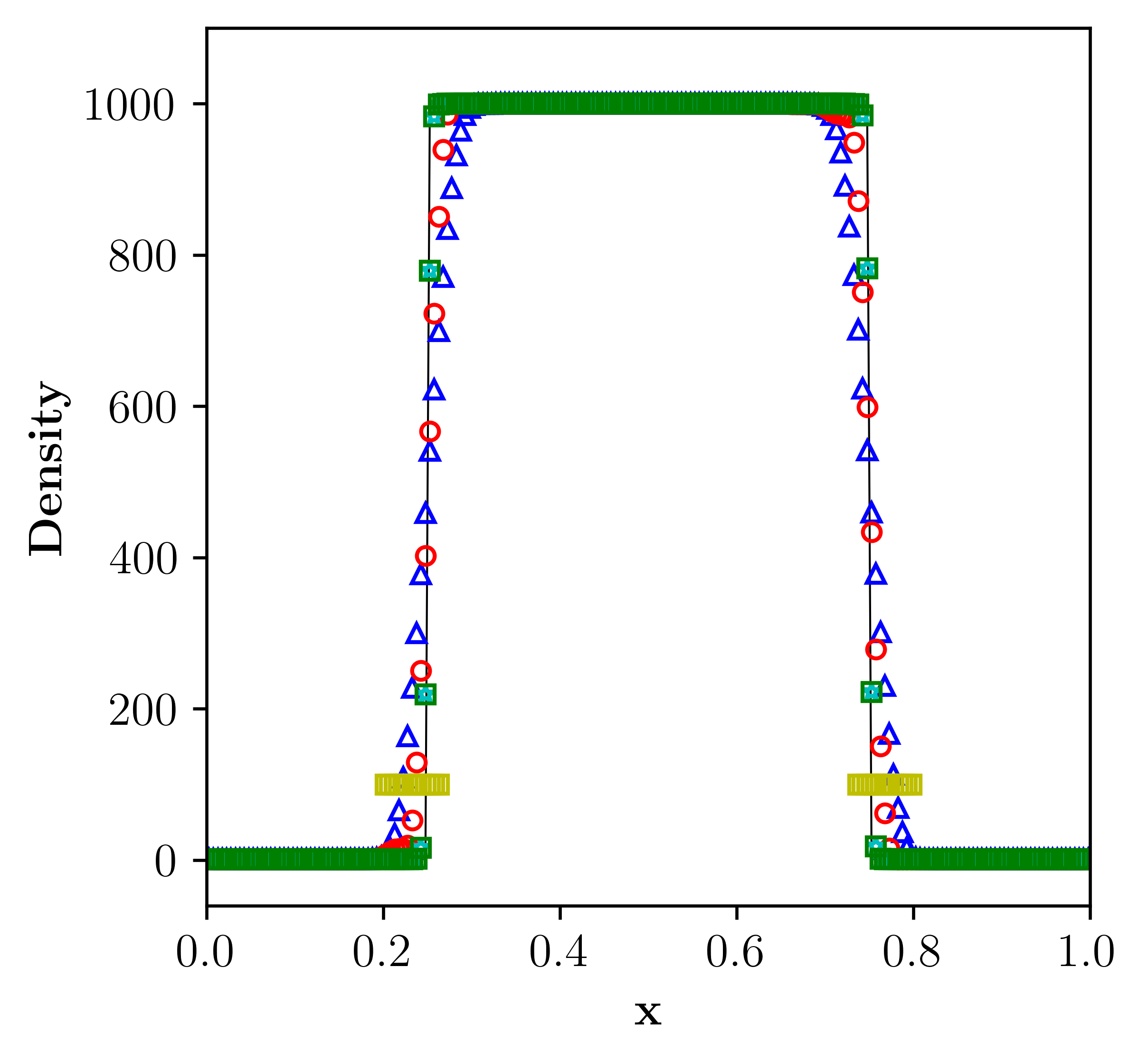}
\label{fig:iso-met_l}}
\subfigure[Local density profile, Example \ref{ex:isol}]{\includegraphics[width=0.45\textwidth]{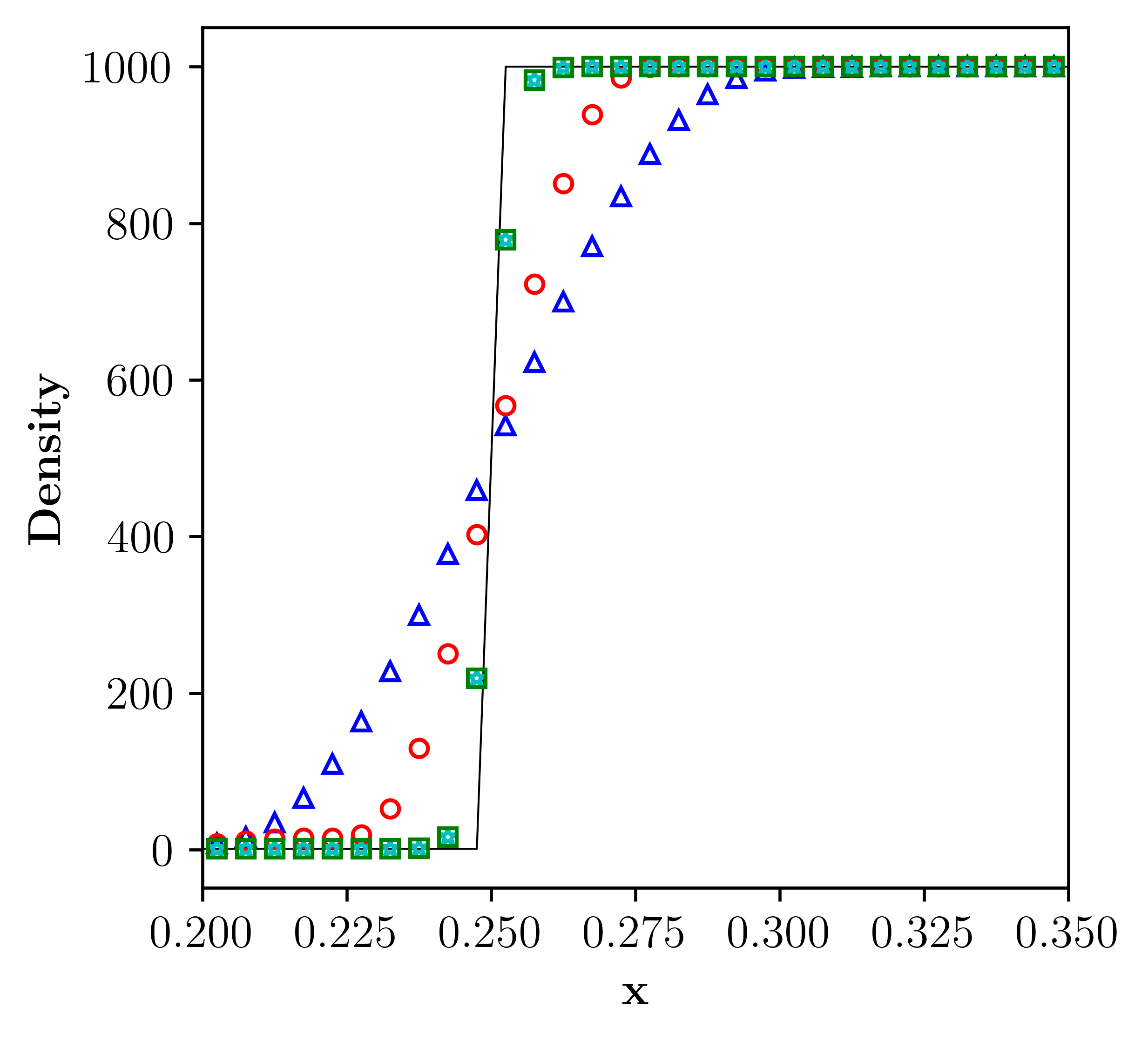}
\label{fig:iso-met2_l2}}
\subfigure[Pressure profile, Example \ref{ex:isol}]{\includegraphics[width=0.45\textwidth]{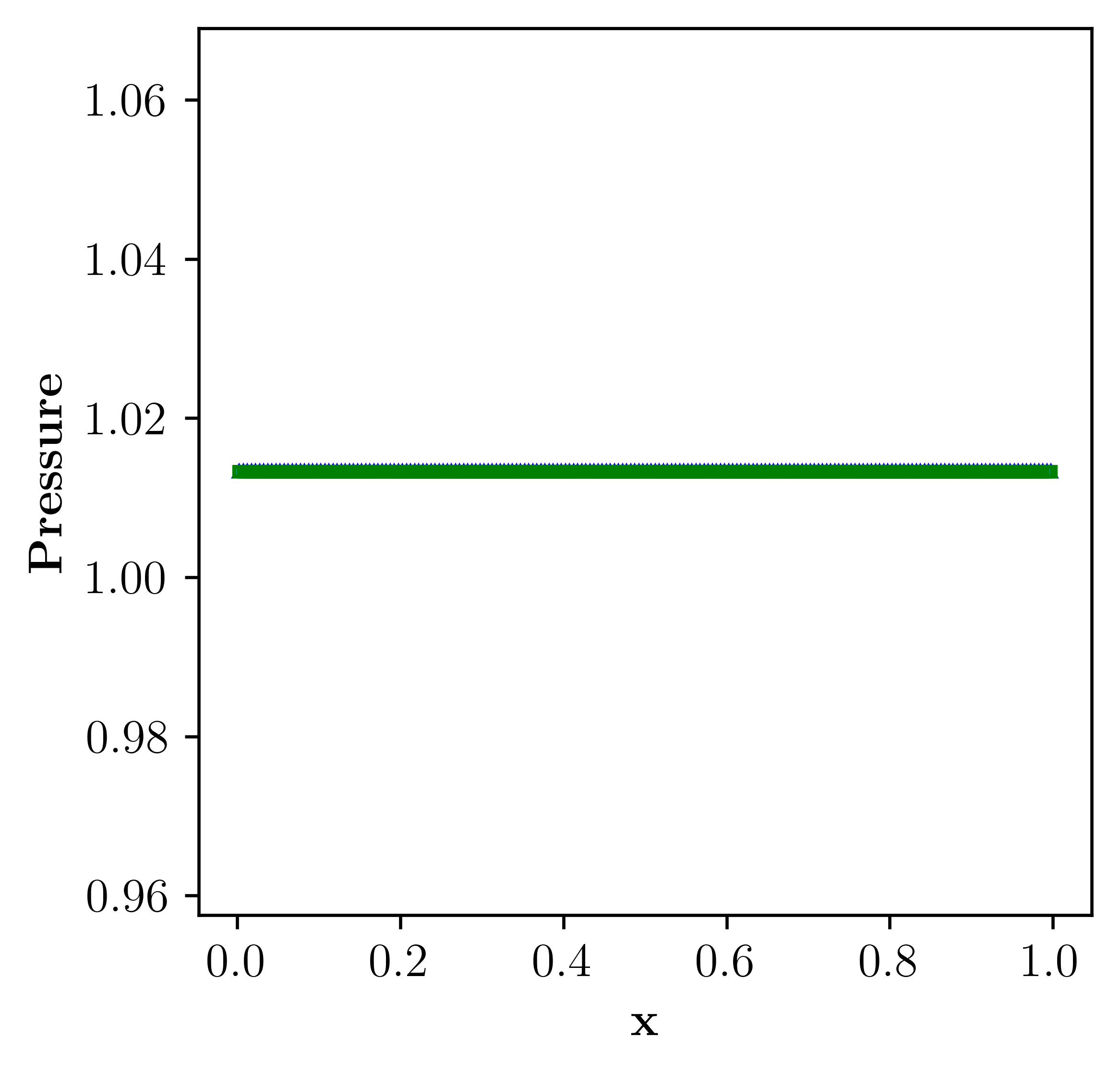}
\label{fig:iso-met_p}}
\subfigure[Density profile (Prim), Example \ref{ex:isol}]{\includegraphics[width=0.45\textwidth]{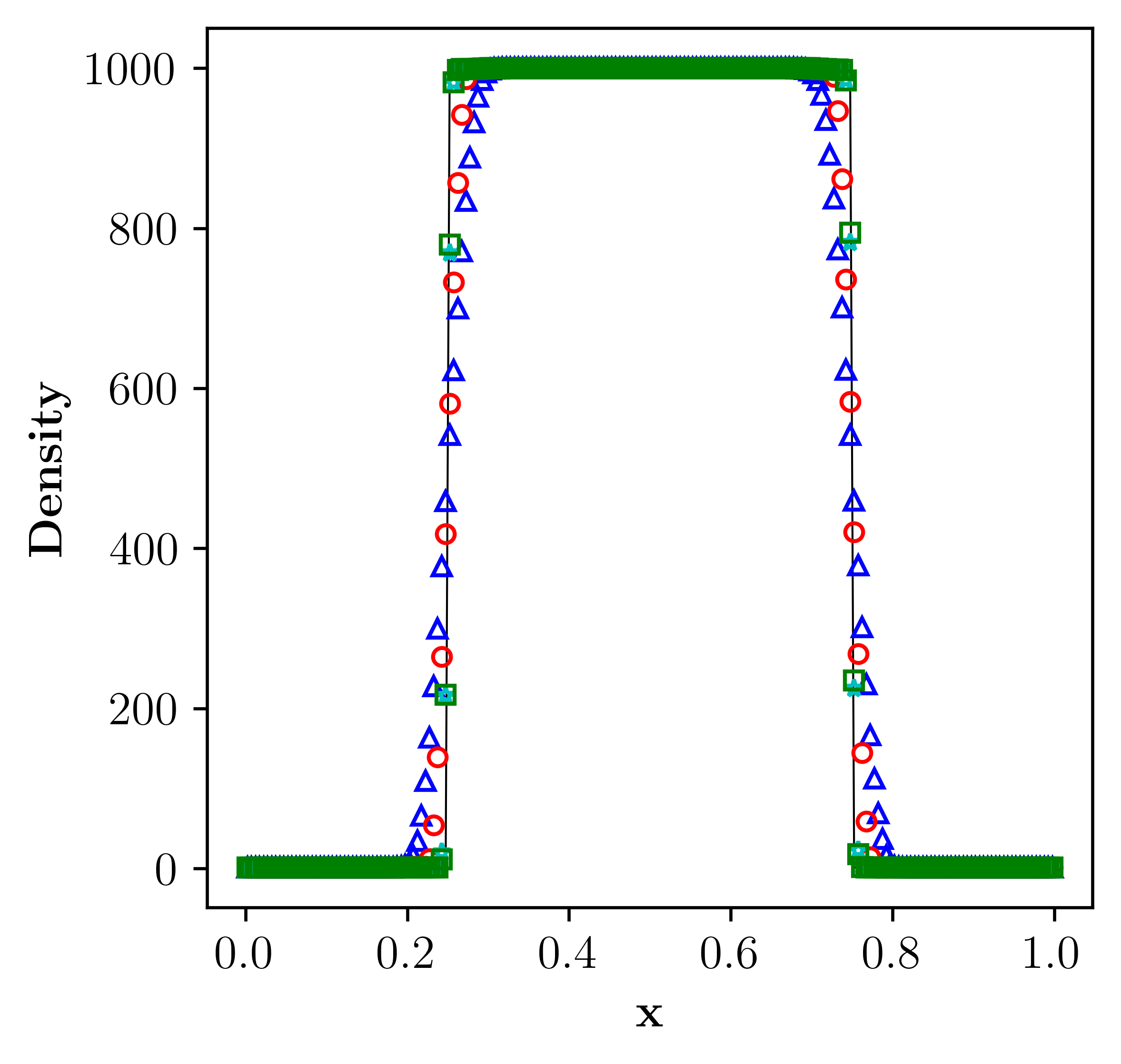}
\label{fig:iso-met2_prim}}
    \caption{Numerical solution for isolated contact test case using $N = 200$ grid points at $t = 0.1$,  Example \ref{ex:isol}, where Solid line: reference solution; blue triangles: MUSCL; red circles: MP5; cyan stars: Wave-MUSCL and green squares: Wave-MP.}
    \label{fig_iso-ml-p}
\end{figure}
\begin{example}{Liquid-gas shock tube}\label{ex:multi-species}
\end{example}
We consider the liquid-gas shock tube problem studied in Ref. \cite{wong2021positivity} in this test case. The initial conditions for this test case are as follows:
\begin{equation}
\begin{aligned}
&\begin{array}{llllcc}
\hline x(\mathrm{~m}) & \rho_1 \alpha_1\left(\mathrm{kgm}^{-3}\right) & \rho_2 \alpha_2\left(\mathrm{kgm}^{-3}\right) & u\left(\mathrm{~ms}^{-1}\right) & p(\mathrm{~Pa}) & \alpha_1 \\
\hline x<0.75 & 1000.0 & 1.0 \times 10^{-8} & 0.0 & 1.0 \times 10^9 & 1.0-10^{-8} \\
\text { Otherwise } & 1.0 \times 10^{-8} & 1.0 & 0.0 & 1.0 \times 10^5 & 1.0 \times 10^{-8} \\
\hline
\end{array}
\end{aligned}
\end{equation}
with fluid properties $\gamma_l=4.4, \pi_{\infty, l}=6 \times 10^{8}, \gamma_g=1.4, \pi_{\infty, g}=0$. The simulation was conducted on a computational domain spanning $x = 0$ to $x = 1$, utilizing  250 uniformly distributed grid points until a final time of $t = 0.00024$. Figure \ref{fig_11} compares the exact solution (computed on a grid with 2000 points using the Wave-MP scheme) to the numerical solutions obtained by various schemes. All the tested schemes accurately captured the material interface without introducing spurious oscillations, except for the MP5 scheme, which failed this test due to the absence of positivity-preserving techniques.
Both the Wave-MUSCL and Wave-MP schemes resolved the material interface with fewer points than the MUSCL scheme, demonstrating the sensor's effectiveness in reliably detecting the interface. There is no discernible difference between the Wave-MUSCL and Wave-MP approaches for this test case.
As illustrated in Figure \ref{fig:multi_11-pres}, the pressure profiles remain free of oscillations. Furthermore, the results obtained using the primitive variable reconstruction (Figure \ref{fig:multi_11-prim}) are nearly identical to those achieved with the adaptive primitive-characteristic variable reconstruction approach.

%Figure \ref{fig_11} compares the exact (computed on a grid size of 2000 points by the Wave-MP scheme) and numerical solution computed by various schemes, which all precisely captured the material interface without any unwanted oscillations. MP5 scheme failed to pass this test case as no positivity-preserving techniques were considered. Both the Wave-MUSCL and Wave-MP schemes captured the material interface within fewer points than the MUSCL scheme, again indicating that the sensor reliably detected the material interface. There is also no difference between the Wave-MUSCL and Wave-MP approach for this test case. As shown in Figure \ref{fig:multi_11-pres}, pressure profiles have no oscillations. Finally, the results obtained by the primitive variable reconstruction, Figure \ref{fig:multi_11-prim}, are nearly identical to that of the adaptive primitive-characteristic variable reconstruction approach.
\begin{figure}[H]
\centering
\subfigure[Density]{\includegraphics[width=0.43\textwidth]{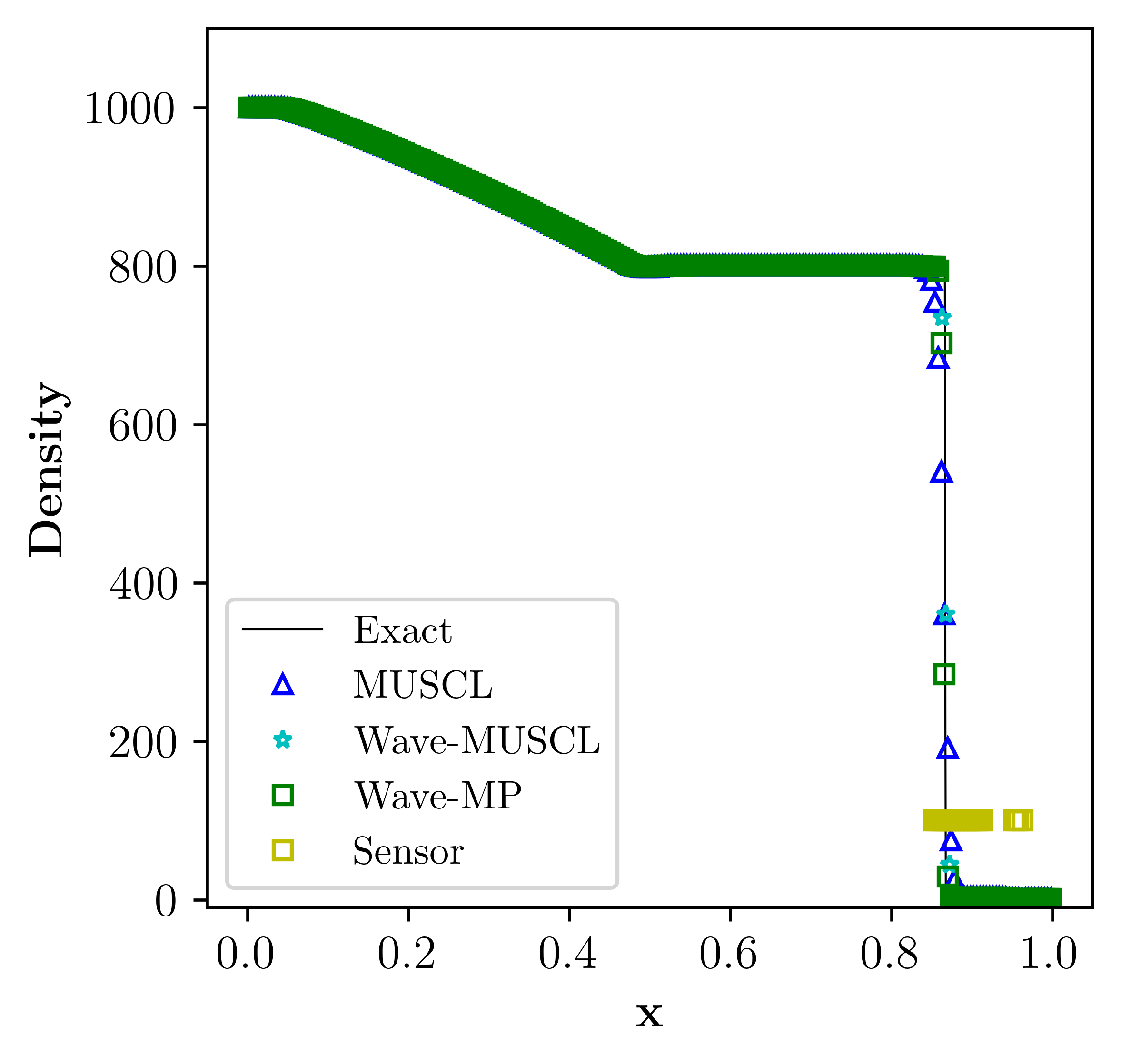}
\label{fig:multi_11-den}}
\subfigure[Density, Local profile]{\includegraphics[width=0.45\textwidth]{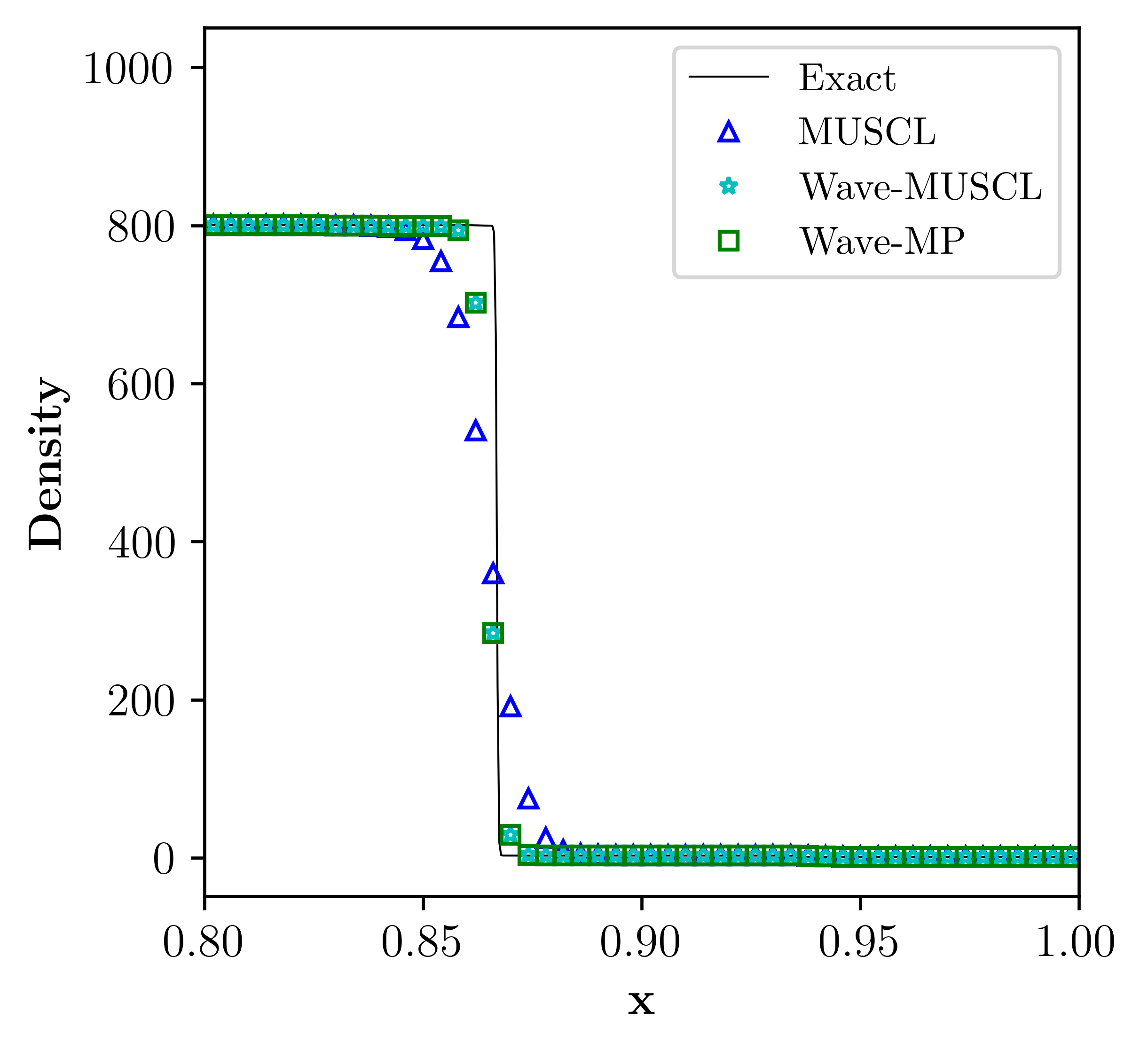}
\label{fig:multi_11-alpha}}
\subfigure[Pressure]{\includegraphics[width=0.415\textwidth]{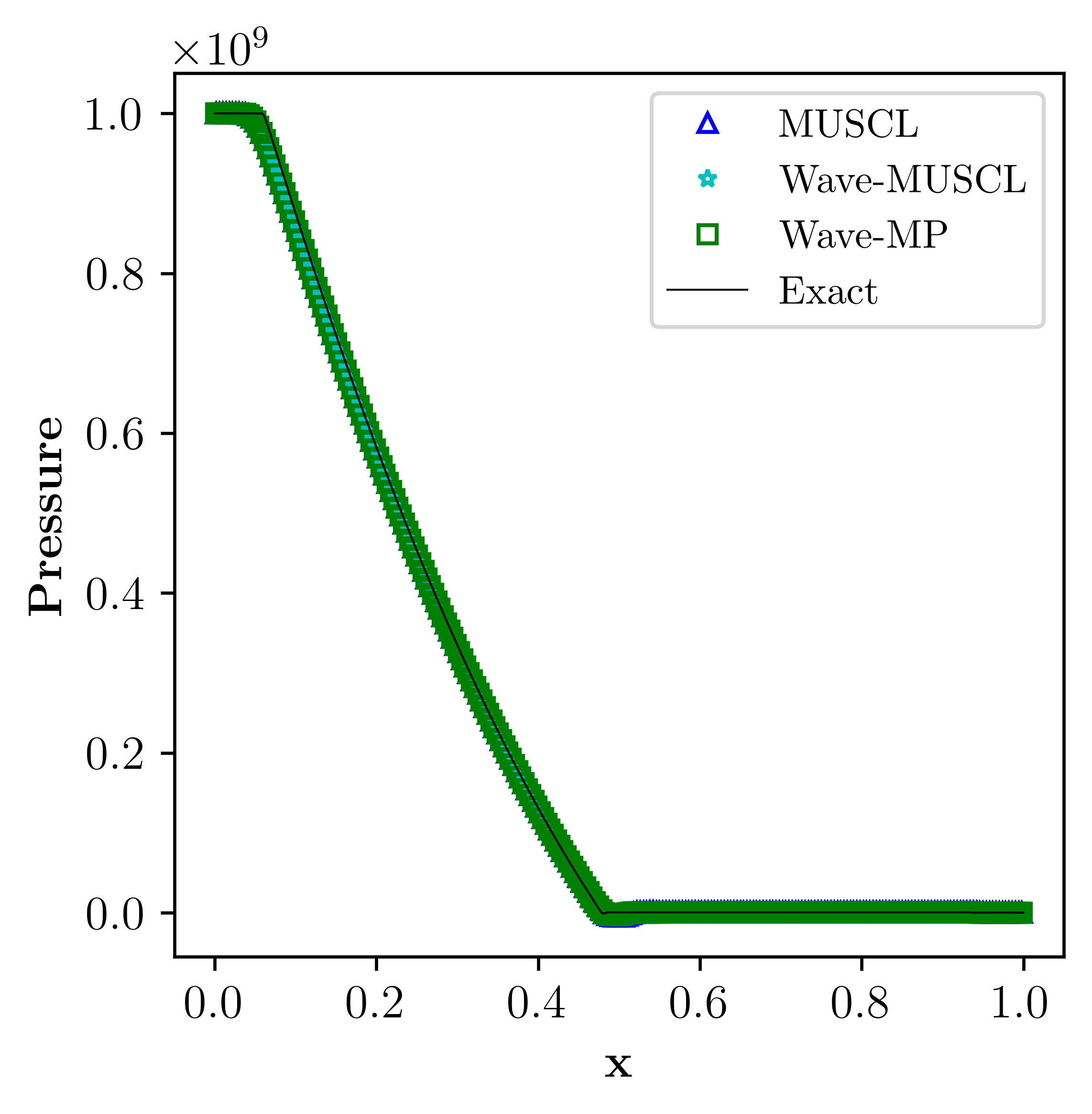}
\label{fig:multi_11-pres}}
\subfigure[Density (Prim)]{\includegraphics[width=0.415\textwidth]{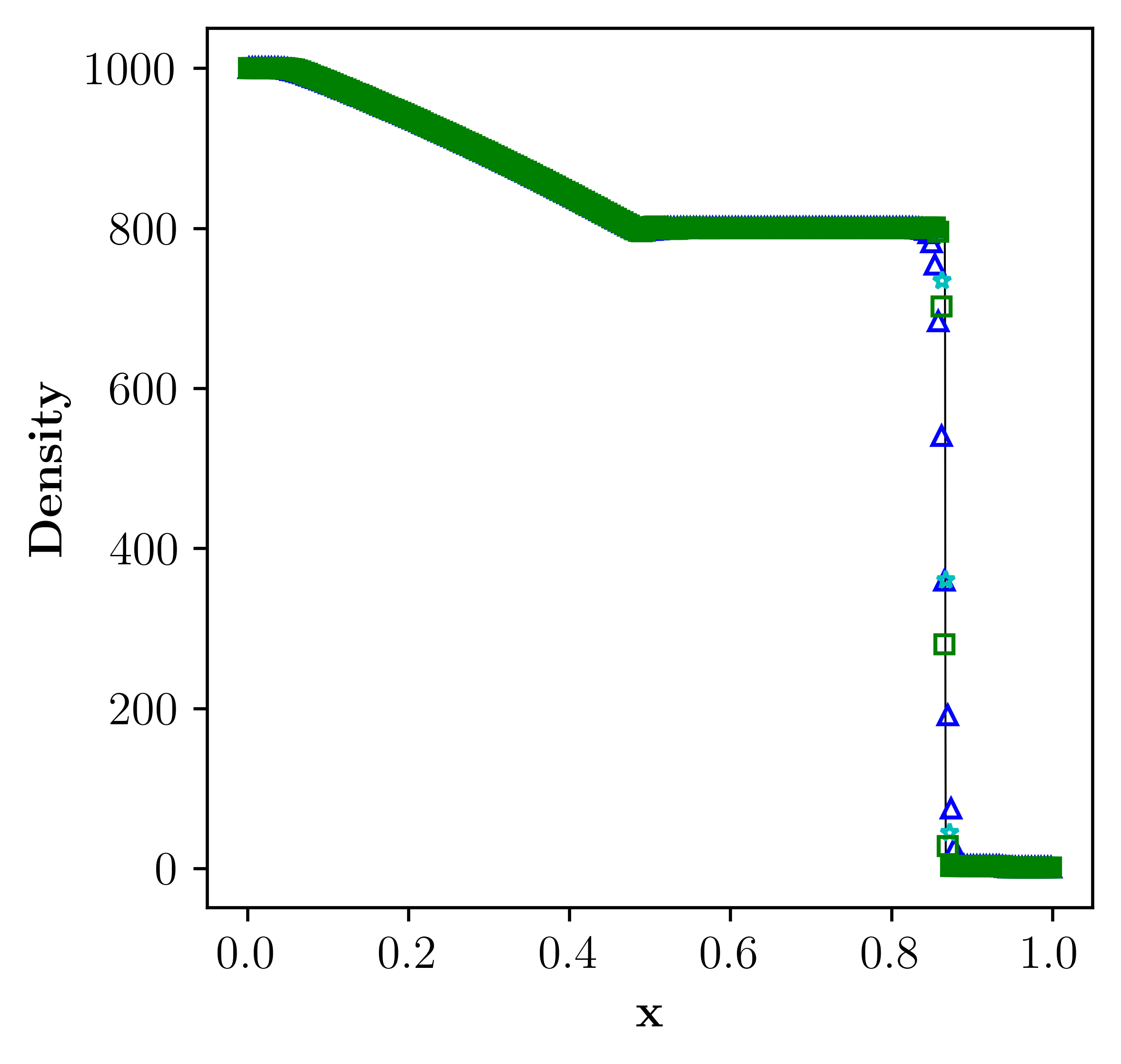}
\label{fig:multi_11-prim}}
\caption{Numerical solution for Liquid-gas shock tube problem in Example \ref{ex:multi-species}  on a grid size of $N=250$, where Solid line: reference solution; Blue triangles: MUSCL; cyan stars: Wave-MUSCL and green squares: Wave-MP.}
\label{fig_11}
\end{figure}

\begin{example}\label{ex:dsl}{Periodic double-shear layer}
\end{example}

This test case studies the impact of computing vorticity waves using a central scheme in an inviscid limit and the advantages of multi-dimensional upwinding. The test involves two initially parallel shear layers that develop into two significant vortices at $t = 1$ \cite{brown1995performance,minion1997performance}. All tests were run with a $N_x \times N_y = 320 \times 320$ grid size. The initial conditions are:
\begin{subequations}
 \begin{align}
        p &= \frac{1}{\gamma \mathrm{Ma}^2_{\infty}},\rho=1,
        u = 
        \begin{cases}
            \tanh \left[ 80 (y-0.25) \right], & \text{ if } (y \leq 0.5), \\
            \tanh \left[ 80 (0.75-y) \right], & \text{ if } (y > 0.5),
        \end{cases},        v = 0.05 \sin \left[ 2 \pi(x+0.25) \right].
    \end{align}
\end{subequations}
It is well known that under-resolved simulations can produce unphysical braid vortices and oscillations along shear layers. Fine grid solution computed on a grid size of 800 $\times$ 800 is shown in Figure \ref{fig:dpsl_fine}. 

\begin{figure}[H]
\centering\offinterlineskip
 \includegraphics[width=0.32\textwidth]{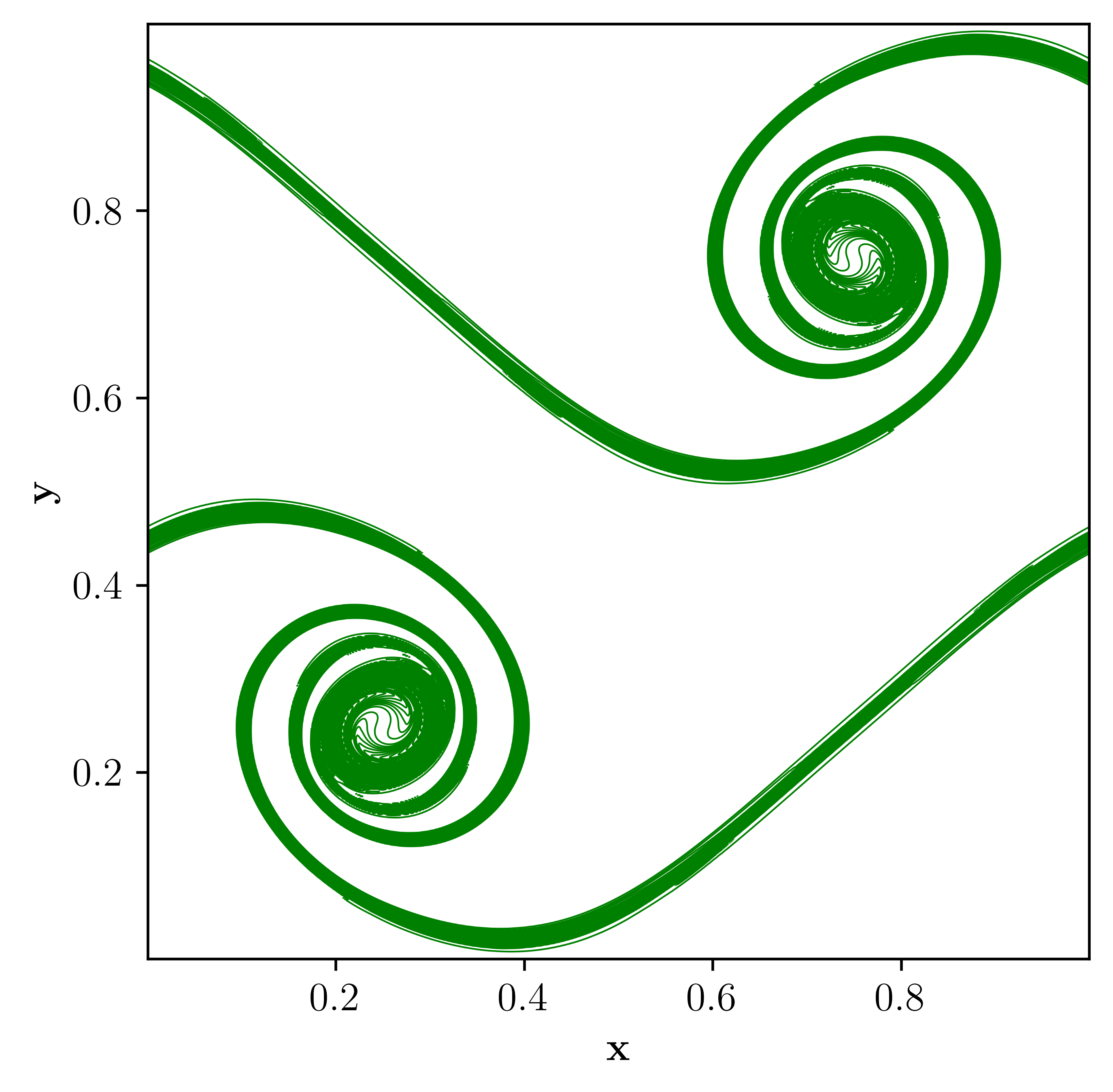}
\caption{$z$-vorticity contours on a grid size of $800^2$, Example \ref{ex:dsl}.}
    \label{fig:dpsl_fine}
%\end{halfspacing}
\end{figure}

Figure \ref{fig:dpsl_96} illustrates the $z$-vorticity computed using the MP5 scheme, the MP scheme (where all waves are handled using the central scheme), and the Wave-MP approach (which incorporates THINC). As shown in Figures \ref{fig:meg-base} and \ref{fig:meg-entr}, the MP5 scheme produces braid vortices, while the MP-Central scheme introduces both vortices and oscillations. In contrast, the Wave-MP scheme—where acoustic waves are computed using the upwind scheme and vorticity waves are handled with the central scheme—eliminates both vortices and oscillations. As discussed in the introduction, the tangential velocities remain continuous due to the presence of artificial viscosity across the material interface \cite{meng2018numerical}. Although this test case does not involve material interfaces, computing vorticity waves using a central scheme effectively prevents spurious vortices. As expected, the THINC scheme has no effect here, given the absence of material interfaces or contact discontinuities.

Similar observations can be made for the results obtained using the primitive variable reconstruction, as shown in Figure \ref{fig:dpsl_prims}. Both the WENO and MP5 schemes exhibit braid vortices (Figures \ref{fig:weno-base} and \ref{fig:mp-basepr}). However, computing tangential velocities using a central scheme in physical space successfully prevents these spurious vortices, as demonstrated in Figure \ref{fig:waverprim}. \textcolor{red}{While the Ducros sensor was employed to compute tangential velocities in \cite{sainadh2024consistent}, this test case shows that the MP criteria alone is sufficient, Equation \ref{eqn:mp5Condition}}. These results further highlight the versatility of the MP scheme \cite{suresh1997accurate}, which has been underutilized in the literature.

For comparison, Figure \ref{fig:teno-adams}, adapted from \cite{feng2024general}, shows simulations performed on a grid of 512 $\times$ 512 points—2.5 times larger than the current approach. Despite this higher resolution and the use of the optimized TENO8 scheme, braid vortices are still present in their results. These results underscore the advantages of incorporating physics-based, wave-appropriate multidimensional upwinding approaches into reconstruction schemes. As discussed in the introduction, schemes designed for the linear advection equation, such as those in \cite{feng2024general}, may yield unexpected results even if they possess optimized dispersion and dissipation properties.
 \begin{figure}[H]
%\begin{halfspacing}
\centering\offinterlineskip
\subfigure[MP5.]{\includegraphics[width=0.322\textwidth]{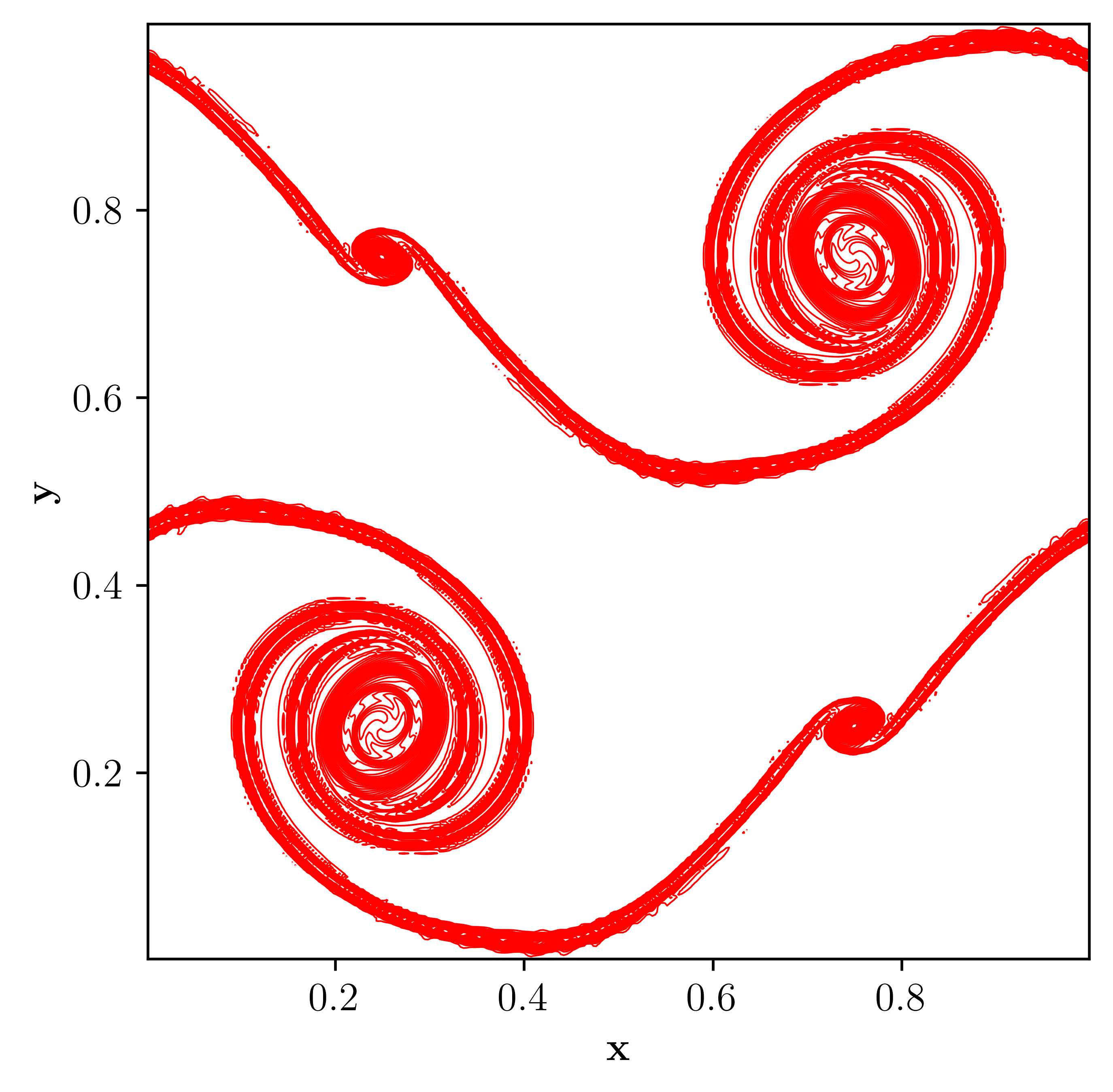}
\label{fig:meg-base}}
\subfigure[MP-Central-all.]{\includegraphics[width=0.322\textwidth]{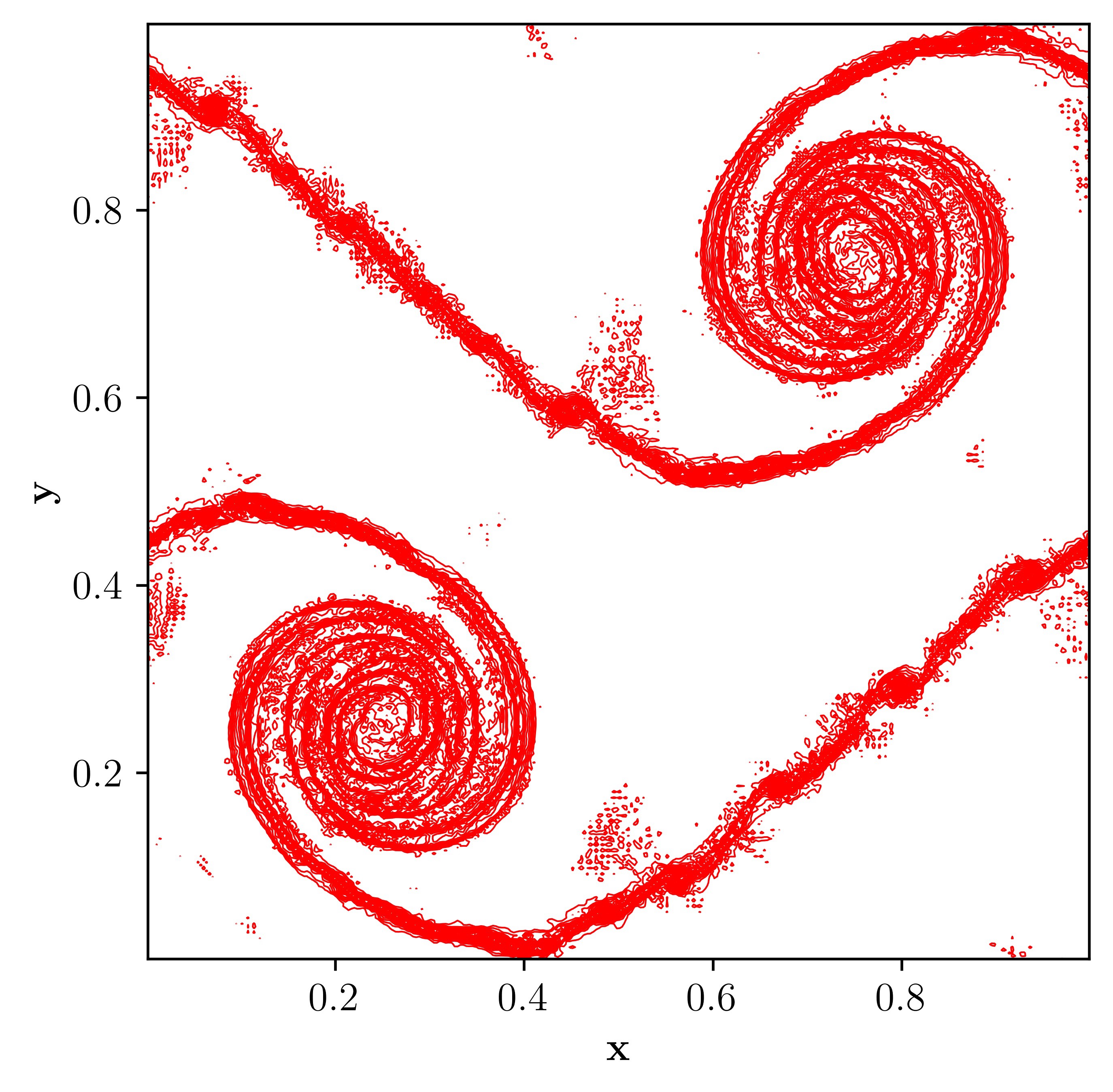}
\label{fig:meg-entr}}
\subfigure[Wave-MP.]{\includegraphics[width=0.322\textwidth]{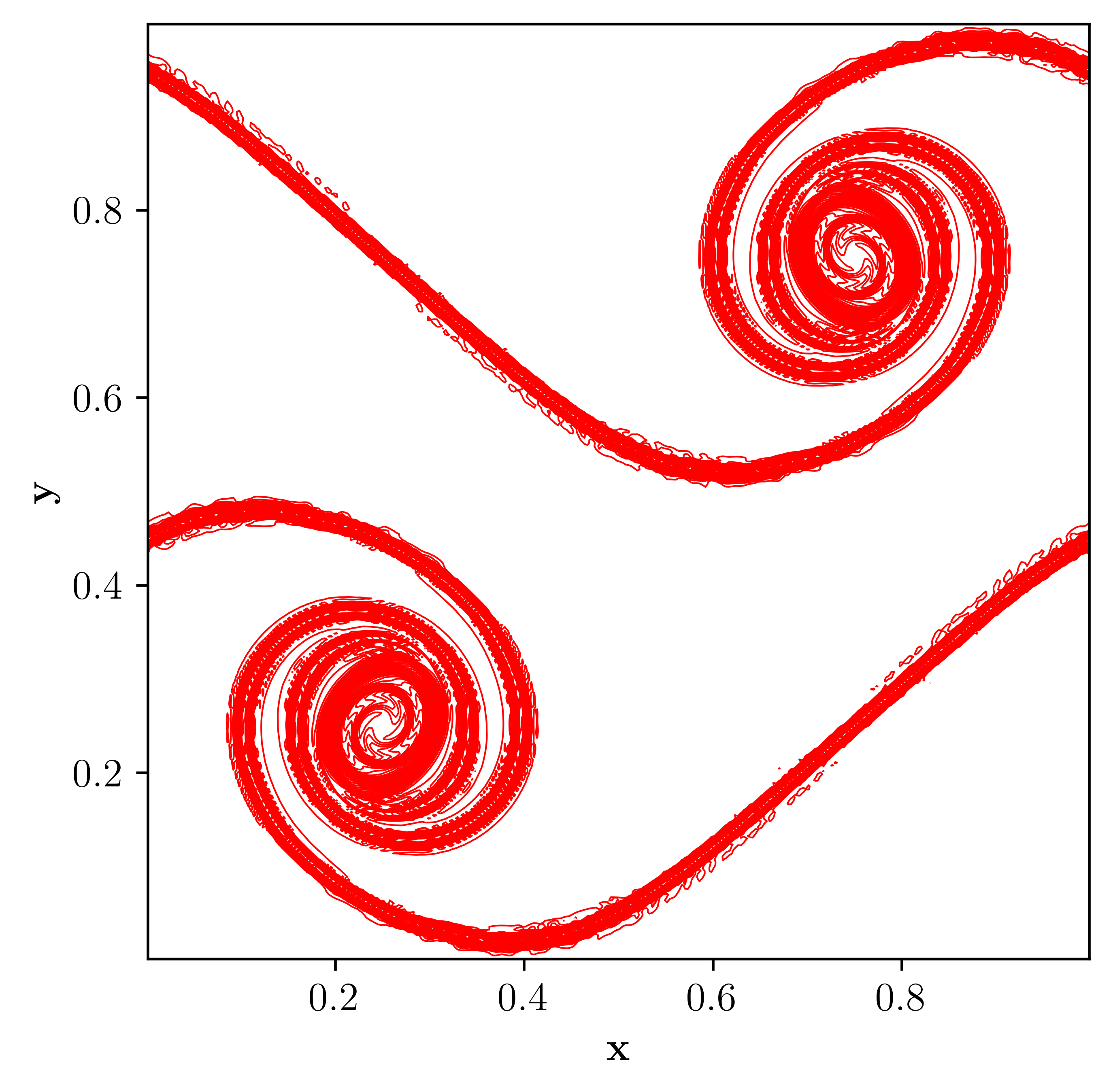}
\label{fig:meg-degen}}
\caption{$z$-vorticity contours of the considered schemes on a grid size of $320^2$, Example \ref{ex:dsl}.}
    \label{fig:dpsl_96}
%\end{halfspacing}
\end{figure}
\begin{figure}[H]
\centering\offinterlineskip
\subfigure[WENO5 (Prim).]{\includegraphics[width=0.322\textwidth]{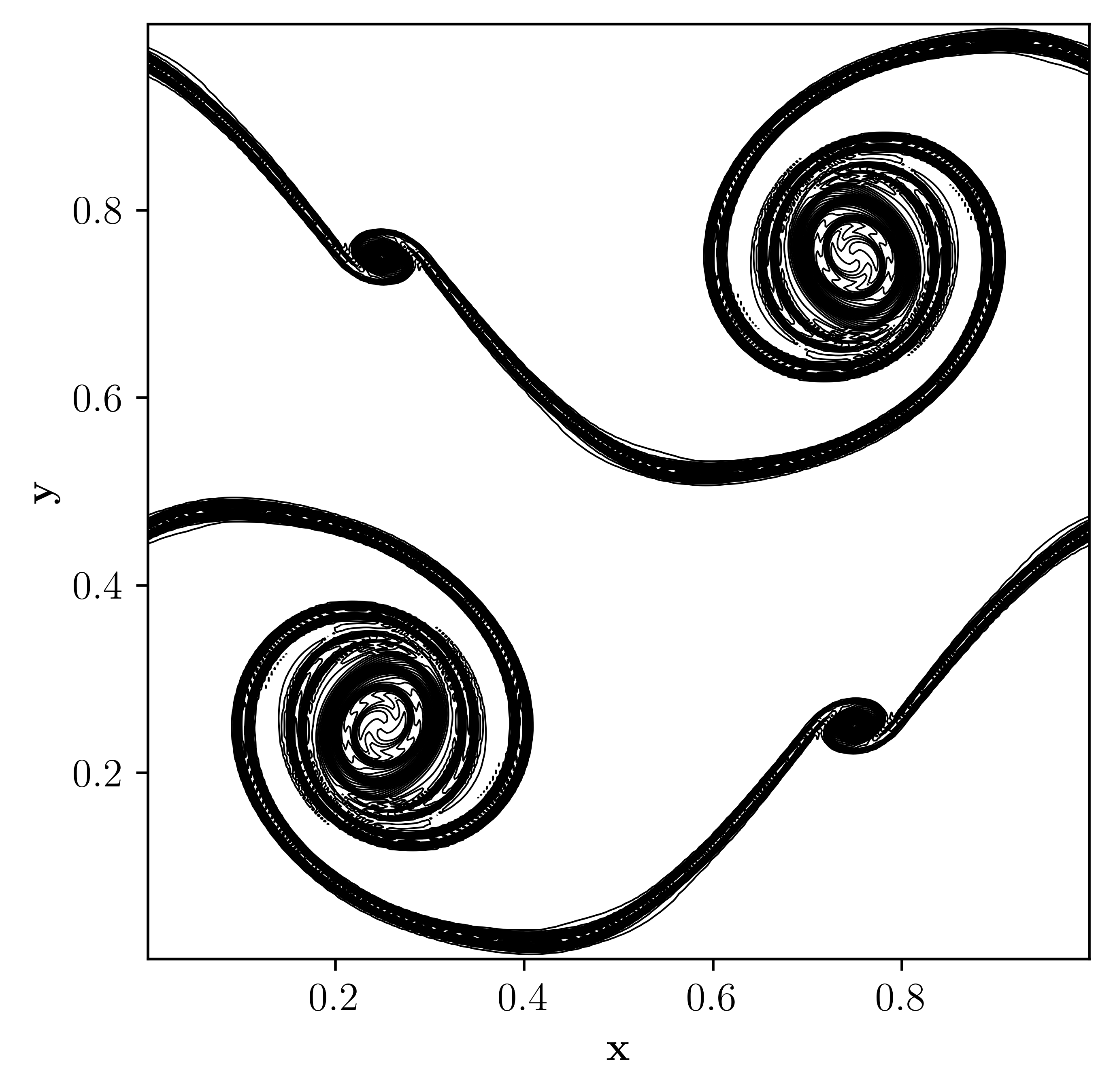}
\label{fig:weno-base}}
\subfigure[MP5 (Prim).]{\includegraphics[width=0.322\textwidth]{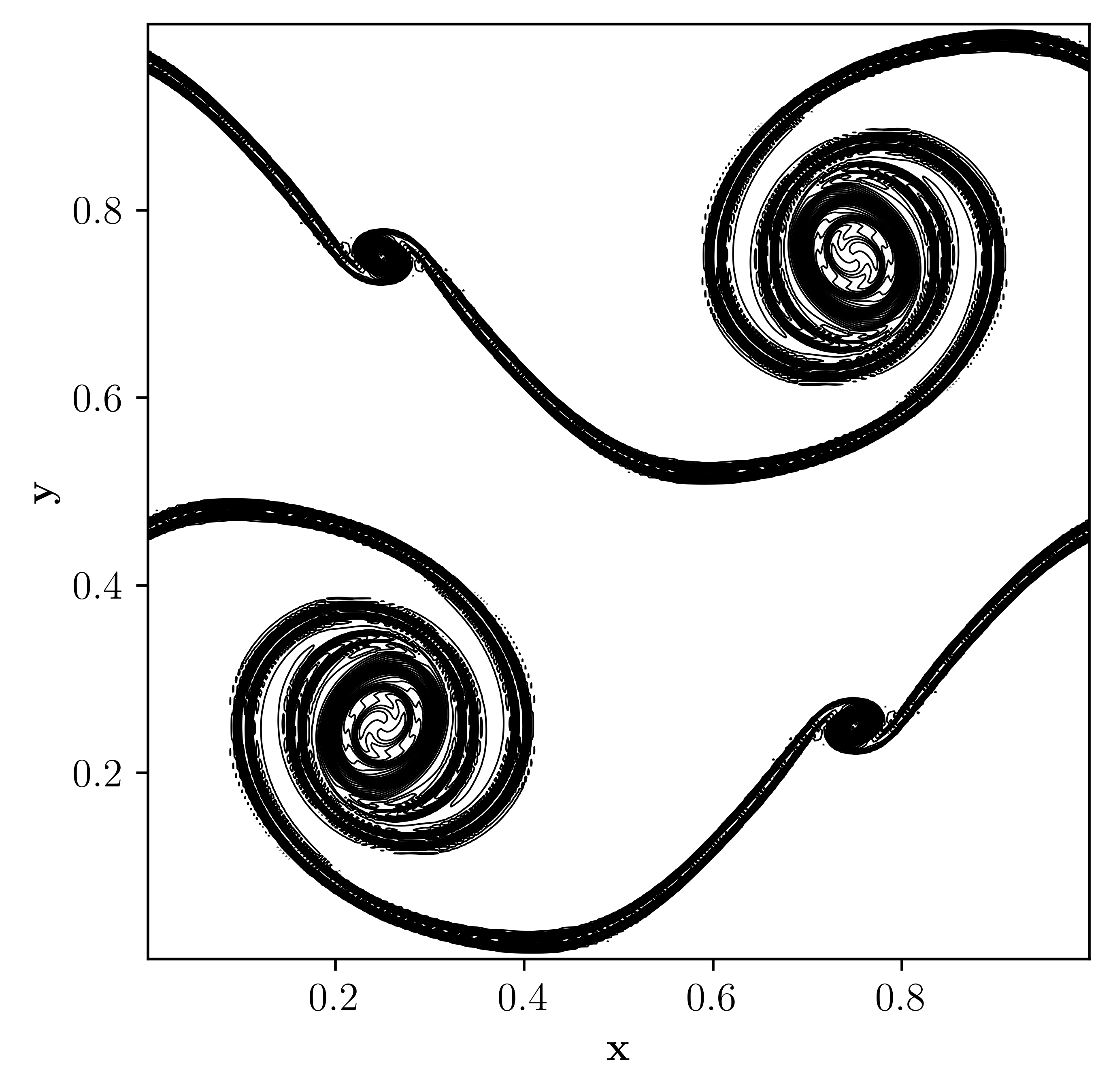}
\label{fig:mp-basepr}}
\subfigure[Wave-MP (Prim).]{\includegraphics[width=0.322\textwidth]{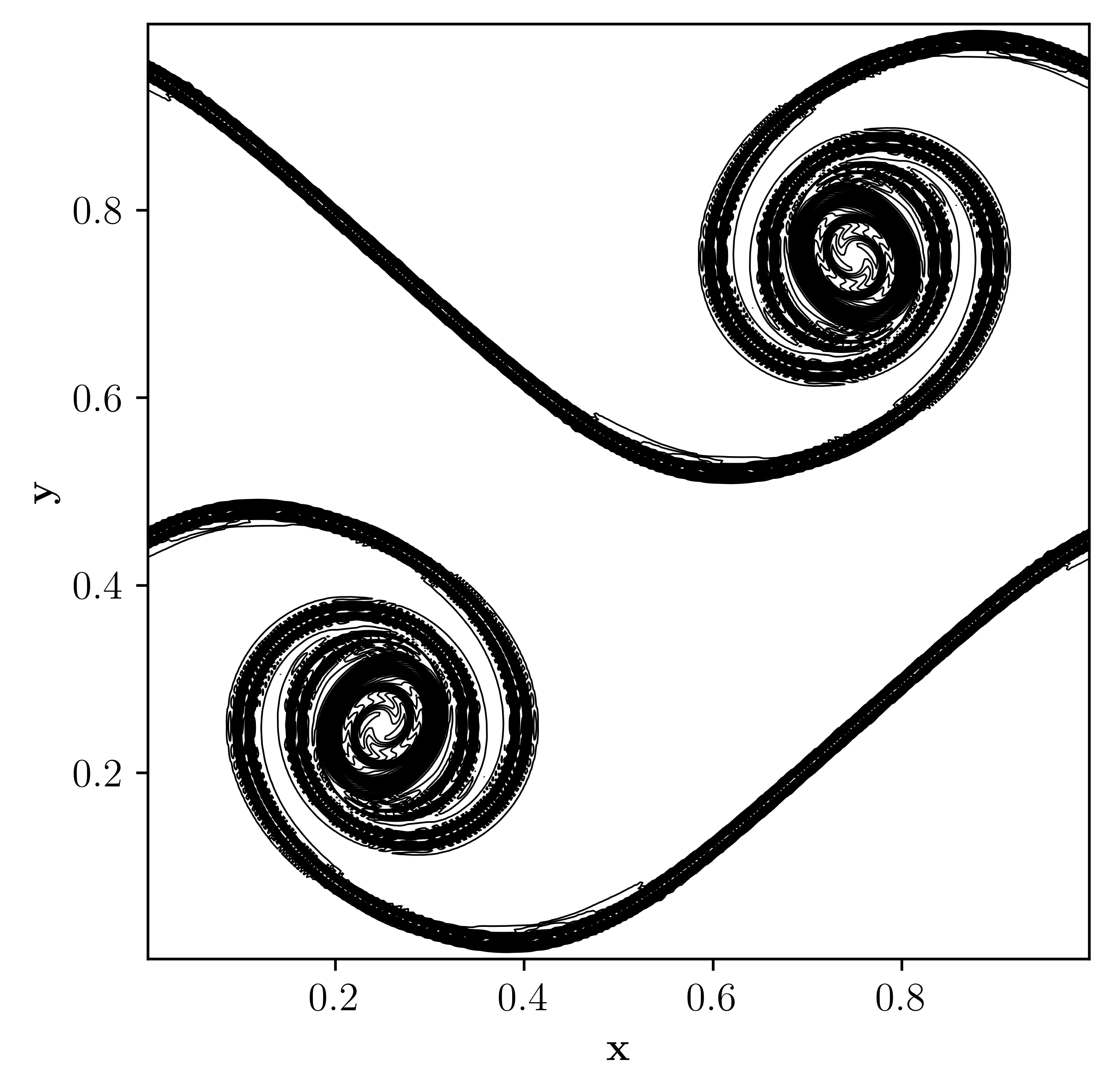}
\label{fig:waverprim}}
\caption{$z$-vorticity contours of the considered schemes (using primitive variables) computed on a grid size of $320^2$, Example \ref{ex:dsl}.}
    \label{fig:dpsl_prims}
%\end{halfspacing}
\end{figure}

\begin{figure}[H]
\centering
 \includegraphics[width=0.99\textwidth]{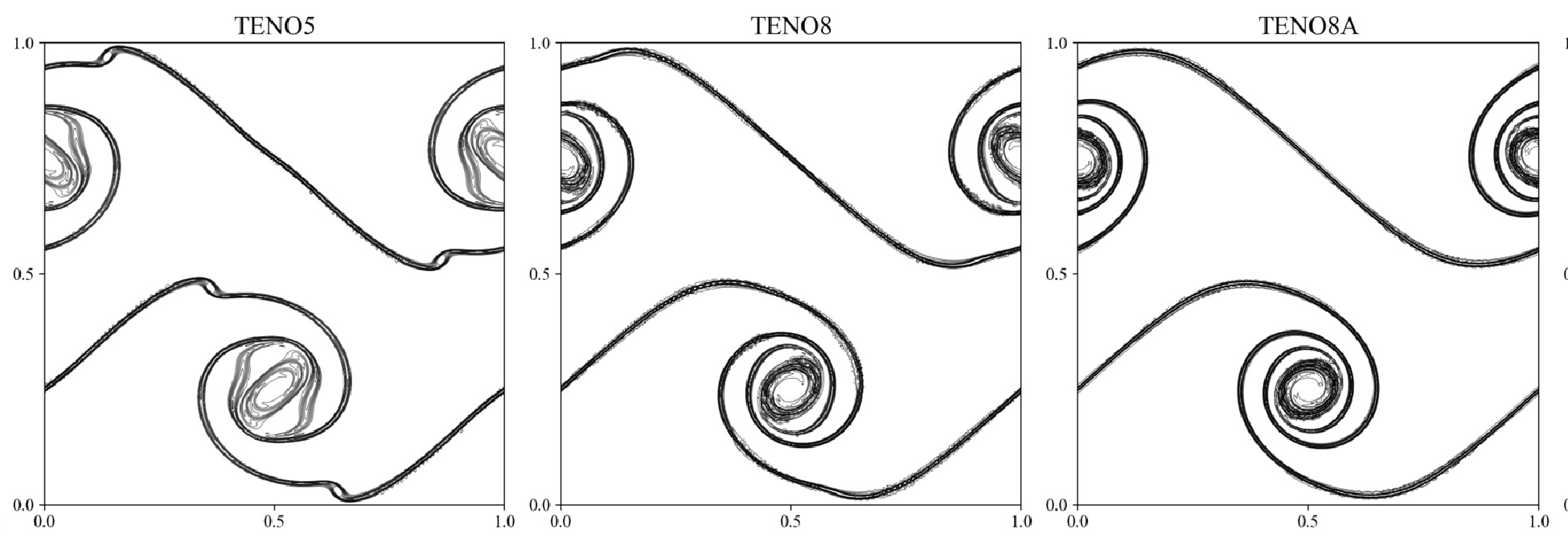}
    \caption{Figure is taken from Reference \cite{feng2024general}, where the simulations are computed on a grid size of $512^2$.}
\label{fig:teno-adams}
\end{figure}

Finally, simulations were performed using the TENO-THINC scheme \cite{takagi2022novel}, which employs a TENO-based discontinuity sensor to detect discontinuities and applies the THINC scheme to all waves at all types of discontinuities. Figure \ref{fig:teno-base} presents the results using the TENO scheme, where unphysical braid vortices are observed. When the THINC scheme is applied only to the entropy wave \textbf{(as in this paper)}, as shown in Figure \ref{fig:teno-deg}, the results are nearly identical to those of the base TENO scheme. However, applying the THINC scheme to all linearly degenerate waves, or all waves—including acoustic waves, as proposed in \cite{takagi2022novel}—yields results that deviate significantly from the base scheme, as illustrated in Figure \ref{fig:teno-all}. These findings suggest that applying the THINC scheme universally to all waves can lead to unexpected results. While this test case does not involve a material interface, it is well established that velocity and pressure remain continuous across material interfaces \cite{abgrall2001computations,batchelor1967introduction,coralic2014finite,shyue2014eulerian,deng2018high}. Thus, using an interface-capturing scheme to compute these variables may explain the observed discrepancies.

\begin{figure}[H]
\centering\offinterlineskip
\subfigure[TENO5.]{\includegraphics[width=0.32\textwidth]{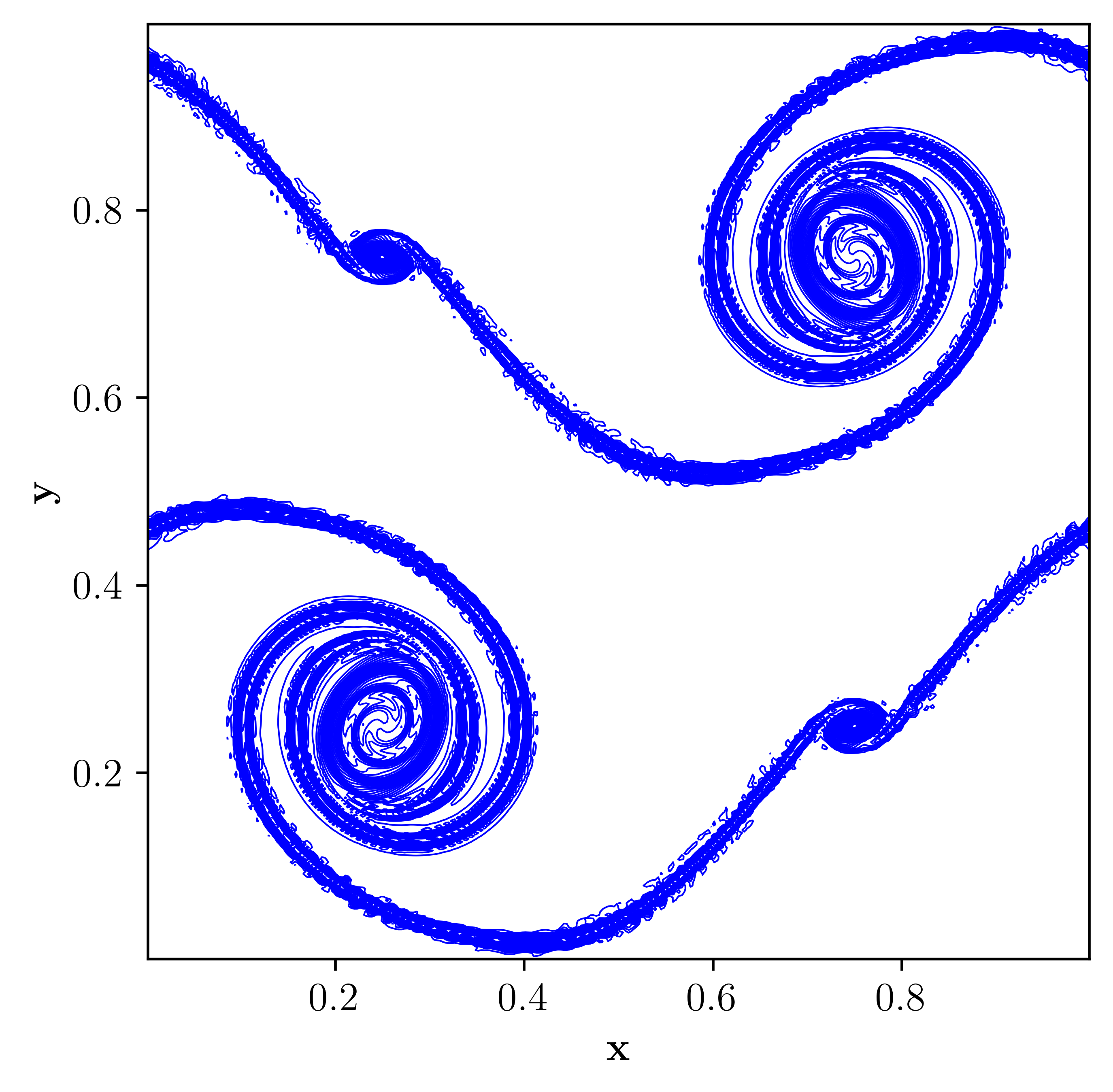}
\label{fig:teno-base}}
\subfigure[TENO5-THINC \cite{takagi2022novel}.]{\includegraphics[width=0.32\textwidth]{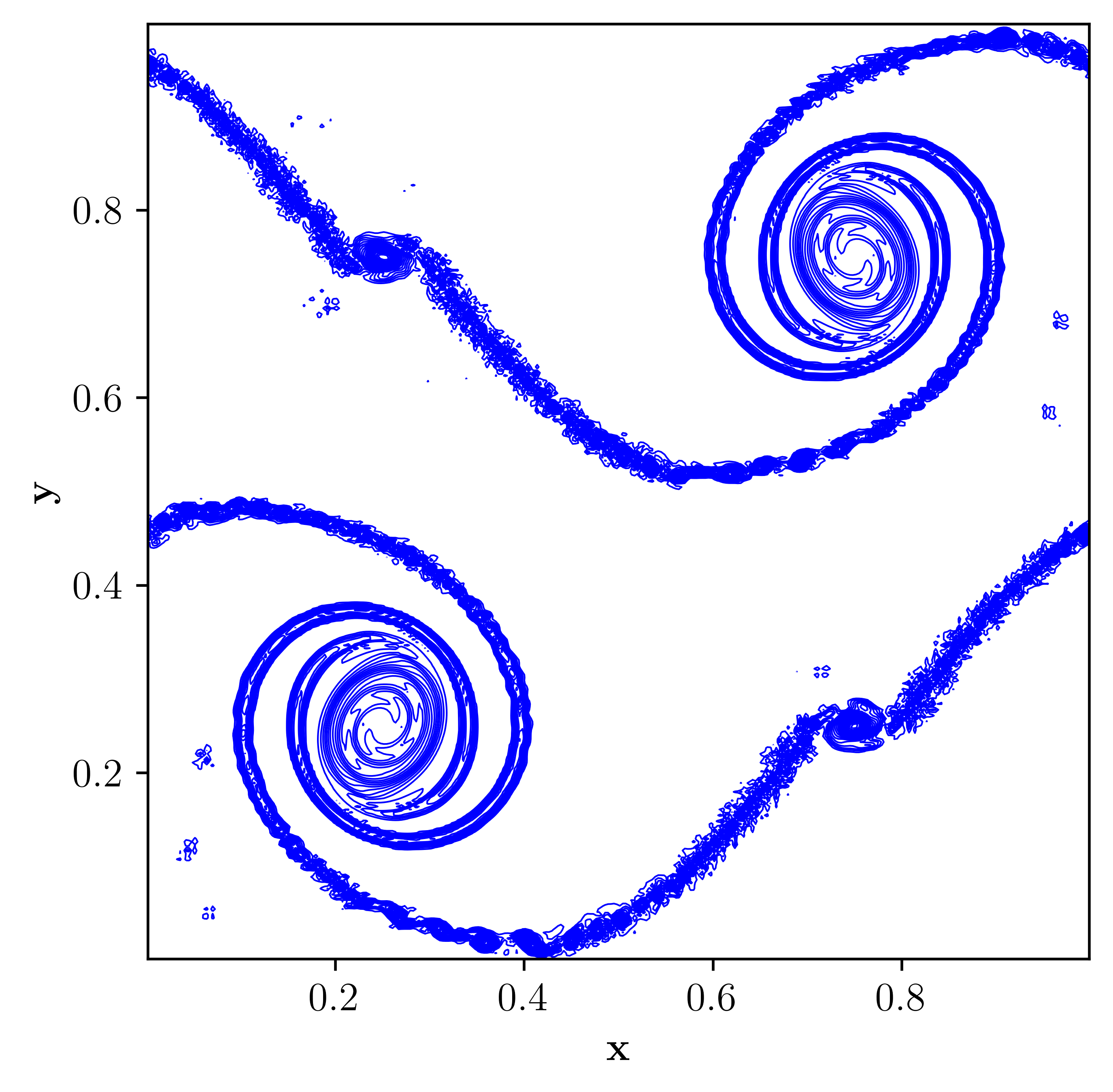}
\label{fig:teno-all}}
\subfigure[TENO5-THINC (entropy wave).]{\includegraphics[width=0.32\textwidth]{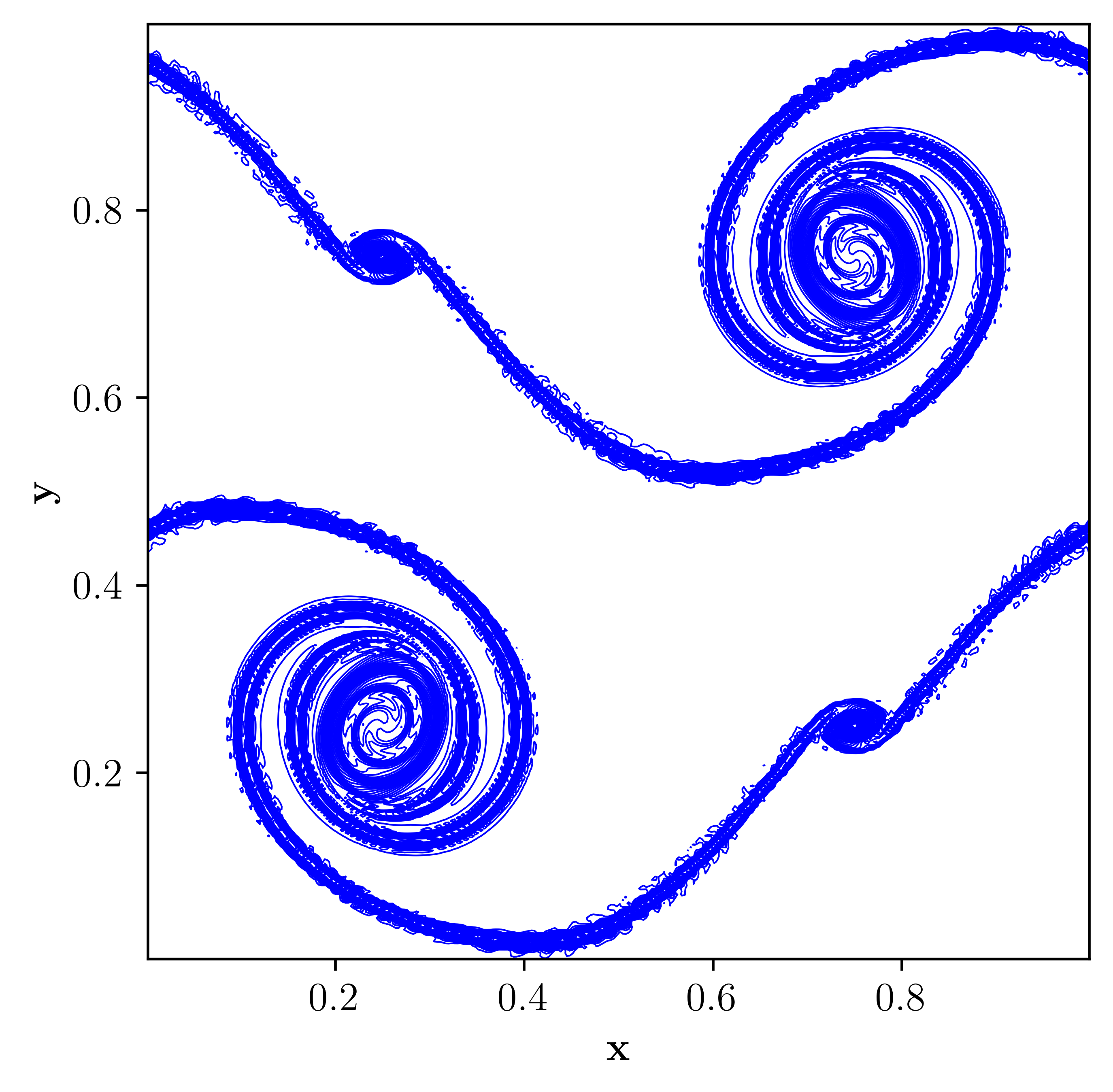}
\label{fig:teno-deg}}
\caption{Figure shows the $z$-vorticity contours for the TENO, TENO-THINC and modified TENO-THINC (THINC for entropy wave only) schemes using a grid size of $320^2$, Example \ref{ex:dsl}.}
    \label{fig:dpsl_teno_96}
%\end{halfspacing}
\end{figure}

\begin{example}{Compressible triple point }\label
{ex:triplet}
\end{example}
This test case examines the multi-species compressible triple-point problem, a two-dimensional Riemann problem involving three states and two distinct materials. \textbf{The primary objective is to demonstrate the benefits of computing vorticity waves using a central scheme.} This test case serves as a benchmark for evaluating the ability of interface-capturing schemes to resolve sharp interfaces and simulate the formation of fine-scale vortical structures along contact discontinuities caused by Kelvin-Helmholtz instabilities. The computational domain is [0, 7] $\times$ [0, 3], with initial conditions defined as follows \cite{chamarthi2023gradient}:
\begin{equation}
(\alpha_1 \rho_1,\alpha_2 \rho_2, u, v, p, \gamma)=\left\{\begin{array}{ll}
(1.0~~~,0.0,0,0,1.0,1.5), & \text { \text{sub-domain} [0, 1]$\times$[0, 3], } \\
(0.0~~~,1.0,0,0,0.1,1.4), & \text { \text{sub-domain} [1, 1]$\times$[0, 1.5], }\\
(0.125,0.0,0,0,0.1,1.5), & \text { \text{sub-domain} [1, 7]$\times$[1.5, 3]}.
\end{array}\right.
\end{equation}
The test case is simulated on two different grid sizes 1792 $\times$ 768 and 3584 $\times$  1536 grid, with reflective boundary conditions and a final simulation time of $t = 5.0$, as in \cite{chamarthi2023gradient}. The simulation results are summarized as follows:
\begin{itemize}
	\item \textbf{Density Gradient Contours at High Resolution:} Figure \ref{fig:fhl} shows the density gradient contours at $t$=5 obtained using the Wave-MP approach on a grid size of 3584 $\times$ 1536. Figure \ref{fig:fhl} depicts various contact discontinuities, denoted by C1, C2 and C3, and shockwaves, denoted by S1, RS1, TS1 and TS2. Figure \ref{fig:fmu} shows the density gradient contours obtained using the WENO scheme on the same grid. The green arrow in both figures indicates that the C1 contact discontinuity is captured sharply in the Wave-MP approach as the THINC scheme computes the material interfaces. The red box in the figures highlights the vortical structures in a particular region of the flow, and they are observed using both the WENO scheme and Wave-MP scheme, but the Wave-MP scheme has more prominent vortices. Vortical structures in the red box are also observed in the published work in the literature, \cite{cheng2021low}.
	\item \textbf{Medium-Resolution Results Without THINC:} Figures \ref{fig:mp} and \ref{fig:mv} show the density gradient contours computed on a grid size of 1792 $\times$ 768 using the MP5 scheme and the Wave-MP scheme without THINC (where tangential velocities are computed using the central scheme, but THINC is not applied to entropy waves). These results demonstrate that computing vorticity waves with a central scheme enhances the vortical structures in the Wave-MP scheme without introducing oscillations.
	\item \textbf{Medium-Resolution Results With THINC:} Figures \ref{fig:fmu2} and \ref{fig:fhl2} show the density gradient contours computed on a grid size of 1792 $\times$ 768 using the MP5-THINC scheme and the Wave-MP scheme (where tangential velocities are computed using the central scheme, and THINC is applied to entropy waves). The Wave-MP approach reveals additional vortical structures in the regions highlighted by the red box, which are dissipated when an upwind scheme is used for vorticity waves. Figures \ref{fig:sx} and \ref{fig:sy} indicate the regions where the material interface detector activates the THINC scheme. Notably, the detector does not activate in the red box region or for shockwaves, confirming that the observed vortices are due to the use of the central scheme for vorticity waves. 
\item \textbf{Primitive variable reconstruction:} Finally, Figures \ref{fig:fmu2p} and \ref{fig:fhl2p} present the density gradient contours computed using the primitive variable reconstruction algorithms, MP5-THINC (Prim) and Wave-MP (Prim). As expected, the deliberate use of a central scheme to compute tangential velocities in the Wave-MP (Prim) approach results in the formation of vortical structures. In contrast, these structures are absent in the MP5-THINC (Prim) approach, where a dissipative upwind scheme is employed for tangential velocity computations. While the direct reconstruction of primitive variables introduces small oscillations, particularly near shockwaves, these are less pronounced in simulations using characteristic variable reconstruction. Nonetheless, both approaches demonstrate the ability to produce vortical structures when tangential velocities are computed using a central scheme.
\begin{figure}[H]
\centering
\subfigure[Wave-MP, 3584 $\times$ 1536.]{\includegraphics[width=0.48\textwidth]{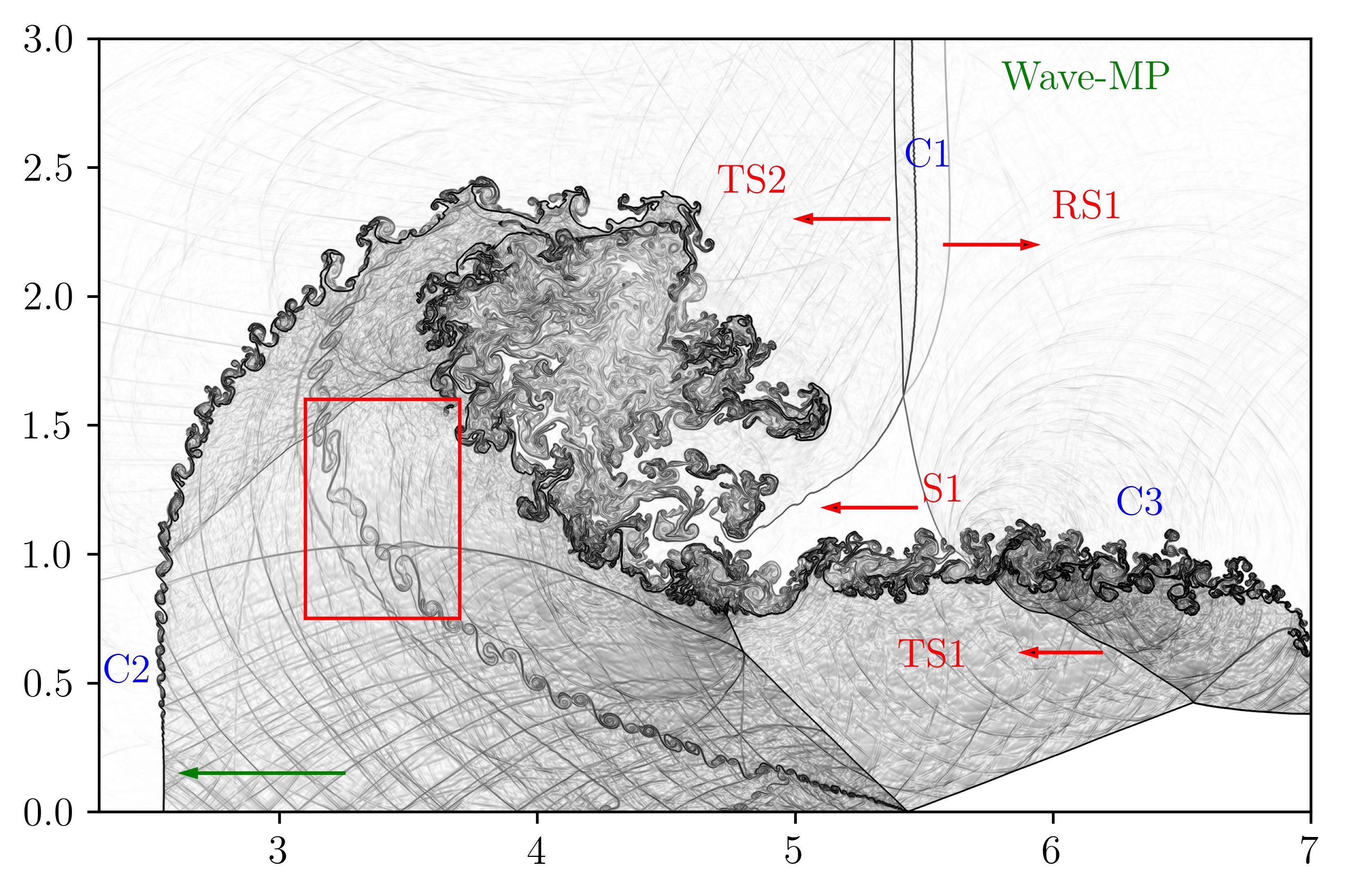}
\label{fig:fhl}}
\subfigure[WENO, 3584 $\times$ 1536.]{\includegraphics[width=0.48\textwidth]{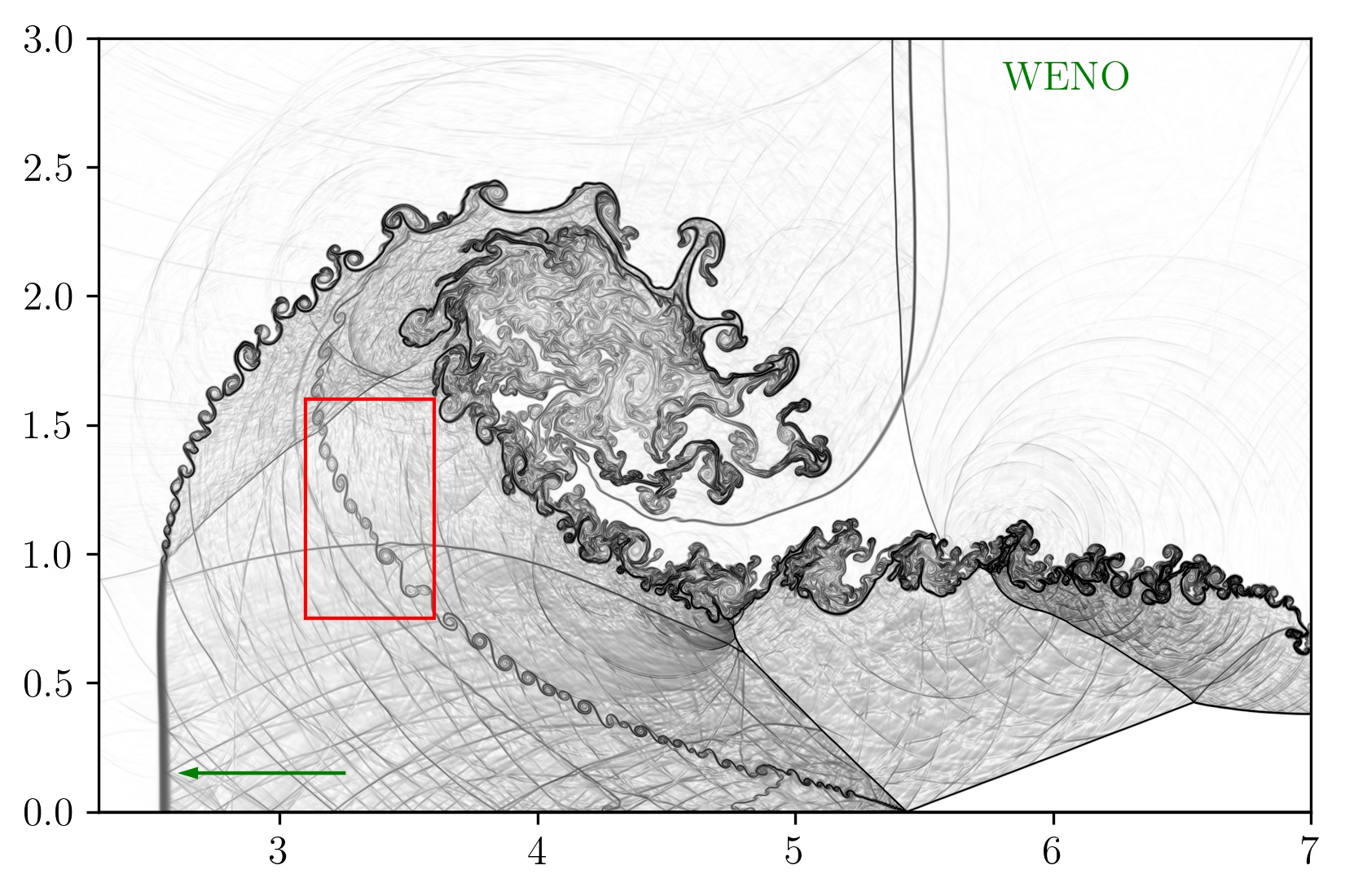}
\label{fig:fmu}}
\subfigure[MP5, 1792 $\times$ 768.]{\includegraphics[width=0.48\textwidth]{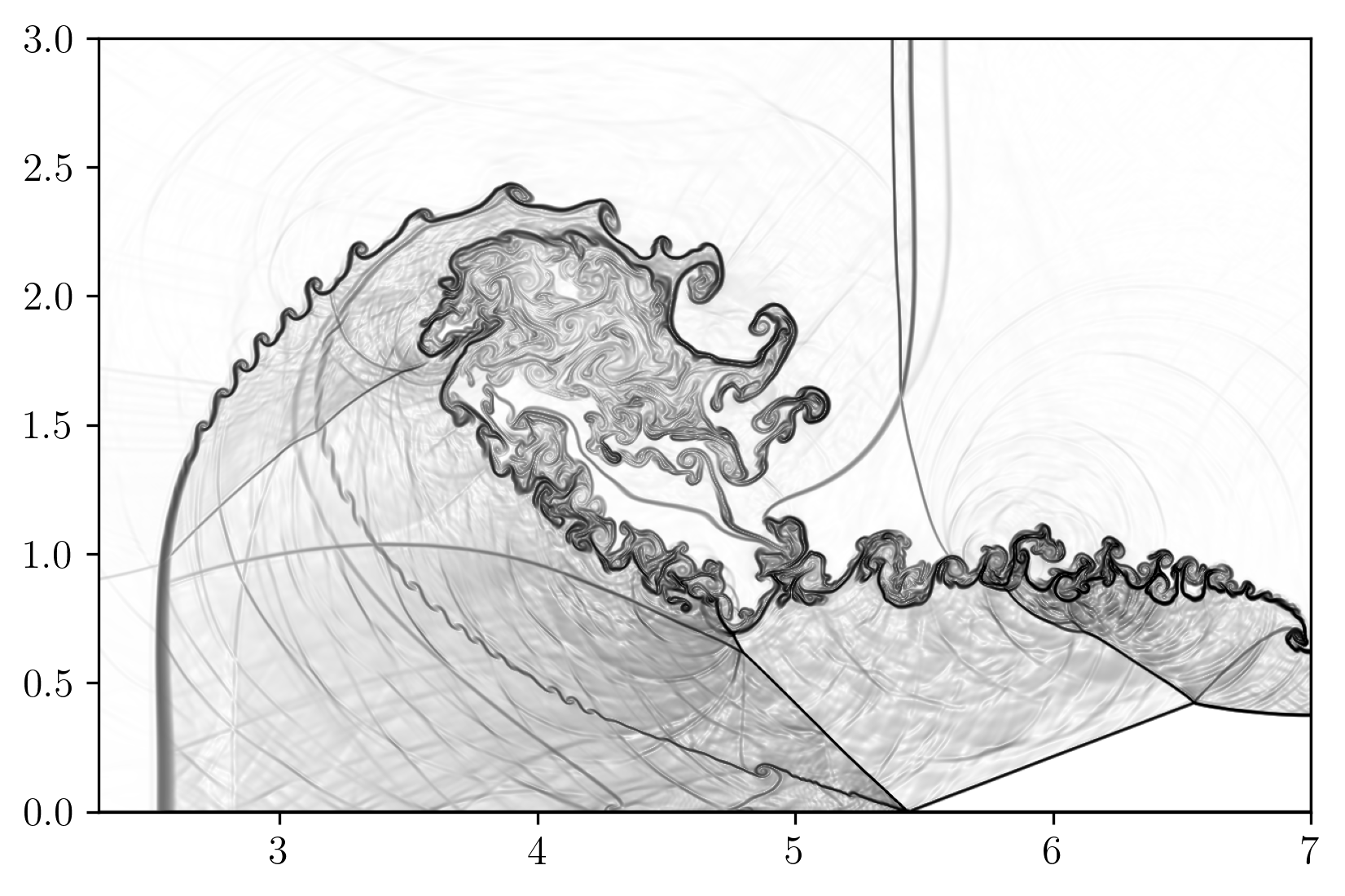}
\label{fig:mp}}
\subfigure[Wave-MP, without THINC, 1792 $\times$ 768.]{\includegraphics[width=0.48\textwidth]{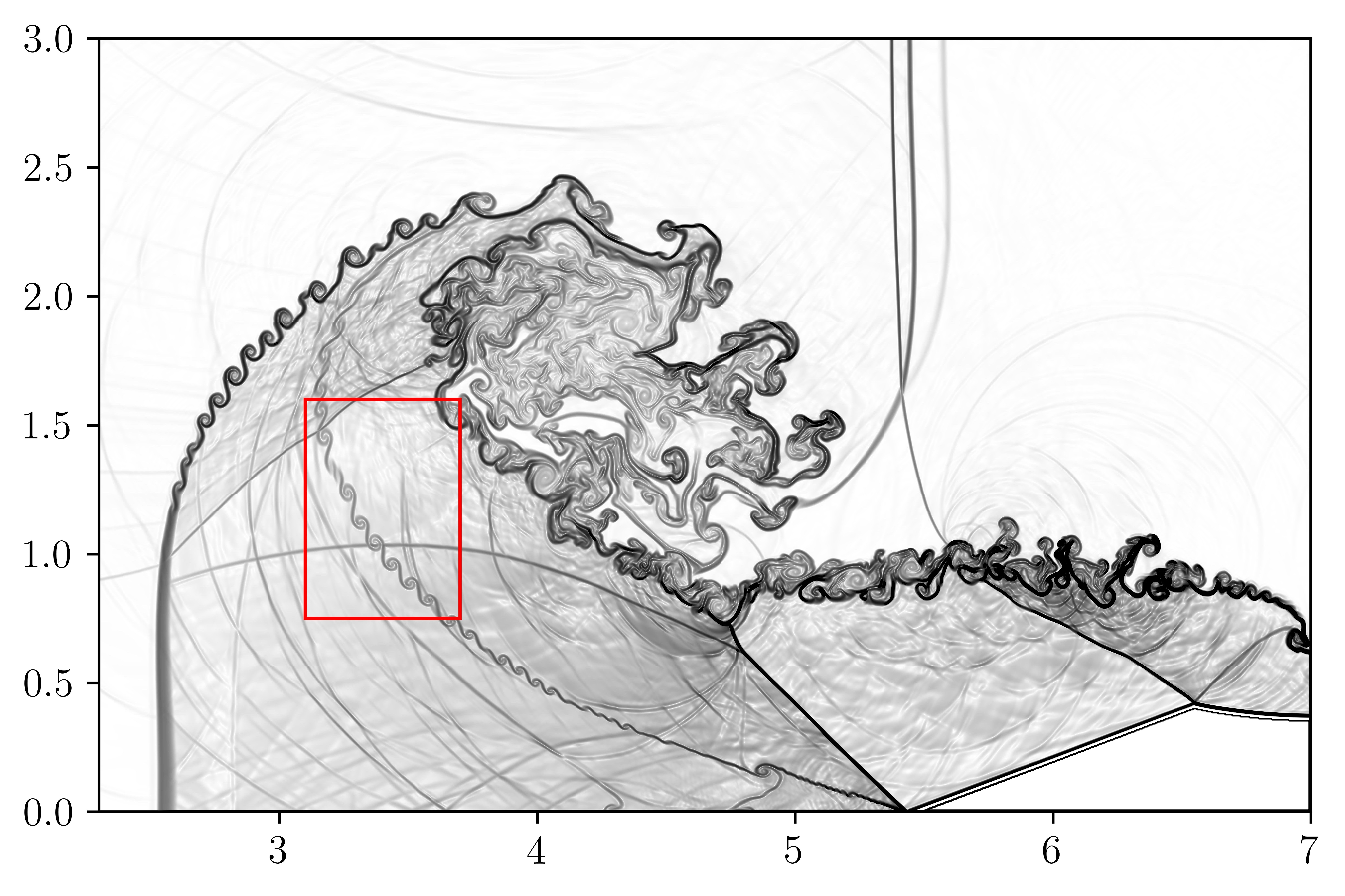}
\label{fig:mv}}
\subfigure[MP5-THINC, 1792 $\times$ 768.]{\includegraphics[width=0.48\textwidth]{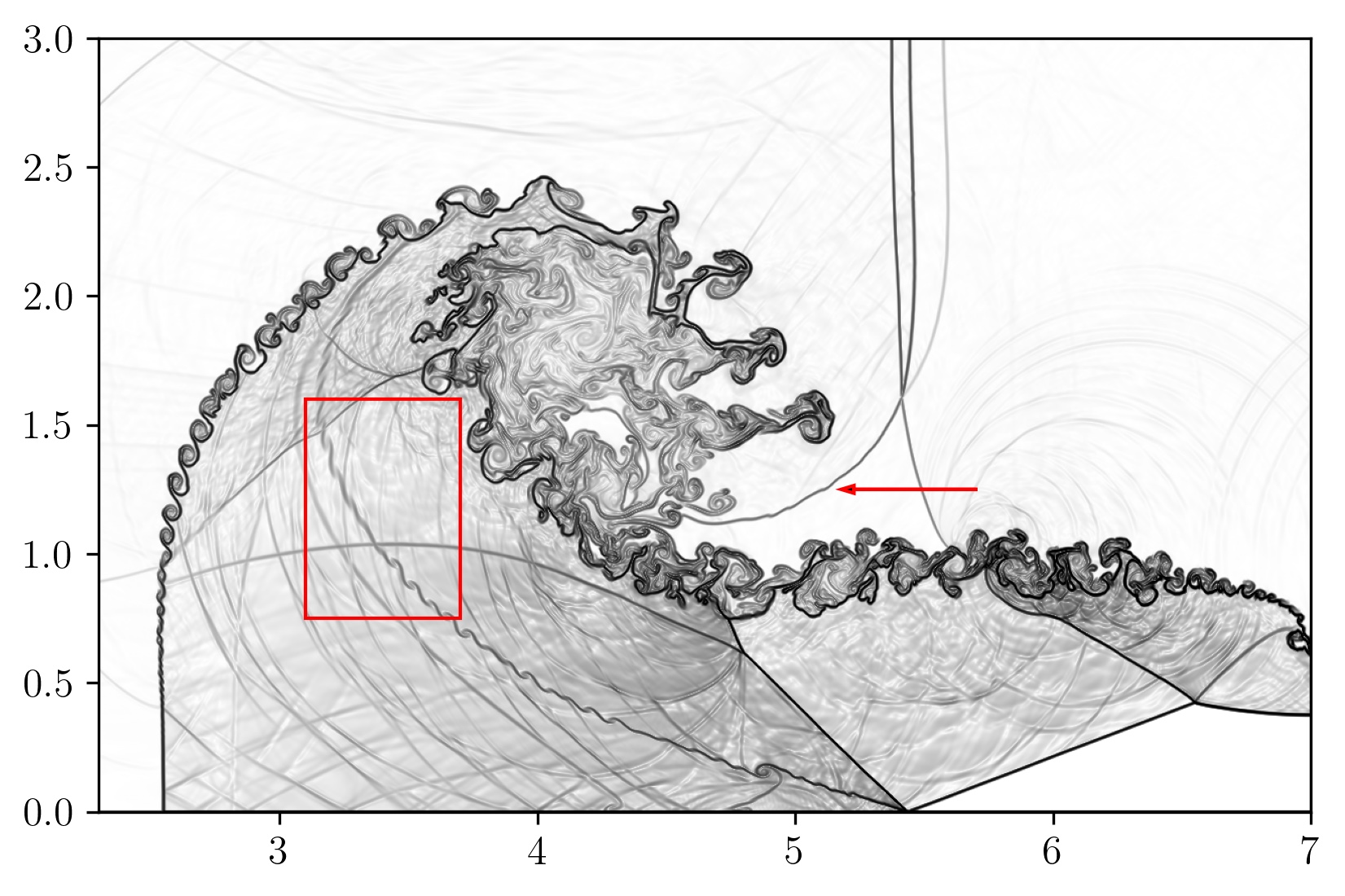}
\label{fig:fmu2}}
\subfigure[Wave-MP, 1792 $\times$ 768.]{\includegraphics[width=0.48\textwidth]{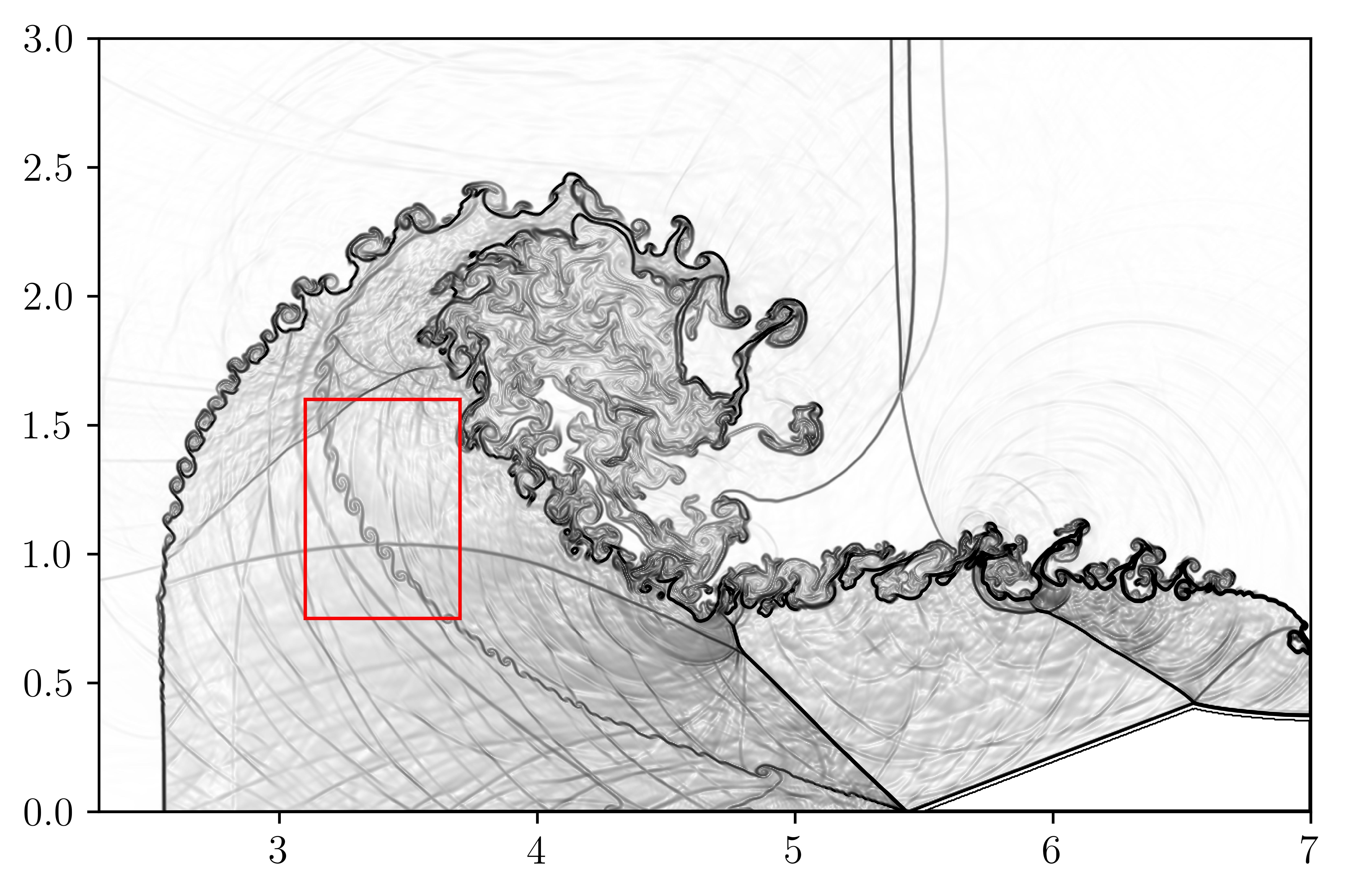}
\label{fig:fhl2}}
    \caption{Density gradient contours at time $t=5$ using various schemes, Example \ref{ex:triplet}.}
 \label{fig_fivetriple}
\end{figure}
\end{itemize}

\begin{figure}[H]
\centering
\subfigure[Sensor location, $x$-direction.]{\includegraphics[width=0.48\textwidth]{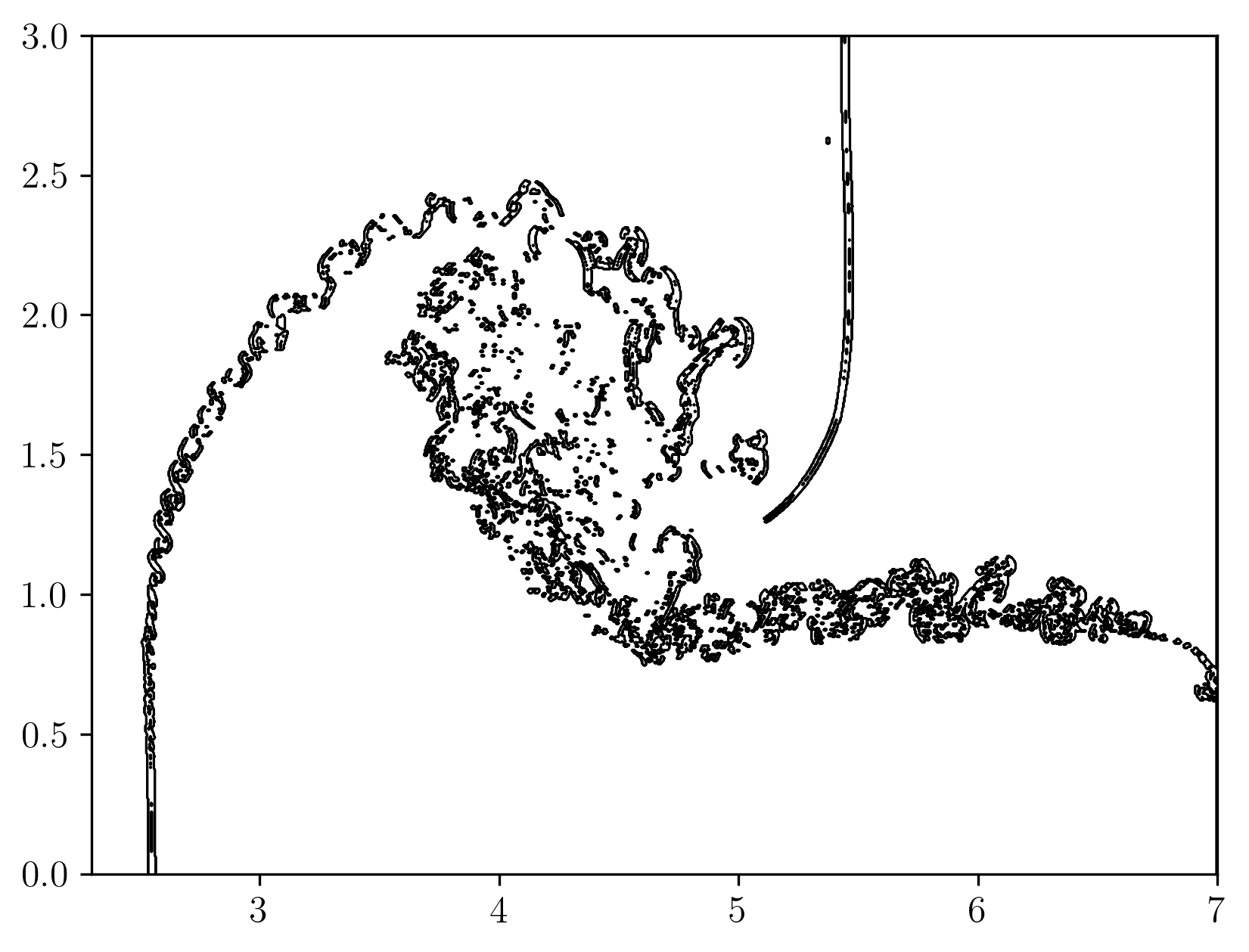}
\label{fig:sx}}
\subfigure[Sensor location, $y$-direction.]{\includegraphics[width=0.48\textwidth]{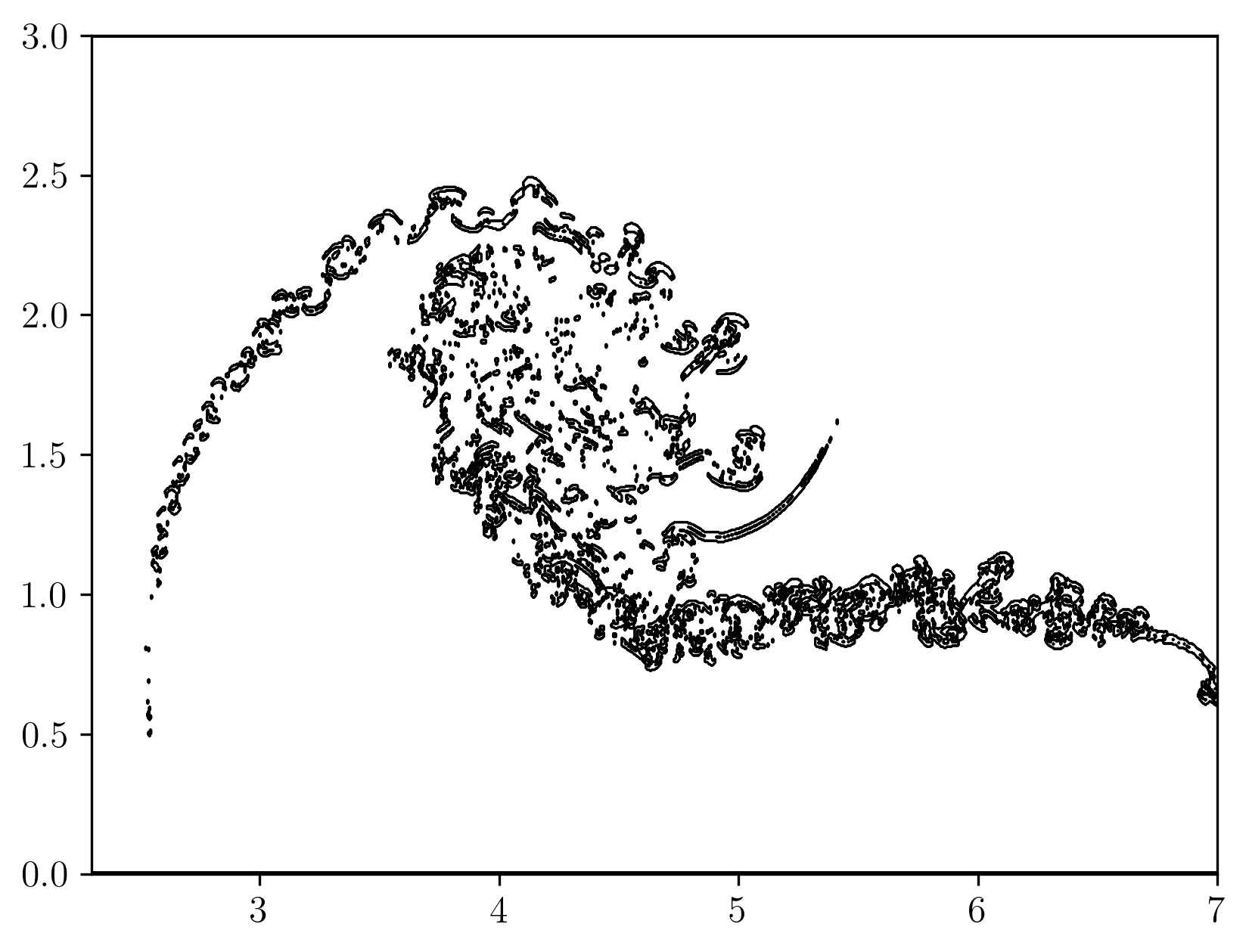}
\label{fig:sy}}
    \caption{ Figures \ref{fig:sx} and \ref{fig:sy} show sensor location regions.}
 \label{fig_fivetriple_cute}
\end{figure}

\begin{figure}[H]
\centering
\subfigure[MP5-THINC (Prim), 1792 $\times$ 768.]{\includegraphics[width=0.48\textwidth]{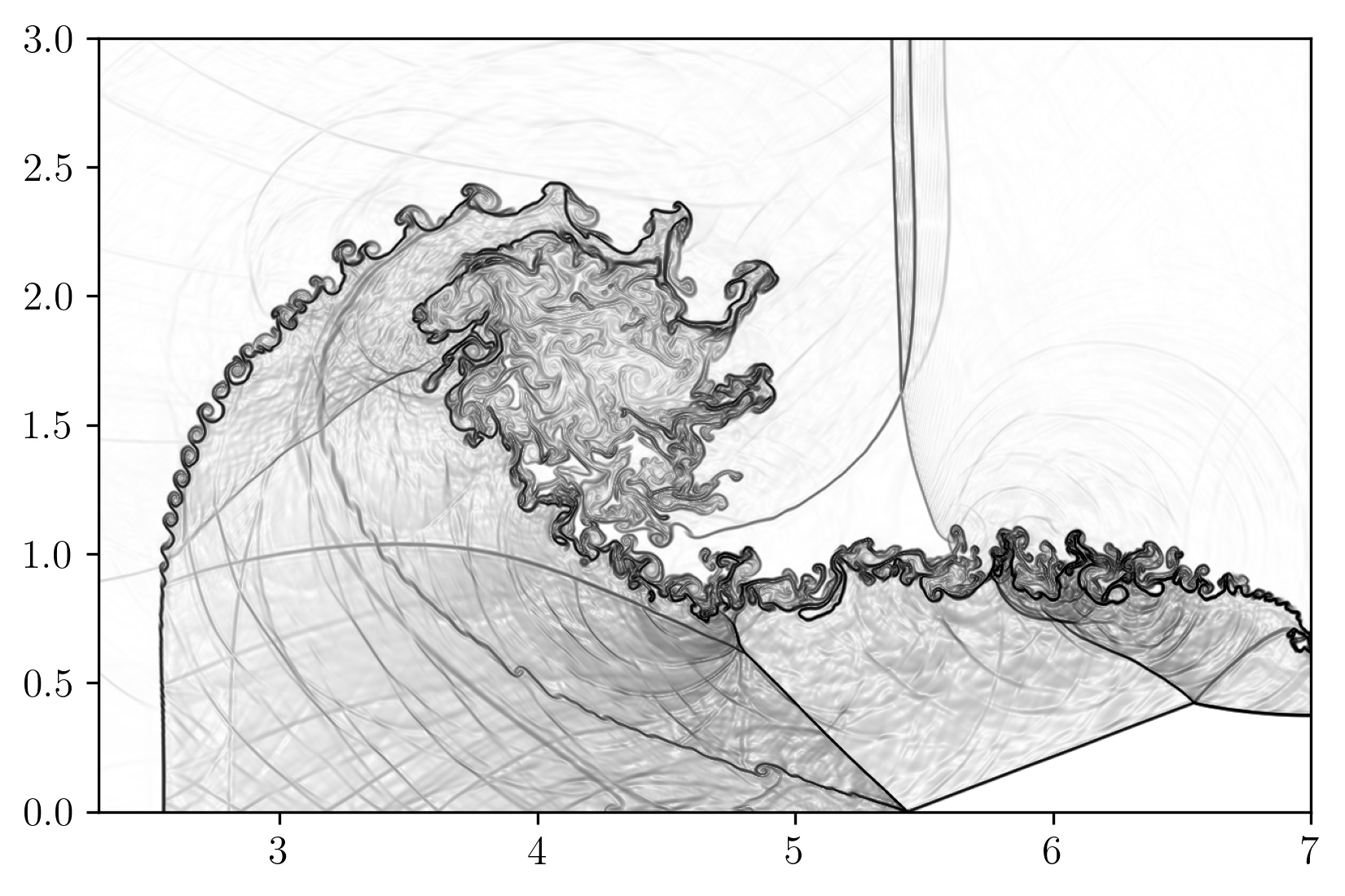}
\label{fig:fmu2p}}
\subfigure[Wave-MP (Prim), 1792 $\times$ 768.]{\includegraphics[width=0.48\textwidth]{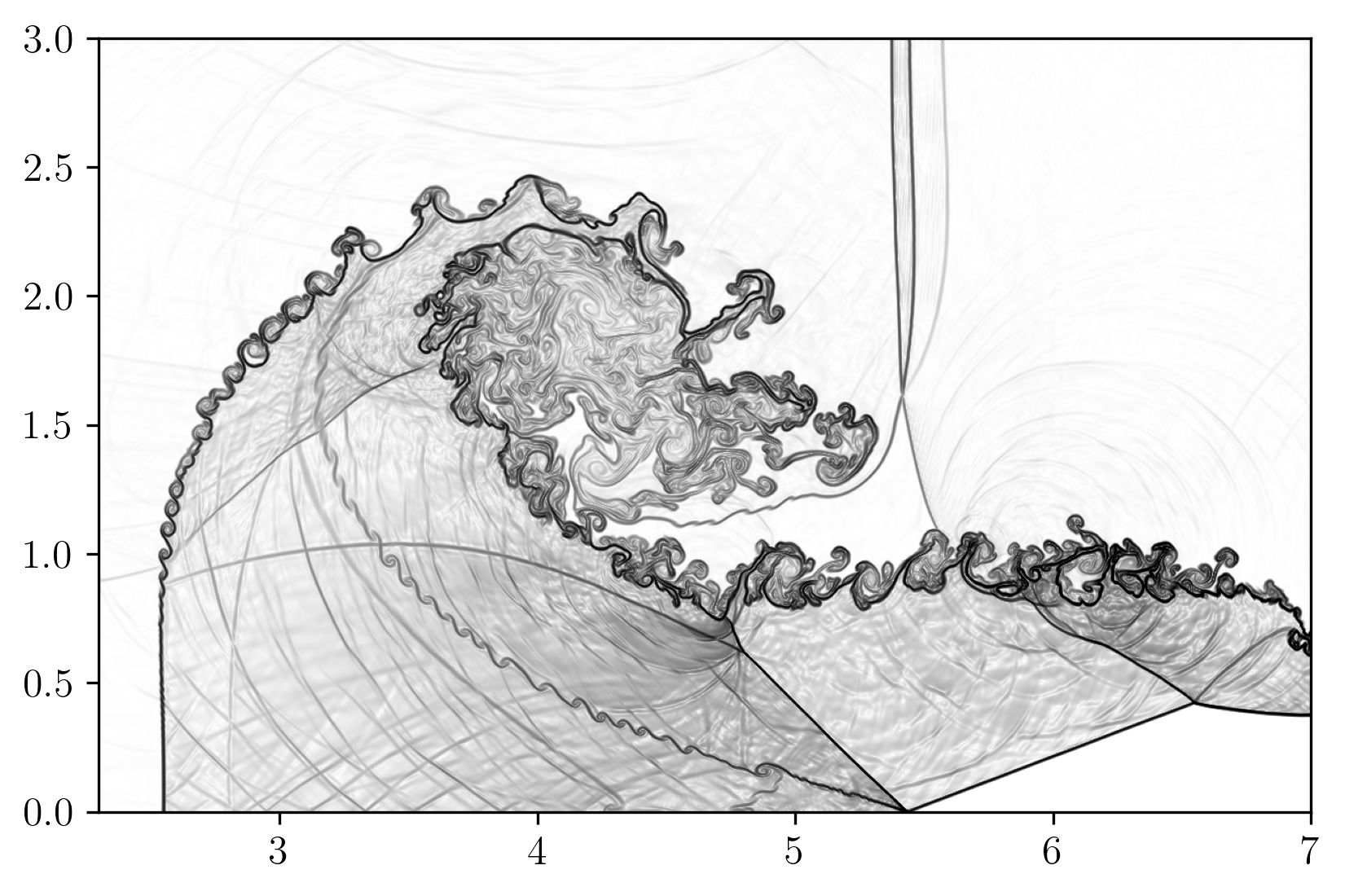}
\label{fig:fhl2p}}
    \caption{Density gradient contours at time $t=5$ using primitive variable reconstruction schemes, Example \ref{ex:triplet}.}
 \label{fig_fivetriple_prim}
\end{figure}

\begin{example}\label{shock-entropy}{2D shock-entropy wave test}
\end{example}
This test case demonstrates the impact of reconstructing the entropy wave on the numerical solution. This test case is the two-dimensional  shock-entropy wave interaction problem, modified for the multi-species case, proposed in \cite{chamarthi2021high} with the following initial conditions:
\begin{align}\label{shock_entropy}
\left(\rho, u, v, p, \alpha_{\mathrm{N}_2}, \alpha_{\mathrm{He}}\right)=
\begin{cases}
&(3.857143, \ \ 2.629369,\ 0, \ 10.3333, \ 1, 0),\quad x<-4,\\
&(1+0.2\sin(10x \cos\theta+10y\sin\theta),\ \ 0,\ 0,\ \ 1, \ 0, 1),\quad otherwise,
\end{cases}
\end{align}
over a domain of $[-5,5]\times [-1,1]$ is considered. The value of $\theta$ = $\pi/6$. Simulations are carried out on a mesh size of $400\times 80$, corresponding to $\Delta x= \Delta y =1/40$. The ``exact'' solution in Figure \ref{fig:SE-fine} is computed on a fine mesh of $1600 \times 320$ using the MP5 scheme. The fine grid solution is shown in Figure \ref{fig:SE-fine} and coarse grid results computed using Wave-MP are shown in \ref{fig:SE-coarse}.

\begin{figure}[H]
\centering\offinterlineskip
\subfigure[Fine grid]{\includegraphics[width=0.48\textwidth]{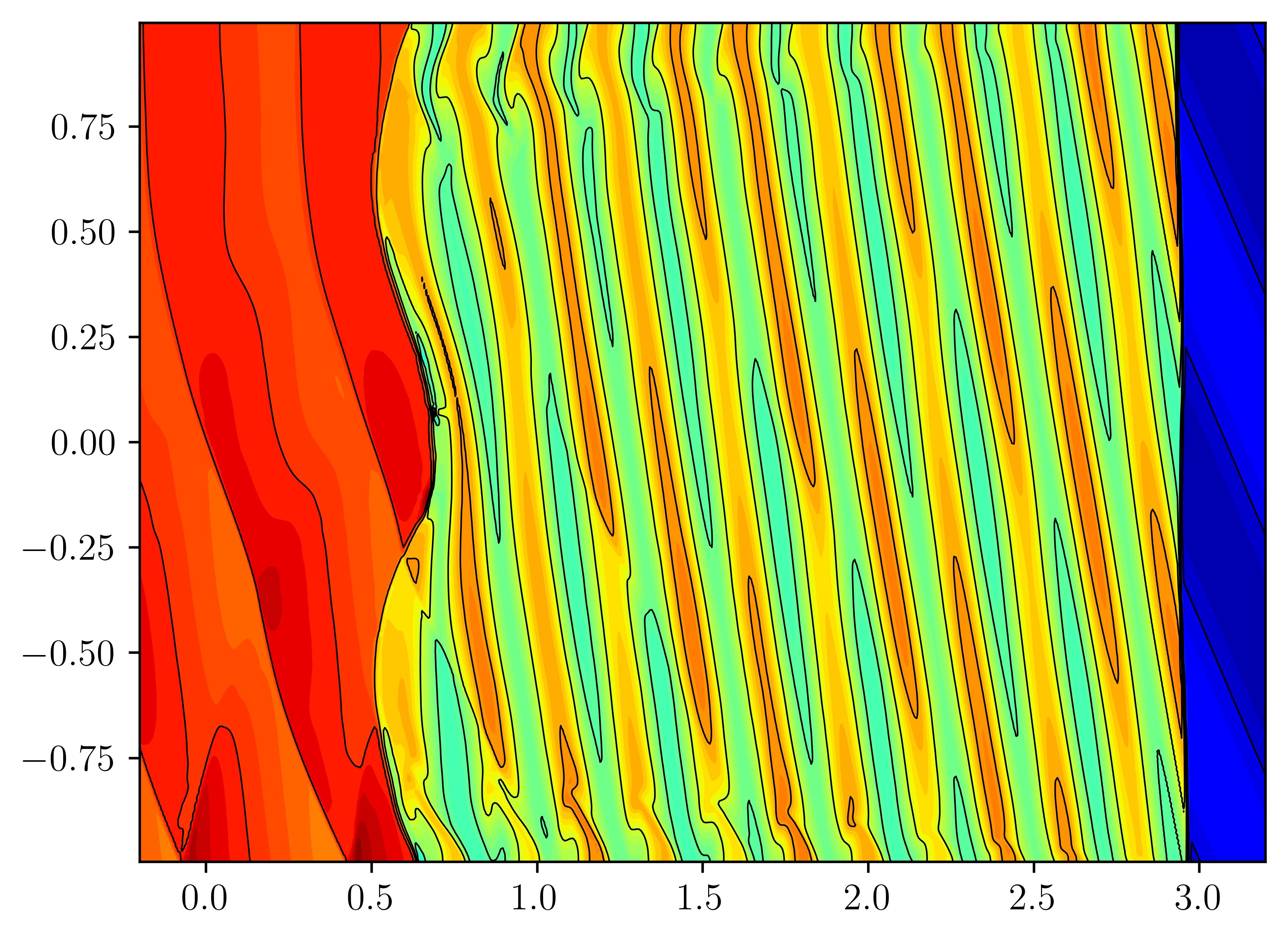}
\label{fig:SE-fine}}
\subfigure[\textcolor{black}{Wave-MP, Coarse grid }]{\includegraphics[width=0.48\textwidth]{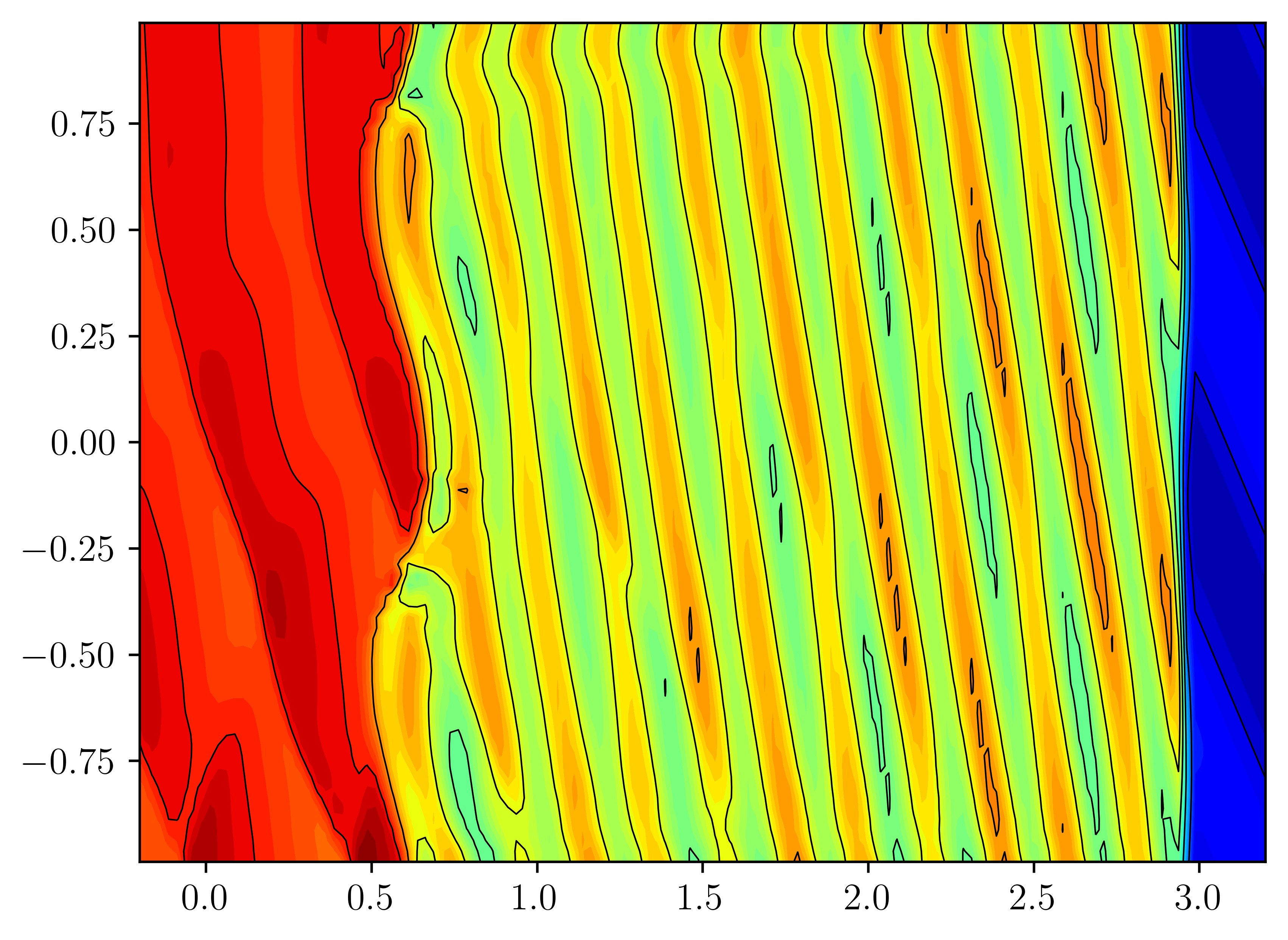}
\label{fig:SE-coarse}}
\caption{The fine grid solution is shown in Figure \ref{fig:SE-fine} and coarse grid results computed using Wave-MP are shown in \ref{fig:SE-coarse}.}
\label{fig_shock_entropy_1}
\end{figure}

 Figure \ref{fig:SE-local} shows the region's local density (along $y=0$) with high-frequency waves for the MP5 and Wave-MP schemes. Computing the vorticity waves using a central scheme did not improve the high-frequency regions of the shock-entropy wave test, as it should be. The obvious question would be how to improve it, and the answer lies in the name of the test case itself - shock-'entropy'' wave. Therefore, computing the entropy waves ${\bm{W}}_{2,3}$ by a central scheme, as shown below in Equation (\ref{eqn:contact_ent}) , did improve the resolution in the high-frequency region, as shown in Figure \ref{fig:SE-local-entropy} - blue circles depict the improved results.
\begin{equation}
    \mathbf{W}^{C}_{i+\frac{1}{2},b} = 
    \left\{
    \begin{array}{ll}
        \text{if } b = 2,3\text{:} & \begin{cases}
            \mathbf{W}^{C,Non-Linear}_{i+\frac{1}{2},b} & \text{if } \left( \mathbf{W}^{C,Linear}_{i+\frac{1}{2}} - \mathbf{W}_i \right) \left( \mathbf{W}^{C,Linear}_{i+\frac{1}{2}} - \mathbf{W}^{C,MP}_{i+\frac{1}{2}} \right) \geq 10^{-40},
            \\[10pt]
            \mathbf{W}^{C,Linear}_{i+\frac{1}{2},b} & \text{otherwise}.
            \\[10pt]
{\bm{W}}_{i+\frac{1}{2},b}^{L, T} & \text{if } \min \left(\psi_{i-1}, \psi_{i},\psi_{i+1}\right)<\psi_{c},
        \end{cases}
    \end{array}
    \right.
    \label{eqn:contact_ent}
\end{equation}
 Likewise, computing phasic densities ($\alpha_1 \rho_1,\alpha_2 \rho_2$) using a central scheme, as shown below in Equation (\ref{eqn:prim_ent}), improves the results in physical space, as shown in Figure \ref{fig:SE-local-entropy} using red stars.

\begin{equation}
    {\bm{U}}^{C}_{i+\frac{1}{2},b} = 
    \left\{
    \begin{array}{ll}
        \text{if } b = 1,2\text{:} & 
        \begin{cases}
            {\bm{U}}^{C,Non-Linear}_{i+\frac{1}{2},b} & \text{if } \left( {\bm{U}}^{C,Linear}_{i+\frac{1}{2}} - {\bm{U}}_i \right) \left( {\bm{U}}^{C,Linear}_{i+\frac{1}{2}} - {\bm{U}}^{C,MP}_{i+\frac{1}{2}} \right) \geq 10^{-40},
            \\[10pt]
            {\bm{U}}^{C,Linear}_{i+\frac{1}{2},b} & \text{otherwise}.
            \\[10pt]
			{\bm{U}}_{i+\frac{1}{2},b}^{L, T} & \text{if } \min \left(\psi_{i-1}, \psi_{i}, \psi_{i+1}\right)<\psi_{c}.
        \end{cases}
    \end{array}
    \right.
    \label{eqn:prim_ent}
\end{equation}
 Results obtained by direct reconstruction of primitive variables are \textit{noisy and not clean}, as expected, but the results are improved nevertheless.

\begin{figure}[H]
\subfigure[Wave-MP, Local profile.]{\includegraphics[width=0.48\textwidth]{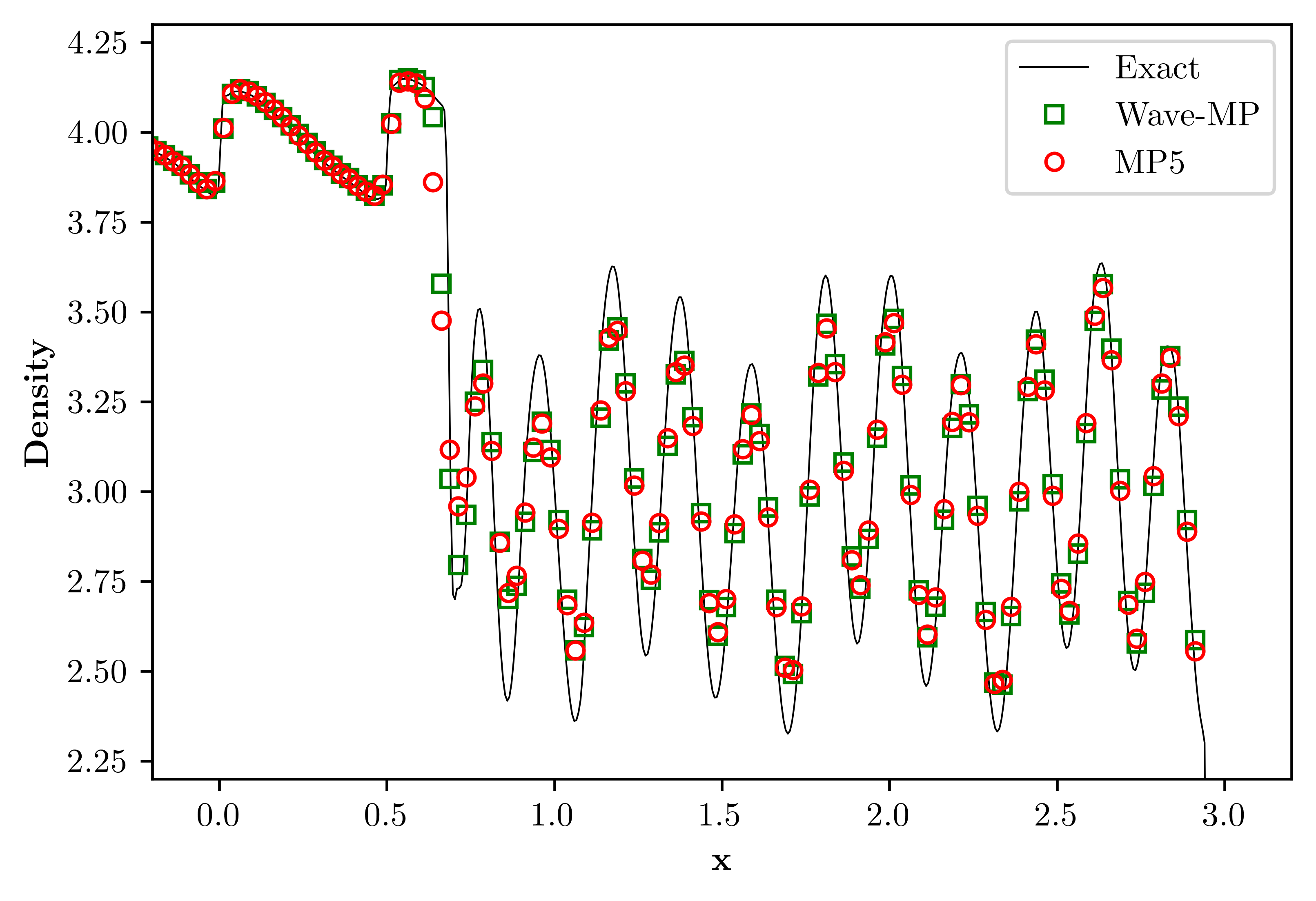}
\label{fig:SE-local}}
\subfigure[Wave-MP (Entropy wave), Local profile.]{\includegraphics[width=0.48\textwidth]{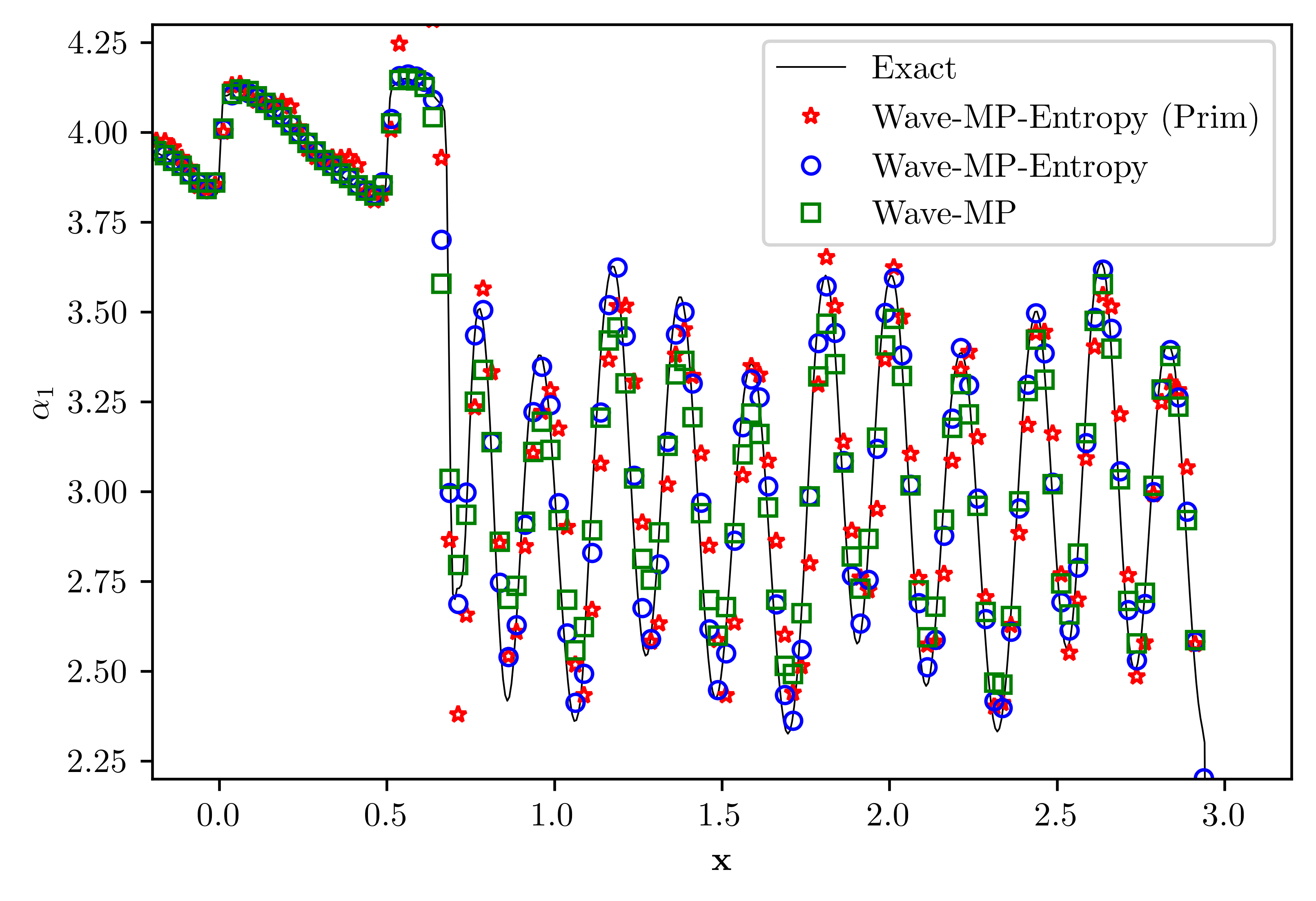}
\label{fig:SE-local-entropy}}
\caption{Figure \ref{fig:SE-local} shows the local density profiles for the MP5 and Wave-MP schemes, and  Figure \ref{fig:SE-local-entropy} shows the local density profiles for the Wave-MP, Wave-MP-Entropy and Wave-MP-Entropy (Prim) schemes.  Solid line: reference solution; red circles: MP5; green squares: Wave-MP; blue circles: Wave-MP-entropy and red stars: Wave-MP-entropy (Prim).}
\label{fig_shock_entropy_2}
\end{figure}

Key observations and insights:
\begin{itemize}
\item Vortical structures improved with a central scheme for vorticity waves in the compressible triple-point test case (Example \ref{ex:triplet}). In this shock-entropy wave test case, high-frequency regions are improved by computing entropy waves with a central scheme, whereas vorticity wave computations did not affect the entropy wave.
\item These results also indicates that the MP criteria, \textcolor{red}{Equation \ref{eqn:mp5Condition}}, can be used for both upwind and central schemes, which is not considered in the literature, and can improve the results. These results show how a reconstruction scheme that considers the physics of the Euler equations can improve the results. Modifying each wave's reconstruction scheme affects the physics of those particular waves. These findings underscore the importance of tailoring reconstruction schemes to the physics of the Euler equations, with modifications affecting specific wave types.
\item Using the central scheme for entropy waves improved the high-frequency regions instead of the upwind scheme. It also explains that not all the regions of the entropy wave are contact discontinuities and are computed using THINC. Contact discontinuities are possibly a subset of the entropy wave. Computing the entropy wave using a central scheme did not cause any oscillations near shockwaves, as shockwaves are part of acoustic waves and not entropy waves.
\item Volume fraction contours are shown in \ref{fig:SE-alpha} and the Figure \ref{fig:SE-alpha_line} shows the local volume fraction at $y/2$ indicating that there are no spurious artefacts in the volume fraction and is crisply captured. Sensor locations are shown in Figure \ref{fig:SE-sensorx} and \ref{fig:SE-sensory}, indicating that the THINC scheme is not used at the high-frequency regions and accurately detects the material interface zone.
\item Figure \ref{fig:shu-tt} shows the discontinuity detection regions of the TENO-THINC scheme \cite{takagi2022novel}, and it can be observed that the TENO-based detector is modifying the high-frequency regions-which indicates the deficiencies of the TENO-based detector. Developers of the TENO scheme also used THINC and TENO in Discontinuous Galerkin methods. Despite using a different algorithm, the THINC scheme was getting activated in the regions of the high-frequency region, and the results were inferior compared to that of the WENO. The authors did mention that their proposed TENO-THINC limiter does not outperform the WENO limiter \cite{huang2025new}. The current algorithm is free of such spurious results as it identifies the material interfaces and not high-frequency regions. \textbf{These comparisons highlight the robustness of the current method while supporting its validity without critiquing prior works.}
\end{itemize}

\begin{figure}[H]
\centering\offinterlineskip
\subfigure[Volume fraction contours]{\includegraphics[width=0.46\textwidth]{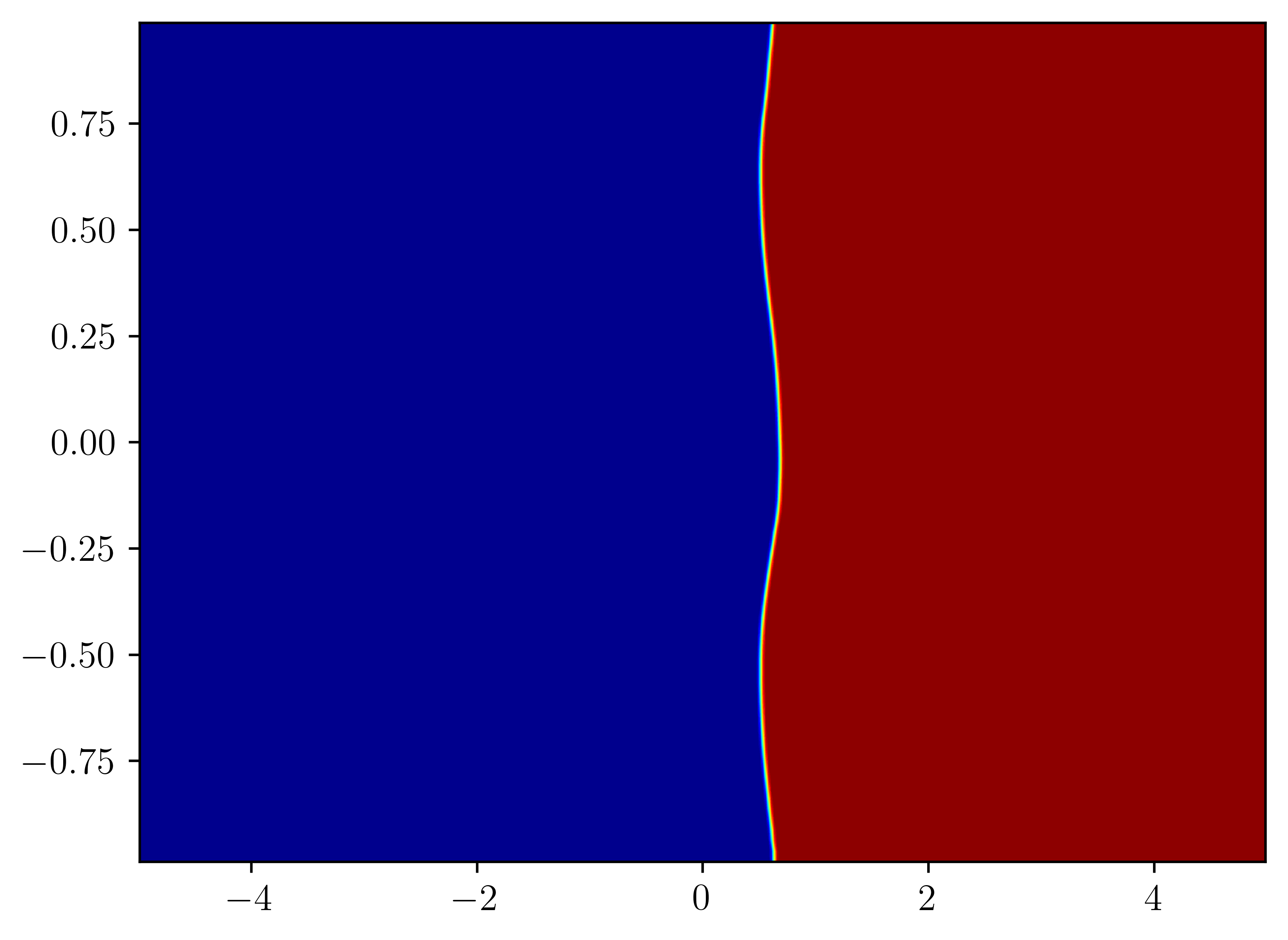}
\label{fig:SE-alpha}}
\subfigure[Volume fraction, $\alpha_1$, at $y$/2]{\includegraphics[width=0.48\textwidth]{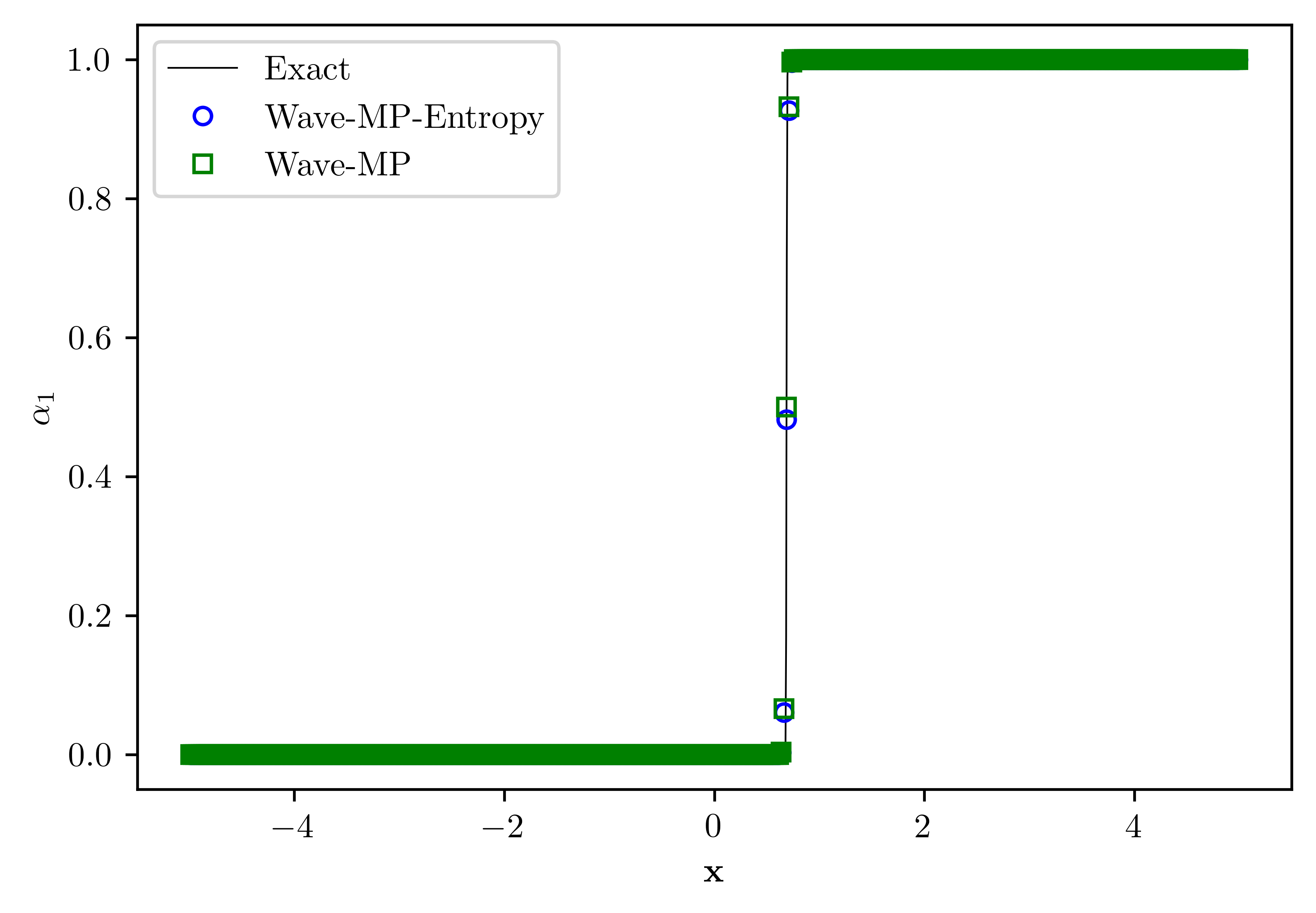}
\label{fig:SE-alpha_line}}
\subfigure[Sensor location, $x$-direction]{\includegraphics[width=0.46\textwidth]{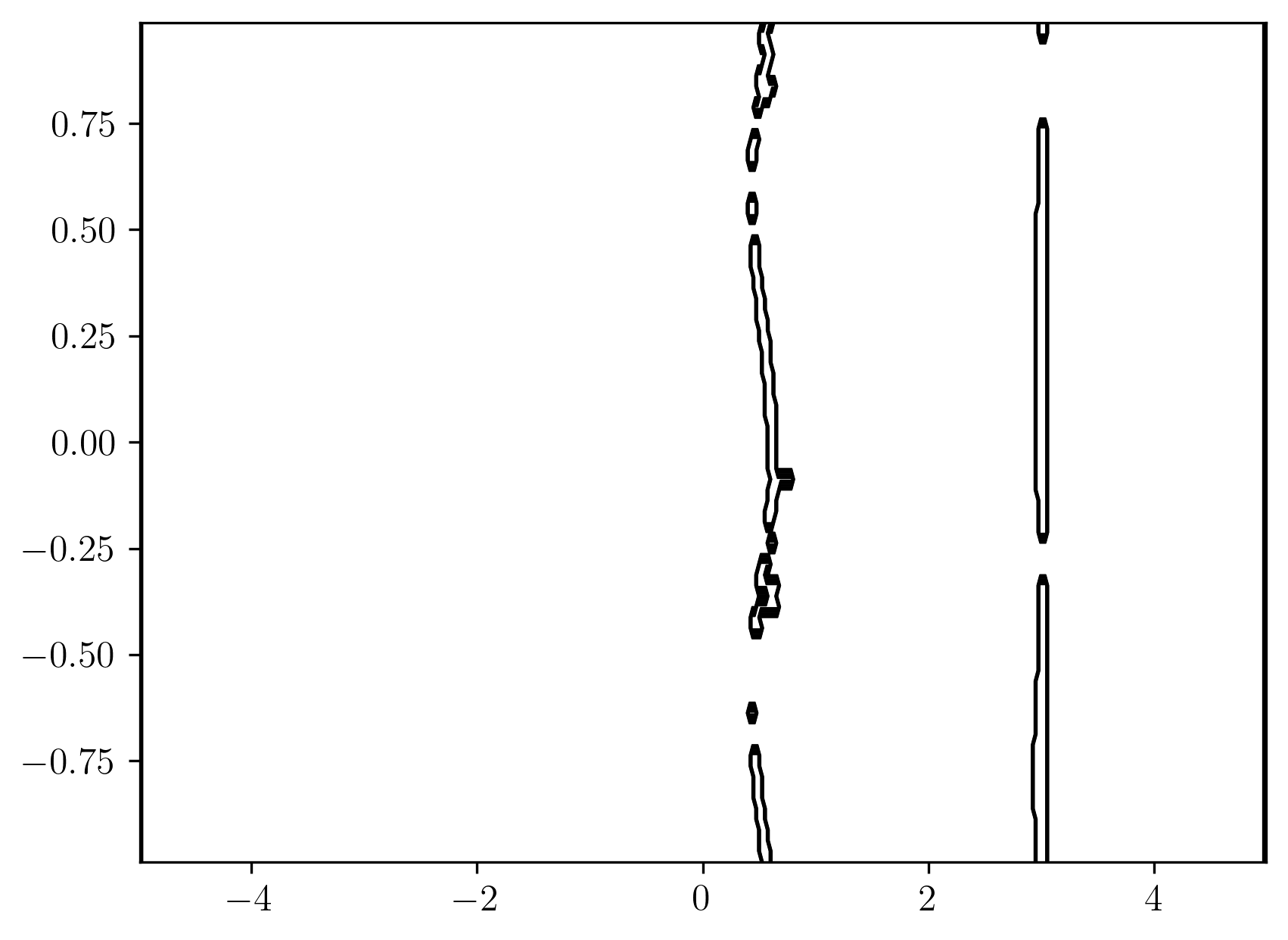}
\label{fig:SE-sensorx}}
\subfigure[Sensor location, $y$-direction]{\includegraphics[width=0.46\textwidth]{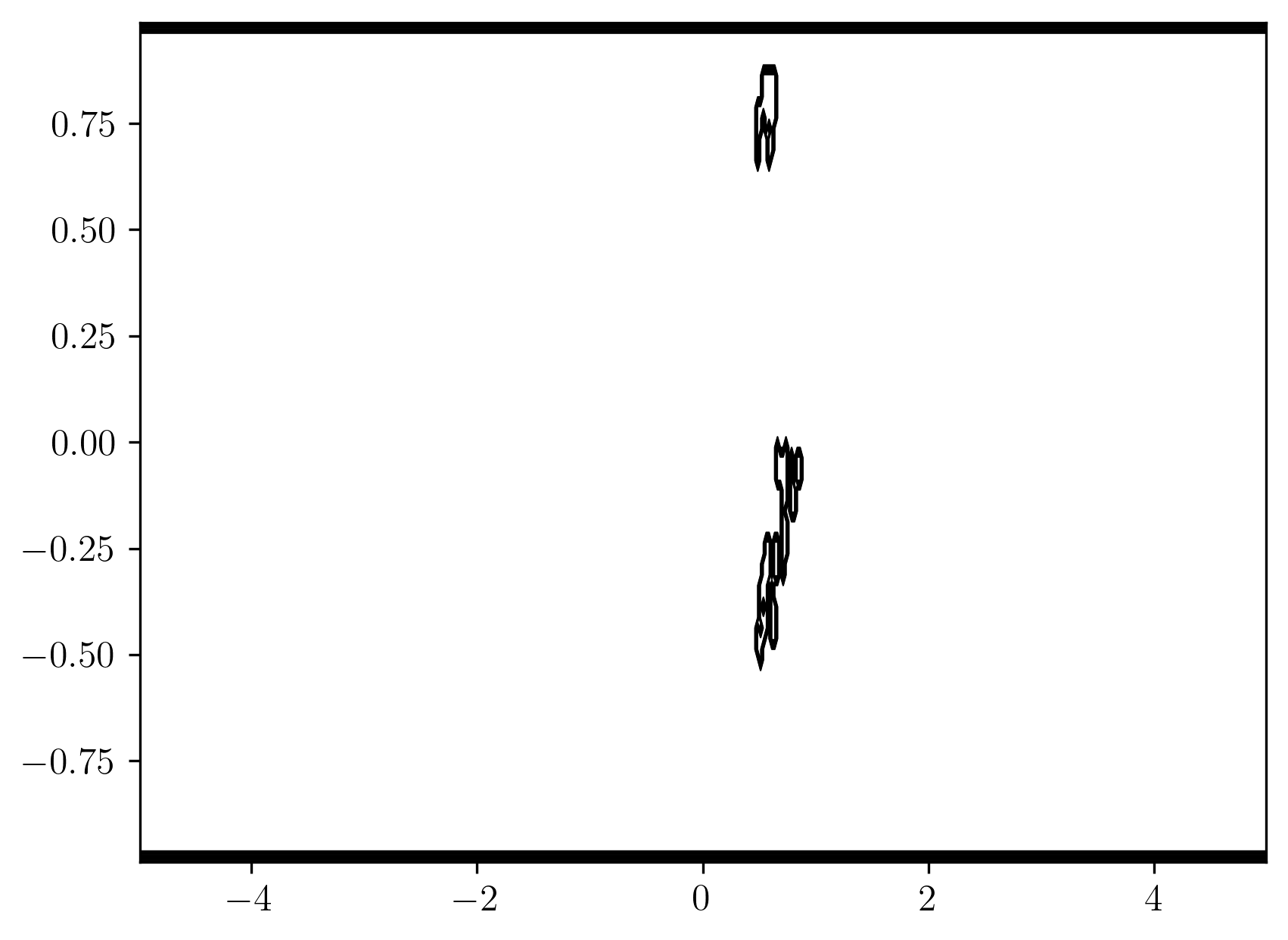}
\label{fig:SE-sensory}}
\caption{Volume fraction contours are shown in \ref{fig:SE-alpha}, Example \ref{shock-entropy}. Figure \ref{fig:SE-alpha_line} shows the local volume fraction at $y/2$. Sensor locations is shown in Figure \ref{fig:SE-sensorx} and \ref{fig:SE-sensory}.}
\label{fig_shock_entropy_3}
\end{figure}

\begin{figure}[H]
\centering
\subfigure[TENO-THINC detector \cite{takagi2022novel}]{\includegraphics[width=0.3\textwidth]{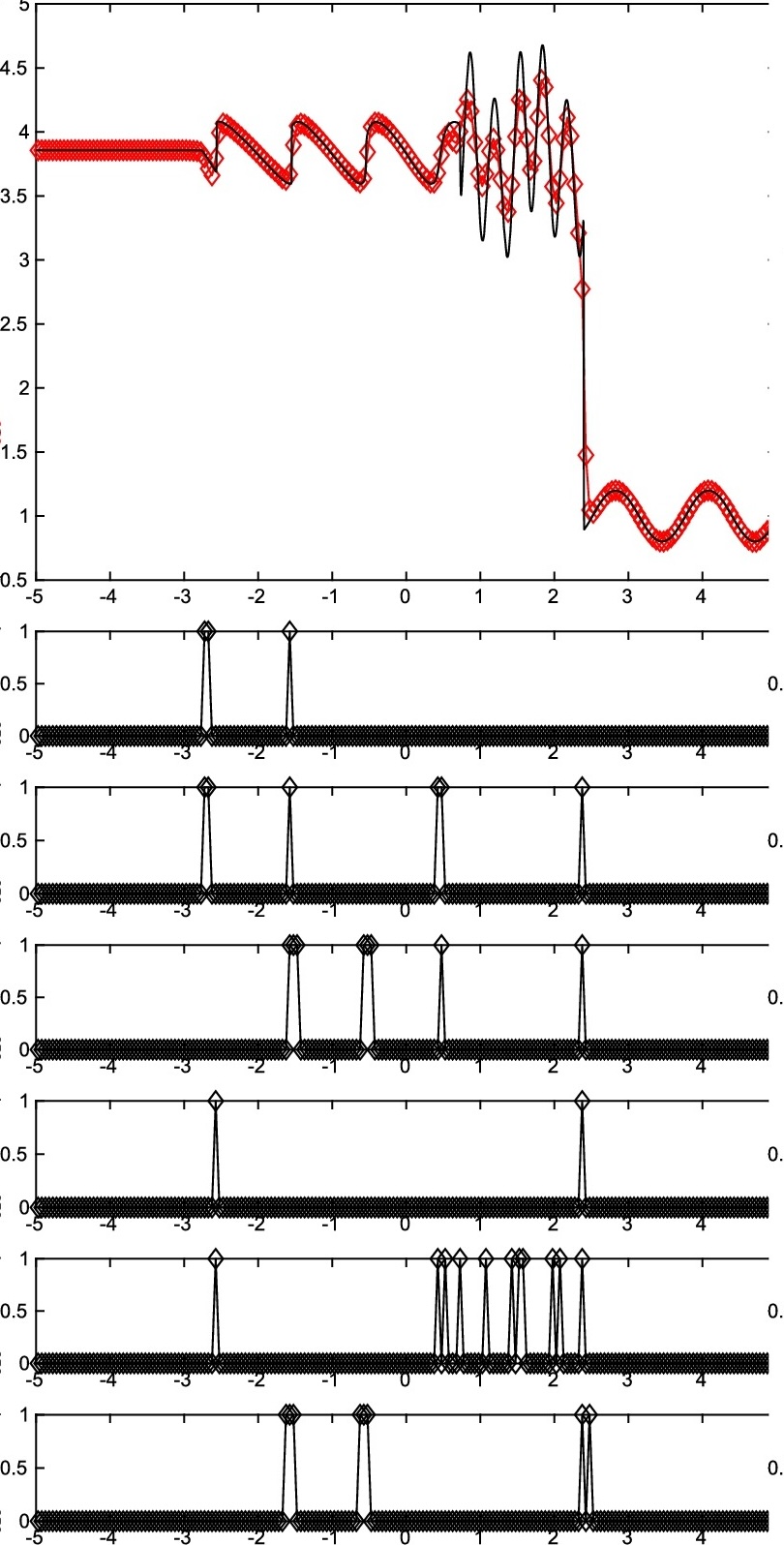}
\label{fig:shu-tt}}
\subfigure[TENO-THINC, RKDG \cite{huang2025new}]{\includegraphics[width=0.4\textwidth]{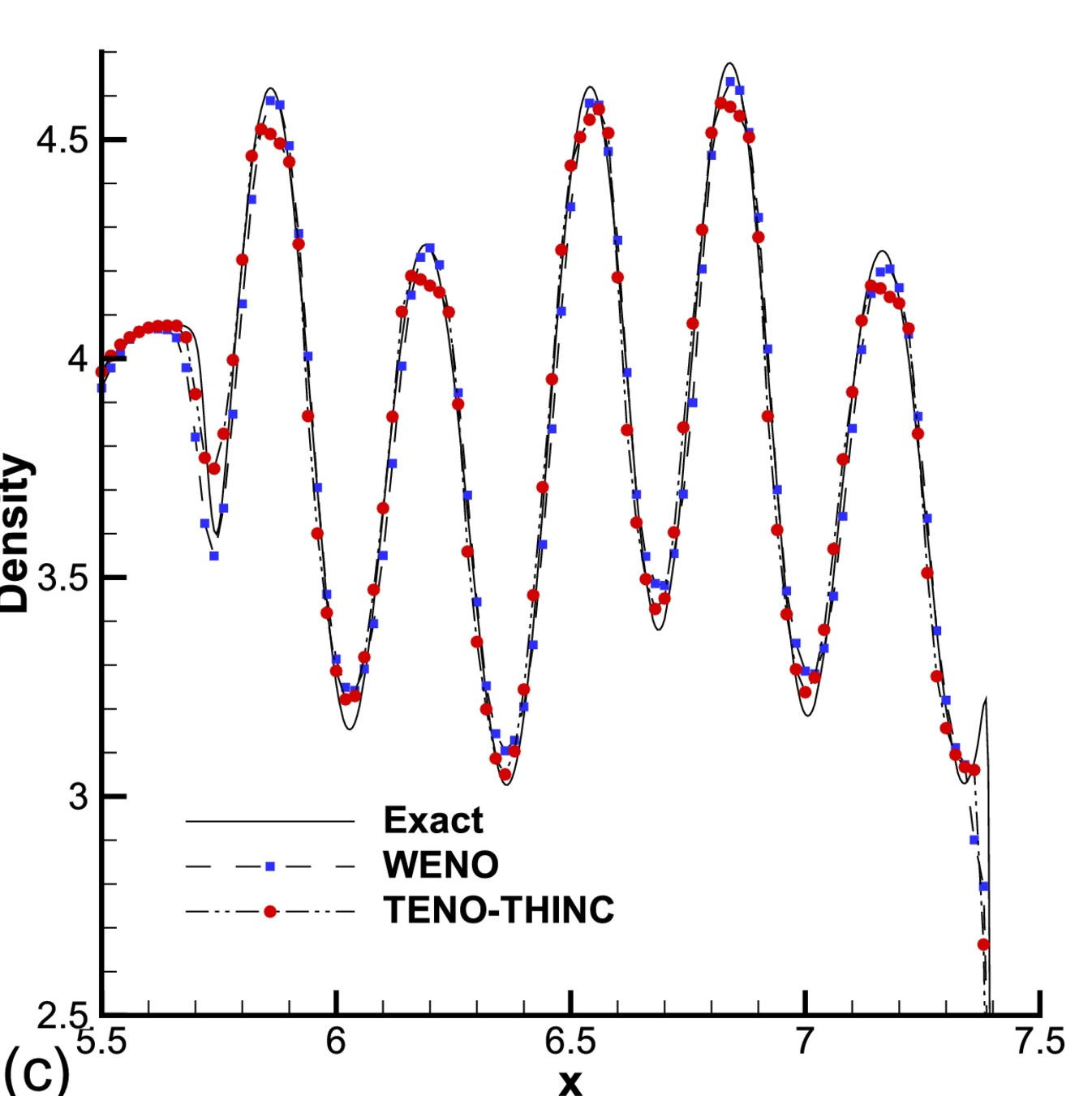}
\label{fig:shu-kriv}}
\caption{Discontinuity detection locations from various papers using THINC from the literature. Figure \ref{fig:shu-tt} is reproduced from \cite{takagi2022novel} with permission from Elsevier BV 2024, License number 5734370293424. Figure \ref{fig:shu-kriv} is reproduced from \cite{huang2025new} with permission from Elsevier BV 2024, License number 5936770943086.}
\label{fig_shu-s}
\end{figure}

\begin{example}{Two-dimensional cylindrical cavity collapse}\label
{ex:airpocket}
\end{example}

In this test case, cylindrical air cavity interaction with a Mach 1.547 shock in water is considered. The initial conditions for the test case are as follows:

\begin{equation}
\left(\alpha_1 \rho_1, \alpha_2 \rho_2, u, v, p, \alpha_1\right)= \begin{cases}(1000,0,0,0,10^5,1), & \text { for Post-shock, } \\ (1219.9,0,424.55,0,10^9,1), & \text { for Pre-shock, } \\ (0,1,0,0,10^5,0), & \text { for Bubble, }\end{cases}
\end{equation}
with fitting parameters $\gamma_1$= 1.4, $\pi_{\infty,1}$=0 for air and $\gamma_2$= 6.12, $\pi_{\infty,2}$=3.43 $\times$ $10^8$ for water. The right-moving planar shock in water is located at x = 4 mm at t = 0 in a square computational domain of 20mm. An 8 mm initial diameter air cavity in a water medium is at the centre of the domain. Non-reflection boundary conditions are applied to all domain boundaries. Simulation is performed on a grid size of 800 $\times$ 800, and the final time of simulation is 7.47 $\mu$s. Figure \ref{cavity-collapse} shows the numerical Schlieren images of the shock-bubble case at 3.25, 5.57, 6.06 and 7.47 $\mu$s computed with the Wave-MP scheme. As a result of the shock impact, the bubble undergoes compression and deformation. The diffraction shock moves along the bubble's surface while the primary jet hits the downstream surface, causing the bubble to split into two parts. This impact creates a water-hammer shock wave, and each split section of the bubble undergoes further compression. The bubble eventually splits into four parts. Complete primitive variable algorithm failed to pass this test case beyond 6.06 $\mu$s.

\begin{figure}[H]
\centering
\includegraphics[width=0.7\textwidth]{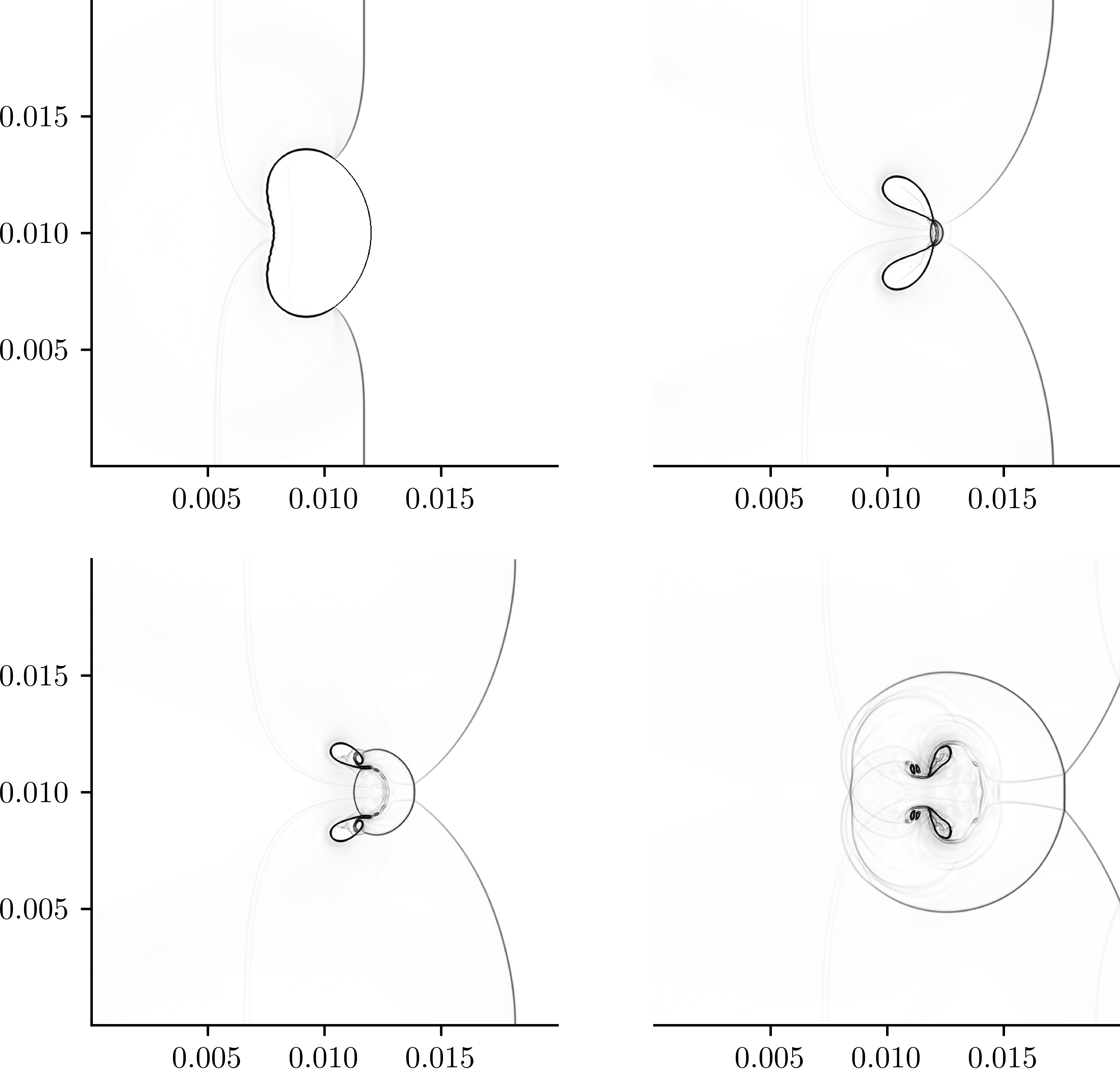}
\caption{Cylindrical cavity collapse, Example \ref{ex:airpocket}, using Wave-MP approach.}
\label{cavity-collapse}
\end{figure}

\begin{example}{Underwater explosion}\label{underoos}
\end{example}
In this example, the explosion of a highly compressed cylindrical air bubble in water under a free surface is investigated \cite{deng2018high,shukla2010interface}. The computational domain spans from [-2, 2] $\times$ [-1.5, 2.5], and reflective boundary conditions are implemented. The solution is evolved on grid sizes of 600 $\times$ 600,  1200 $\times$ 1200 and 2400 $\times$ 2400 up until the time t = 0.19. The air-water free surface lies at $y$ = 0 and the center of the air bubble is $(x_c, y_c)$ = (0.0,-0.3) with a radius r = 0.12. The initial conditions of this test case are as follows:
\begin{equation}
\left(\rho_1 \alpha_1, \rho_2 \alpha_2, u, v, p\right)= \begin{cases}(0,1.225,0,0,1.01325) & \text { if } y>0 \\ \left(0,1250,0,0,10^5\right) & \text { if } r<0.12 \\ (1000,0,0,0,1.01325) & \text { otherwise },\end{cases}
\end{equation}
with fitting parameters $\gamma_1$= 1.4, $\pi_{\infty,1}$=0 for air and $\gamma_2$= 4.4, $\pi_{\infty,2}$=6000 for water. Figures \ref{fig:under00} and \ref{fig:under_prim} show the normalized density gradient magnitudes computed using various numerical schemes. The results effectively capture the transmitted shock in water, the reflected shock in air, and the subsequent rarefaction waves. Figures \ref{fig:uw_musl}, \ref{fig:uw_weno}, \ref{fig:uw_wav-muscl}, and \ref{fig:uw_wavemp} present the numerical results computed using the MUSCL, WENO, Wave-MUSCL, and Wave-MP schemes, respectively.

The results from the MUSCL and WENO schemes exhibit excessive dissipation of the material interface, whereas the Wave-MUSCL and Wave-MP schemes capture the interfaces sharply, as the THINC scheme is used for interface capturing. There is no noticeable difference between the results from the Wave-MUSCL and Wave-MP schemes, as both use the MUSCL/THINC scheme near the gas-liquid interface, correctly identifying the liquid interface based on the stiffened gas parameter. The thin water bridge between the expanding bubble and the surrounding air remains intact even in the later stages of the process with the Wave-MP scheme. Material interface sensor locations are shown in Figures \ref{fig:sens_x} and \ref{fig:sens_y}.
\begin{figure}[H]
%\begin{halfspacing}
\centering\offinterlineskip
\subfigure[\textcolor{black}{MUSCL}]{\includegraphics[width=0.4\textwidth]{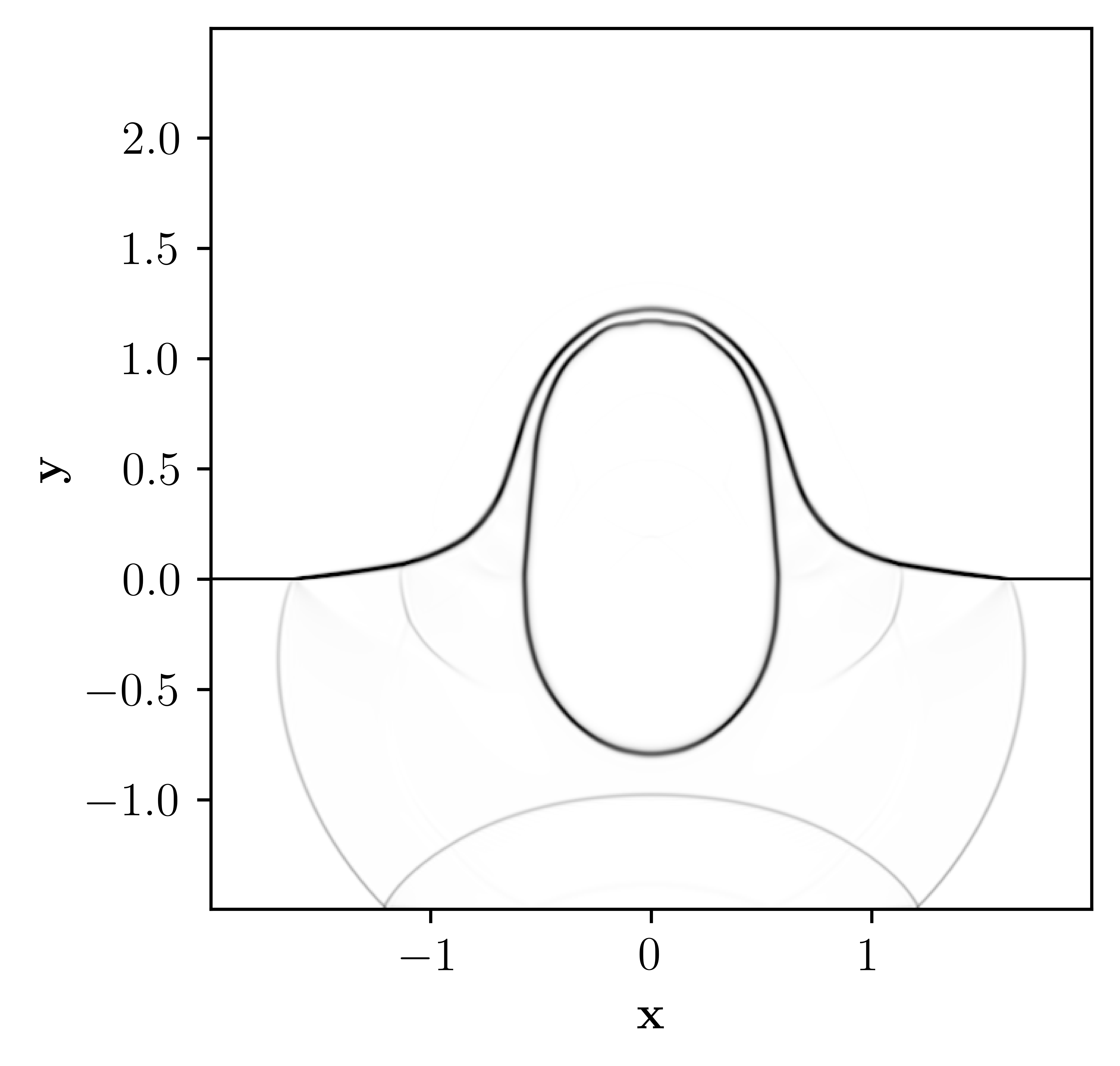}
\label{fig:uw_musl}}
\subfigure[\textcolor{black}{WENO}]{\includegraphics[width=0.4\textwidth]{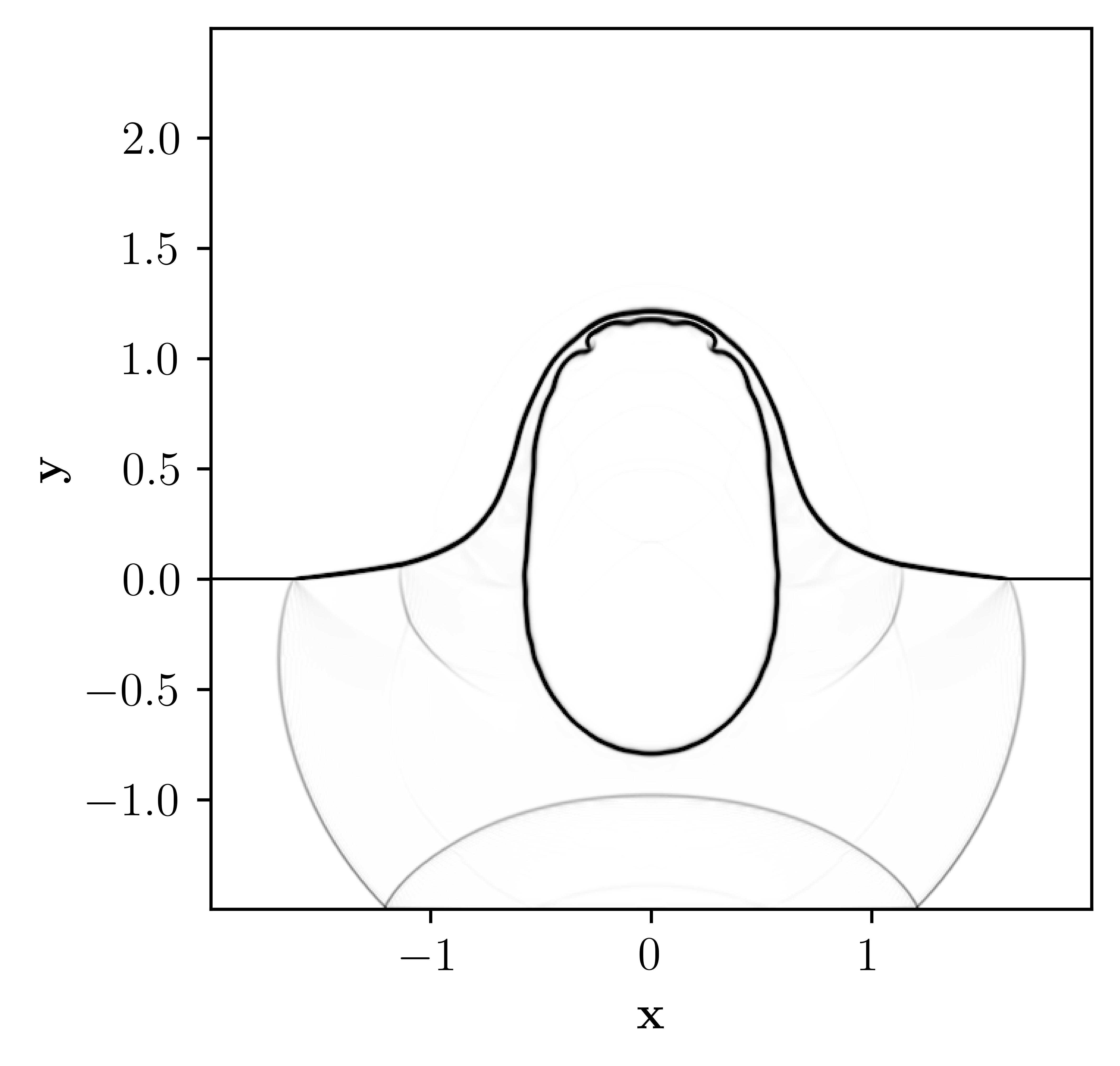}
\label{fig:uw_weno}}
\subfigure[\textcolor{black}{Wave-MUSCL}]{\includegraphics[width=0.4\textwidth]{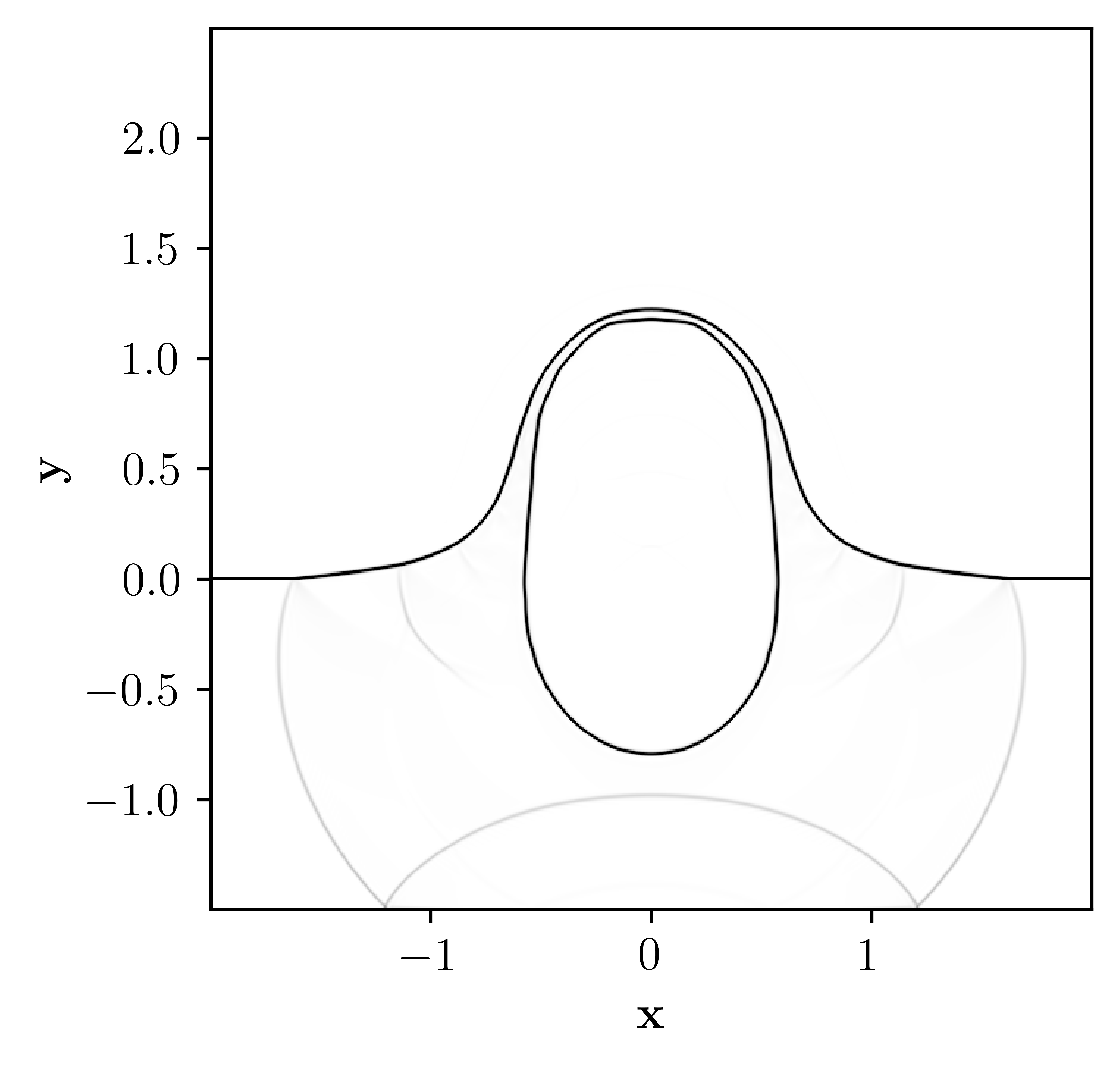}
\label{fig:uw_wav-muscl}}
\subfigure[\textcolor{black}{Wave-MP}]{\includegraphics[width=0.4\textwidth]{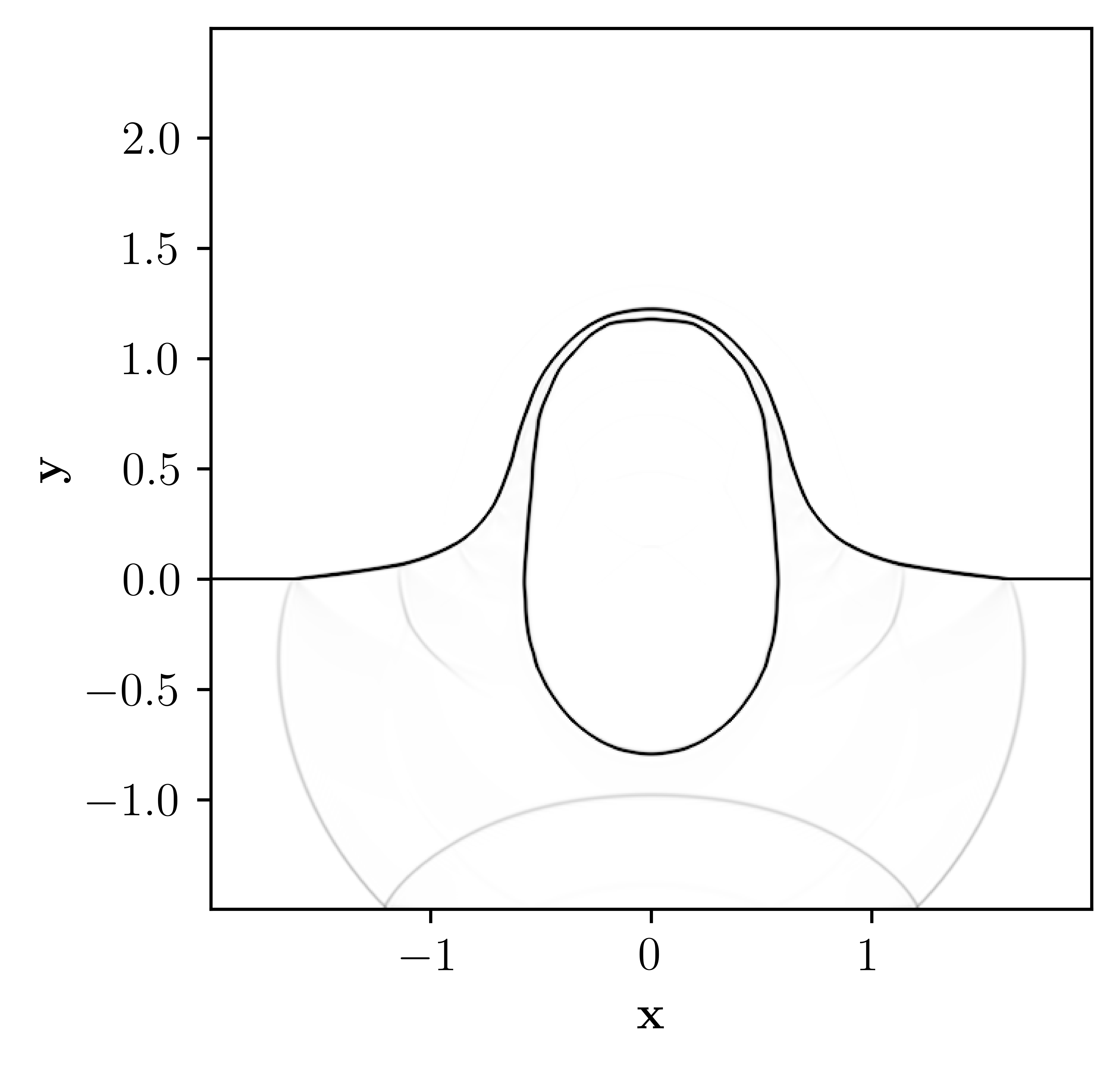}
\label{fig:uw_wavemp}}
\subfigure[\textcolor{black}{Sensor location, $x$-direction}]{\includegraphics[width=0.4\textwidth]{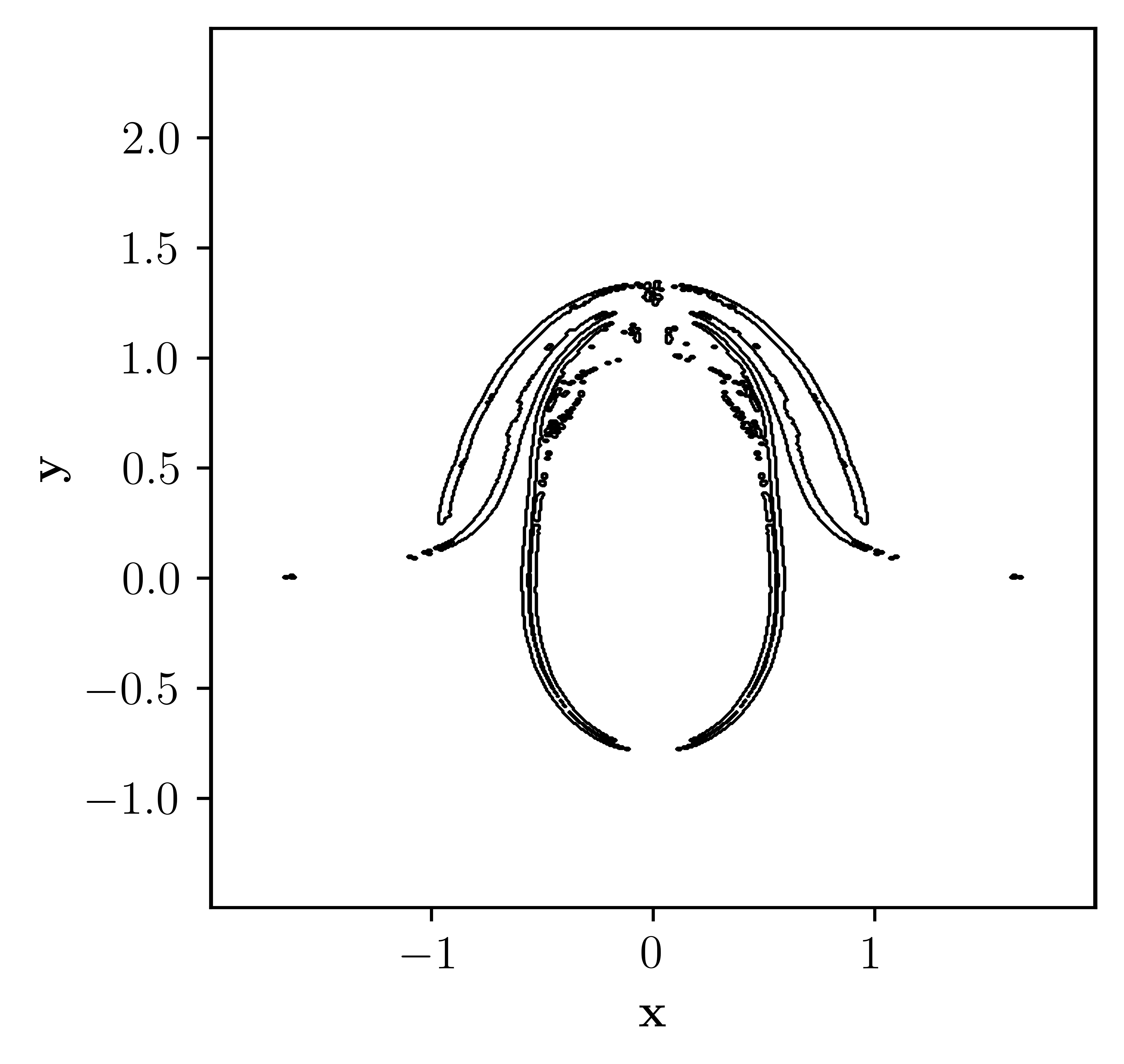}
\label{fig:sens_x}}
\subfigure[\textcolor{black}{Sensor location, $y$-direction}]{\includegraphics[width=0.4\textwidth]{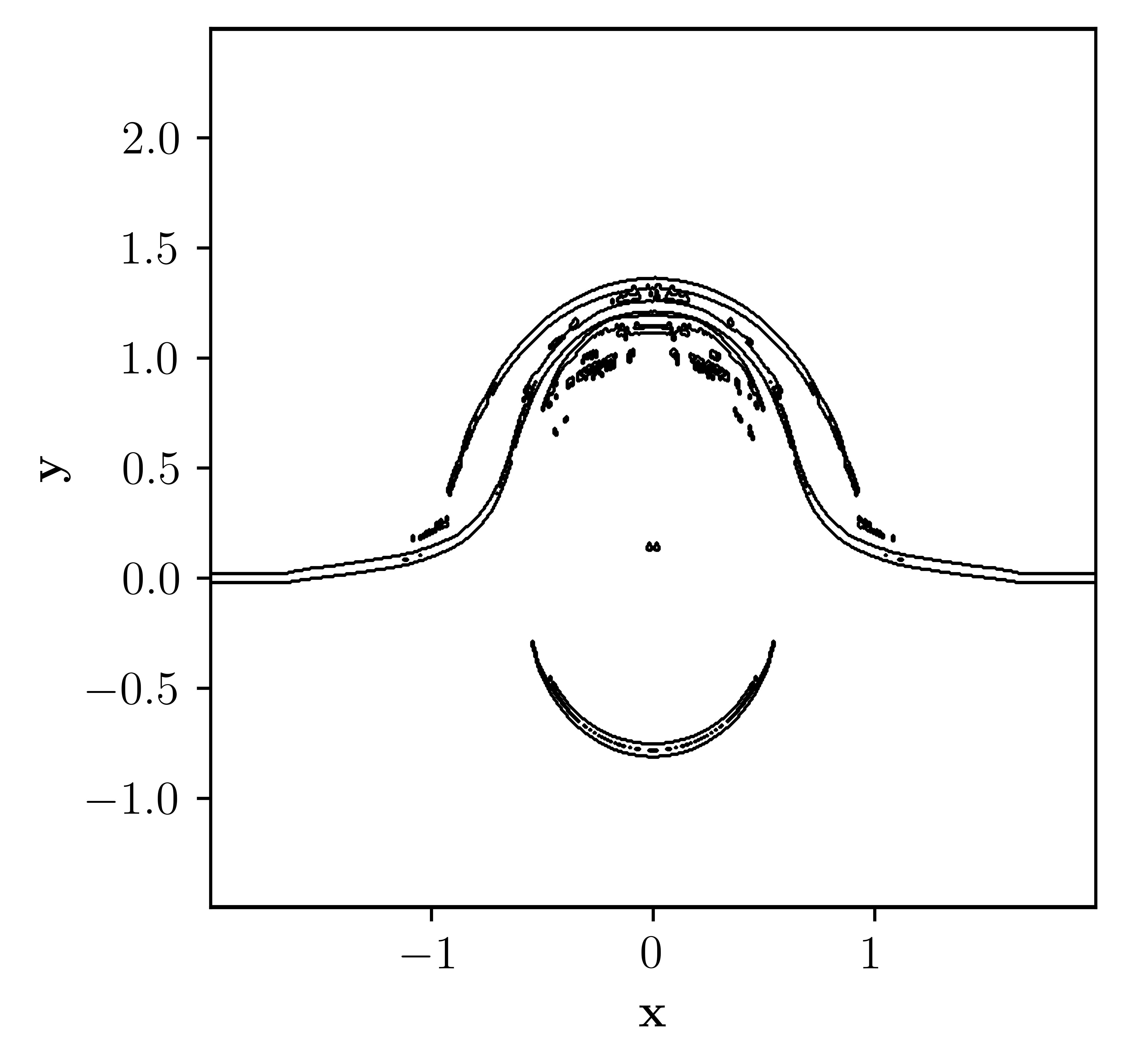}
\label{fig:sens_y}}
    \caption{Nonlinear function of normalized density gradient magnitude, $\phi=\exp \left(|\nabla \rho| /|\nabla \rho|_{\max }\right)$, computed by various schemes for underwater explosion problem, Example \ref{underoos}, on a 600 $\times$ 600 mesh, at t=0.19. Contours are from 1.0 to 1.7. Sensor locations are shown in Figure \ref{fig:sens_x} and \ref{fig:sens_y}.}
    \label{fig:under00}
\end{figure} 
Figures \ref{fig:wmpp} and \ref{fig:wmppf} present the numerical results computed by the Wave-MP scheme using the primitive variable algorithm on two different grid sizes. While the interface computed by the adaptive primitive-characteristic scheme (Figure \ref{fig:uw_wavemp}) appears smooth upon visual inspection, the computations performed with the Wave-MP (Prim) scheme are not as smooth and appear jagged, as the high-order scheme is used throughout (Figure \ref{fig:wmpp}).
As the grid is refined to 1200 $\times$ 1200, vortical structures begin to emerge, as shown in Figure \ref{fig:wmppf}. Figure \ref{fig:wfinest} shows the results computed on a 2400 $\times$ 2400 grid using the adaptive primitive-characteristic scheme, where small vortical structures also begin to appear. The differences between the adaptive primitive-characteristic scheme and the primitive variable approach can largely be attributed to the use of the MUSCL scheme near the interface in the latter, which is inherently dissipative and low-order. In contrast, the primitive variable approach employs the high-order MP scheme, where tangential velocities are computed using a central scheme.
It is challenging to determine which approach better represents the interfacial structures definitively. The current work does not account for viscous effects, surface tension, phase changes, or the equation of state, all of which could influence the results. Nevertheless, the simulations align with published results from other studies \cite{deng2018high, barton2019interface}.

\begin{figure}[H]
\centering\offinterlineskip
\subfigure[\textcolor{black}{Wave-MP (Prim), 600 $\times$ 600}]{\includegraphics[width=0.32\textwidth]{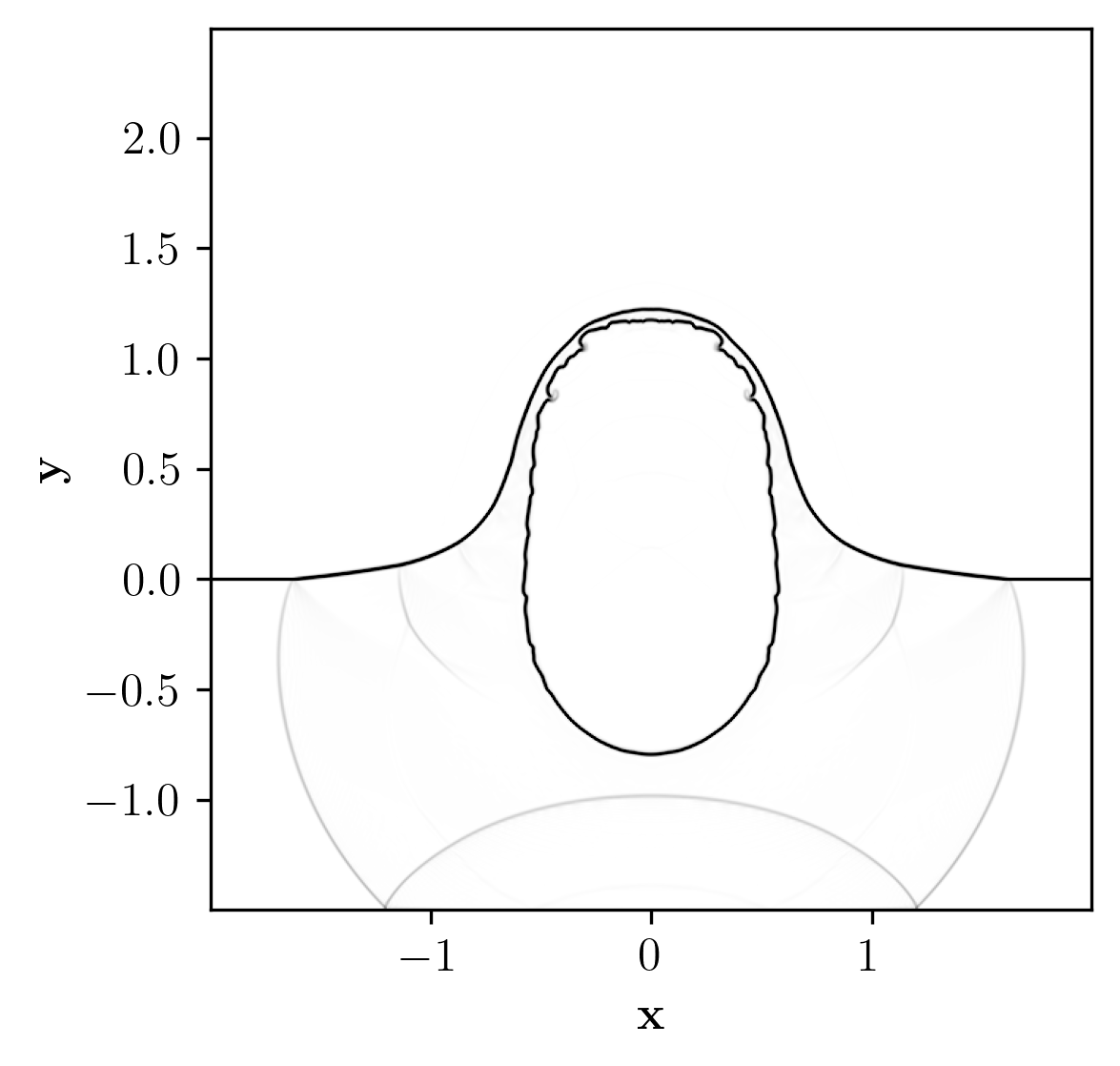}
\label{fig:wmpp}}
\subfigure[\textcolor{black}{Wave-MP (Prim), 1200 $\times$ 1200}]{\includegraphics[width=0.32\textwidth]{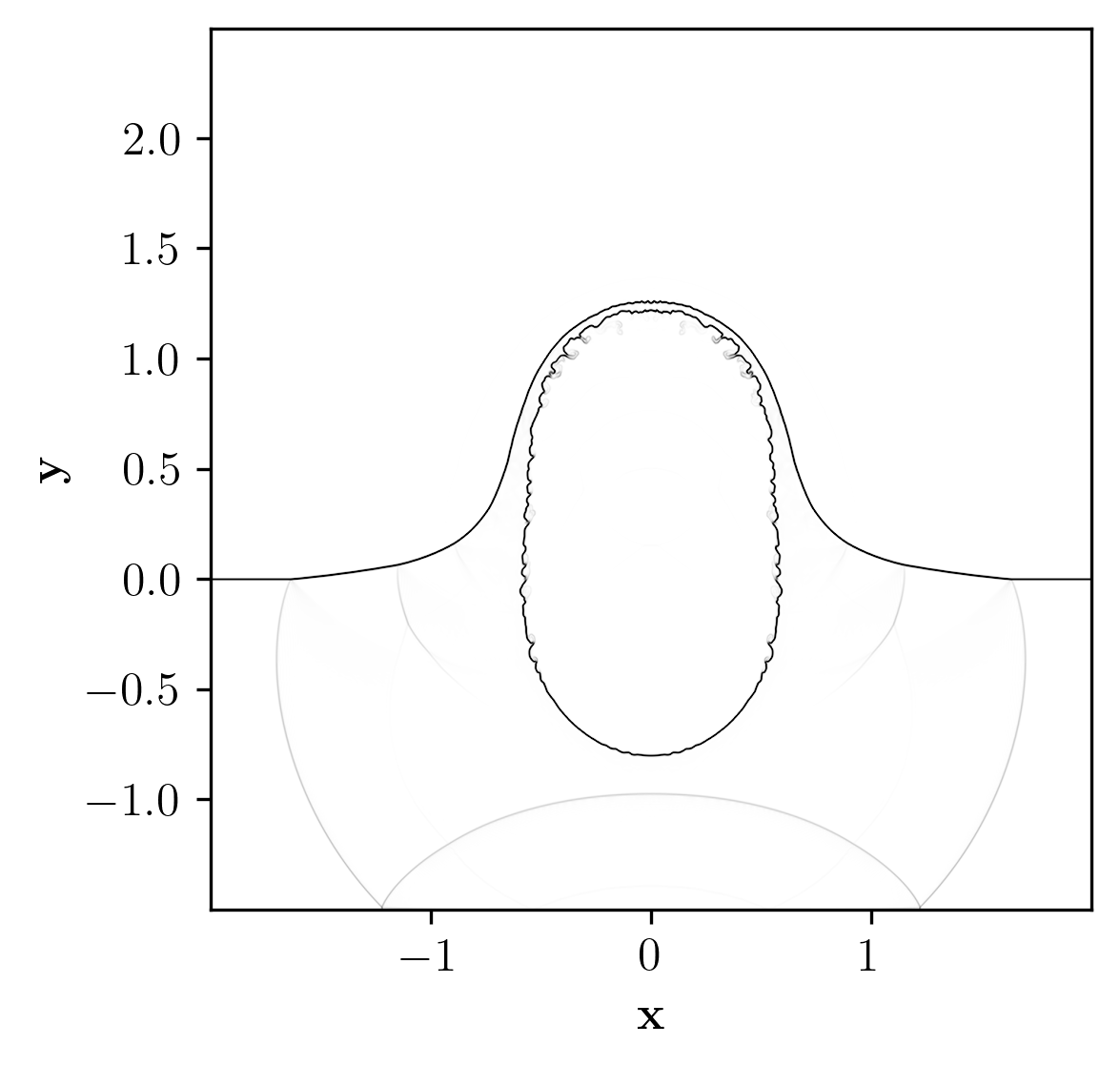}
\label{fig:wmppf}}
\subfigure[\textcolor{black}{Wave-MP, 2400 $\times$ 2400}]{\includegraphics[width=0.32\textwidth]{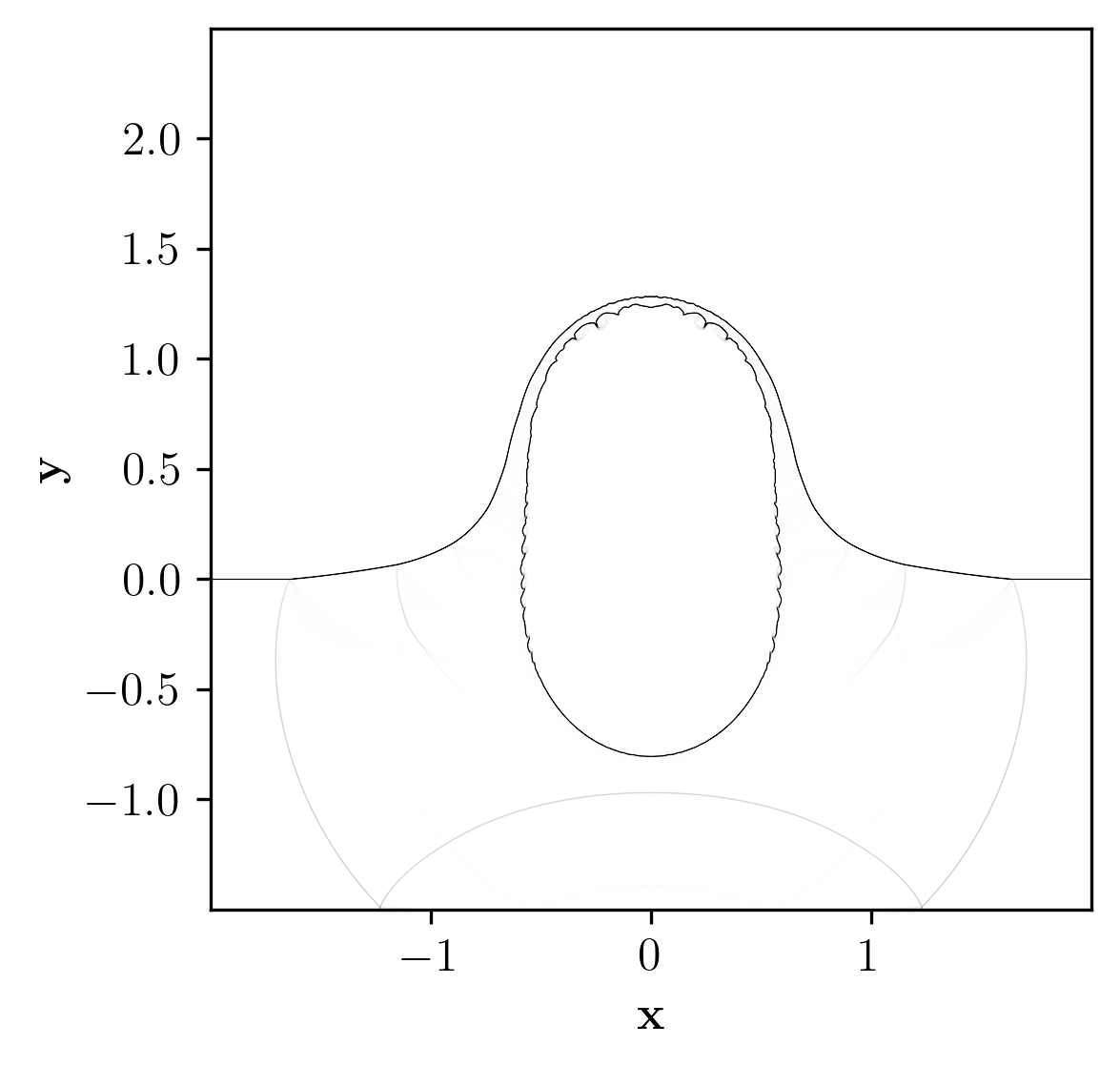}
\label{fig:wfinest}}
    \caption{Nonlinear function of normalized density gradient magnitude, $\phi=\exp \left(|\nabla \rho| /|\nabla \rho|_{\max }\right)$, computed by various schemes for underwater explosion problem, Example \ref{underoos} at t=0.19. Contours are from 1.0 to 1.7.}
    \label{fig:under_prim}
\end{figure} 

\begin{example}{Interaction between $Ma$=2.4 shockwave and a water cylinder}\label{water-24}
\end{example}
This test case simulates the interaction of a Mach 2.4 planar shock with a water cylinder to evaluate the reliability of the proposed algorithms for simulating multi-dimensional, two-phase flows with shocks. The computational domain is [0, 0.111] $\times$ [0, 0.074] $m^2$, with a water cylinder of diameter D = 0.022 m located at [0.04, 0.037] m. The incident shock is positioned at $x$ = 0.029 m. The simulation is performed on a 3072 $\times$ 2048 mesh using the MUSCL, Wave-MUSCL, Wave-MP, and Wave-MP (Prim) schemes, with reflective boundary conditions applied to the top and bottom boundaries of the domain. The fitting parameter $\gamma$ for air is 1.4, and for water, it is 6.12. The parameter $\pi_{\infty}$= 0 for air, and $\pi_{\infty}$ = 3.43 $\times$ $10^8$ for water.

The experimental Schlieren images, shown in Figure \ref{fig:water_exp} \cite{sembian2016plane}, are compared with the density gradients computed from the simulation results in Figure \ref{fig:water_num}. The interaction of the incident shock wave with the air-water interface generates a reflected shock in the upstream air and a transmitted shock that propagates into the water. Due to water’s significantly higher acoustic impedance than air, the transmitted shock travels faster than the incident shock. As the incident shock strikes the interface, an increased angle of reflection develops, resulting in Mach reflection. When the transmitted shock reaches the downstream side of the interface, it generates a rarefaction wave within the water cylinder, which focuses and continues to rebound inside the cylinder. Eventually, the two Mach stems converge downstream of the water cylinder, forming a pair of vortices behind the cylinder. 

Barrett, Subbareddy, and Candler also developed a low-dissipation method for multiphase flow simulations \cite{barrett2024development}. It was mentioned that the vortices pointed with the red arrow in Figure \ref{fig:water_exp2} are contact wave vortices which exist in the experimental results and are reproduced in the present numerical simulation (they also observed in their simulation but are not as prominent).

\begin{figure}[H]
%\begin{halfspacing}
\centering
\subfigure[Experimental result \cite{sembian2016plane}.]{\includegraphics[width=1.0\textwidth]{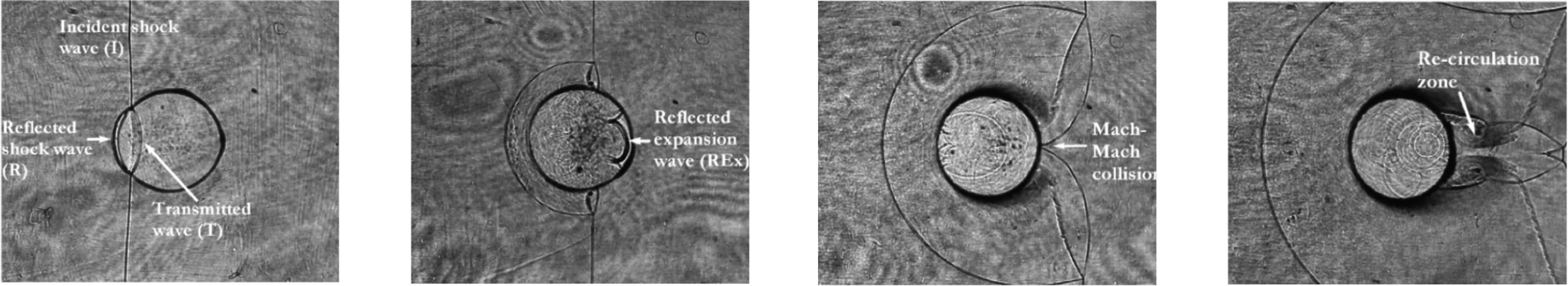}
\label{fig:water_exp}}
\subfigure[Numerical simulation using Wave-MP scheme.]{\includegraphics[width=1.0\textwidth]{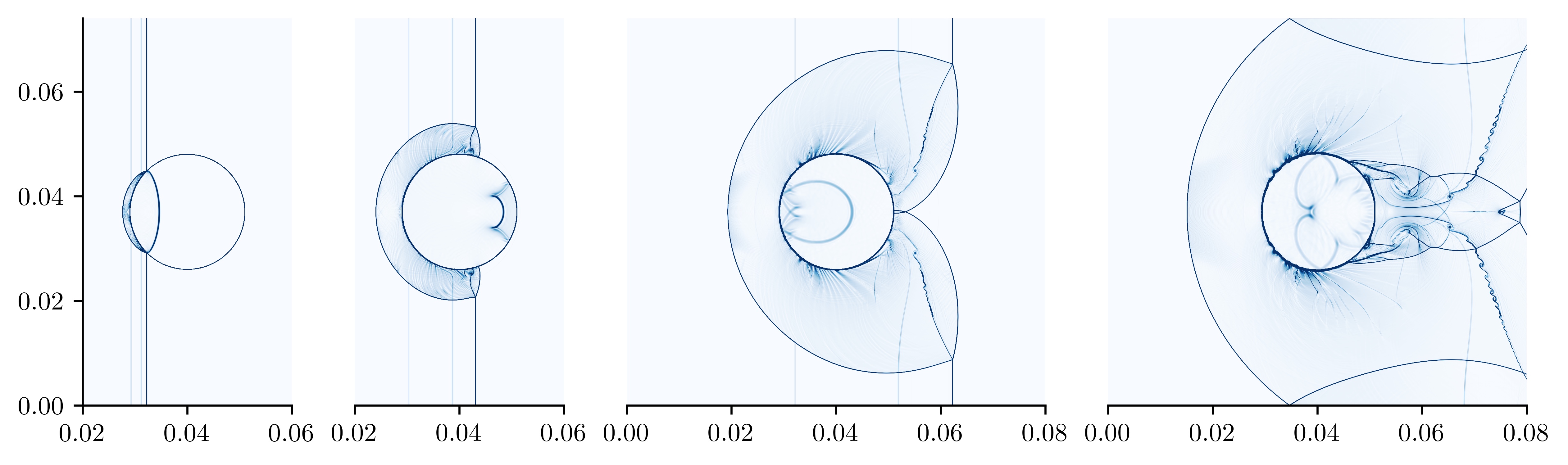}
\label{fig:water_num}}
    \caption{Shock-water cylinder interation, Example \ref{water-24}, comparison of numerical results with experiment.}
    \label{fig:w_exp2}
%\end{halfspacing}
\end{figure} 

\begin{figure}[H]
\centering
 \includegraphics[width=0.95\textwidth]{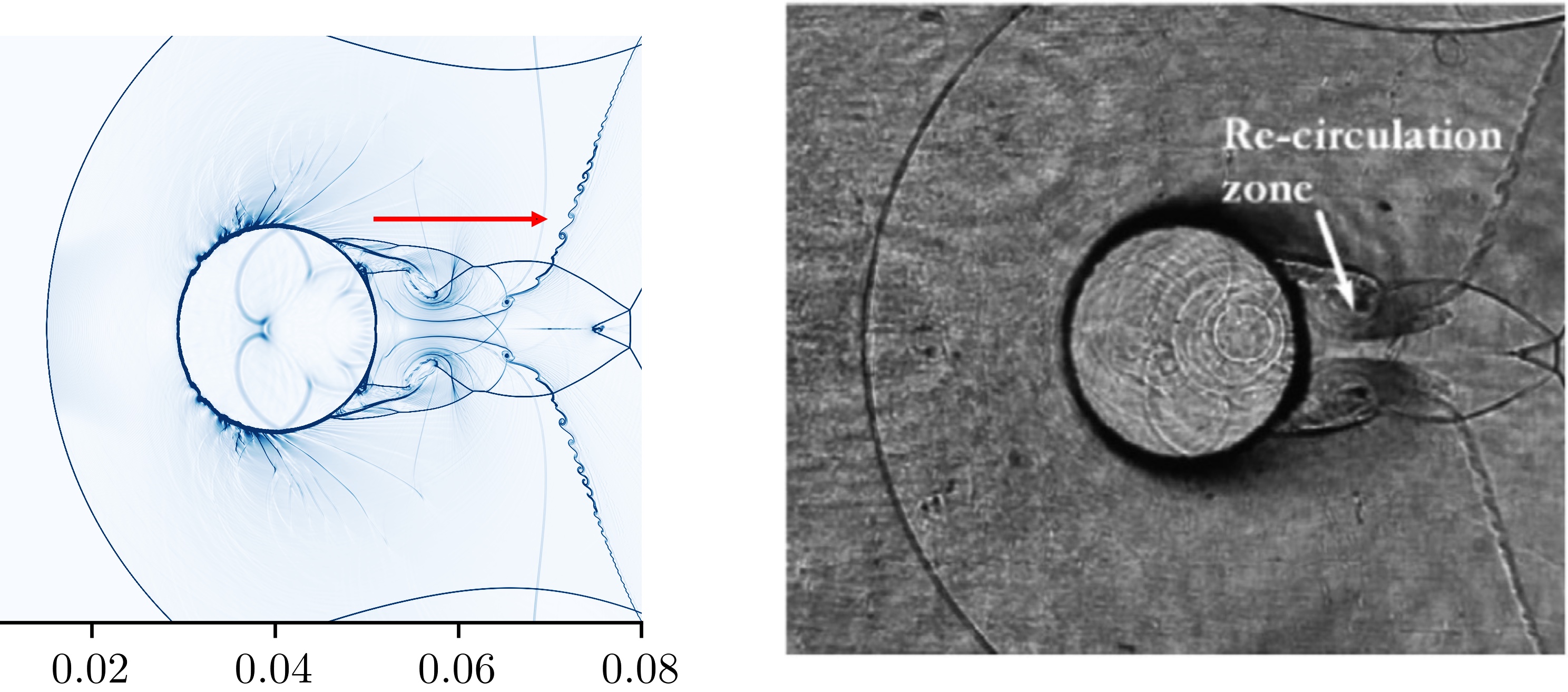}
    \caption{Shock-water cylinder interaction, Example \ref{water-24}. Numerical and experimental result at $t$=67$\mu$s.}
\label{fig:water_exp2}
\end{figure}

The advantages of the proposed adaptive primitive-characteristic reconstruction methodology are evident in Figure \ref{fig:w_adapt}. Figure \ref{fig:mus-water} shows the simulation results obtained using the MUSCL scheme, where the liquid droplet interface is smeared and thicker due to numerical dissipation. Figure \ref{fig:mut-water} presents results from the Wave-MUSCL scheme, where vortical structures observed in the experimental results are absent. However, the droplet interface is thinner, due to the use of the THINC scheme for material interfaces. Figure \ref{fig:up_water} shows the results from the MP-THINC approach, where the interface is similar to that of the Wave-MUSCL scheme, with some roll-up vortices appearing due to the use of the fifth-order upwind scheme in the gas phase. These results suggest that the MP scheme is active in the gas region, while the MUSCL scheme is used in the liquid region. The adaptive primitive-characteristic approach, utilizing the stiffened gas parameter ($\Pi$) to identify liquid regions, has proven effective. Finally, Figure \ref{fig:wave_water} shows the results from the Wave-MP approach, which successfully reproduces the vortical structures in the gas region that are absent in the MP-THINC results. These vortical structures arise from the central scheme used for vorticity waves in the Wave-MP scheme, which is the key difference between Wave-MP and MP-THINC. As observed in the compressible triple point test, vortical structures emerge from the central scheme’s computation of vorticity waves.
\begin{figure}[H]
%\begin{halfspacing}
\centering\offinterlineskip
\subfigure[\textcolor{black}{MUSCL.}]{\includegraphics[width=0.48\textwidth]{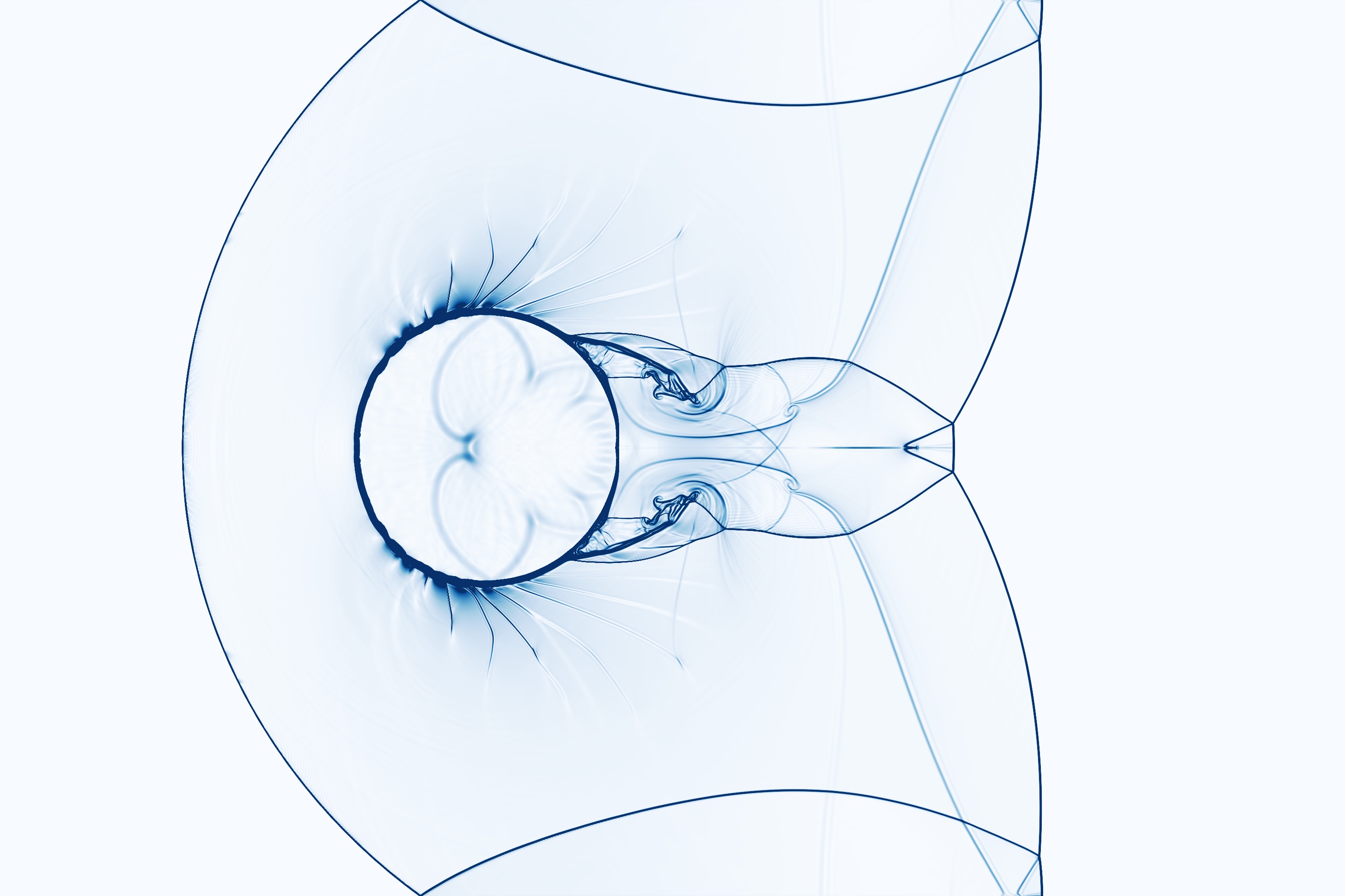}
\label{fig:mus-water}}
\subfigure[\textcolor{black}{Wave-MUSCL.}]{\includegraphics[width=0.48\textwidth]{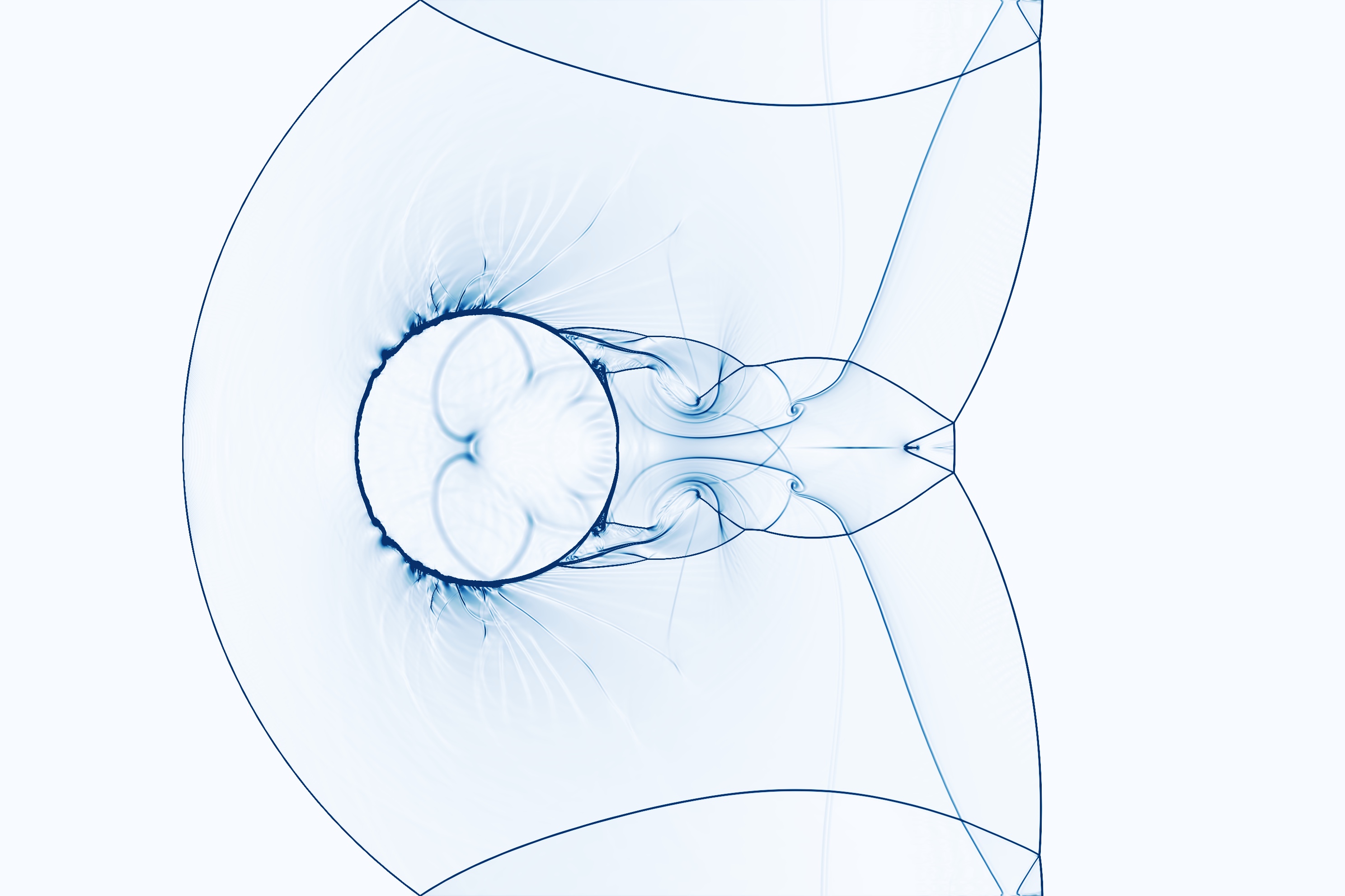}
\label{fig:mut-water}}
\subfigure[\textcolor{black}{MP-THINC.}]{\includegraphics[width=0.48\textwidth]{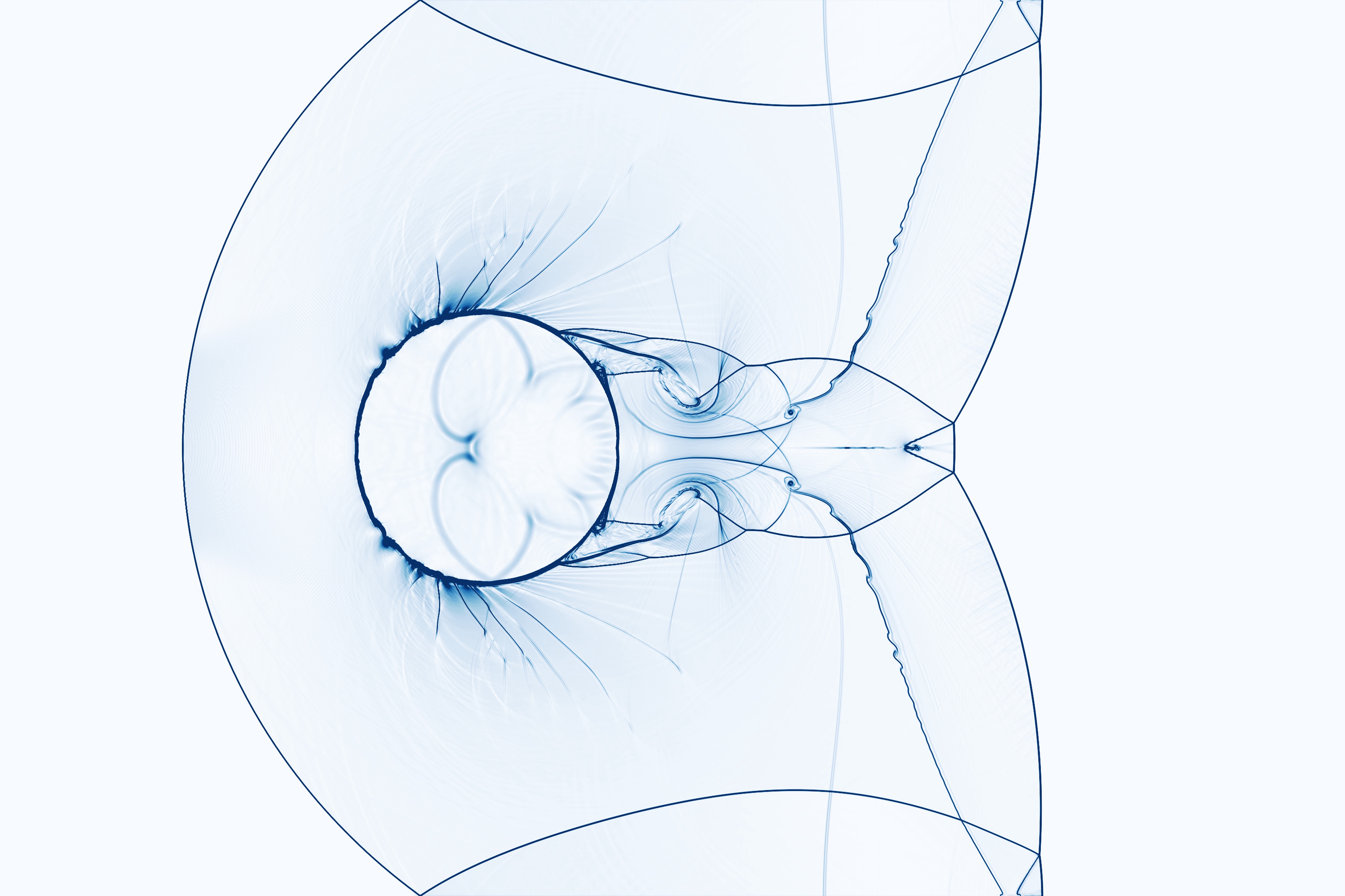}
\label{fig:up_water}}
\subfigure[\textcolor{black}{Wave-MP.}]{\includegraphics[width=0.48\textwidth]{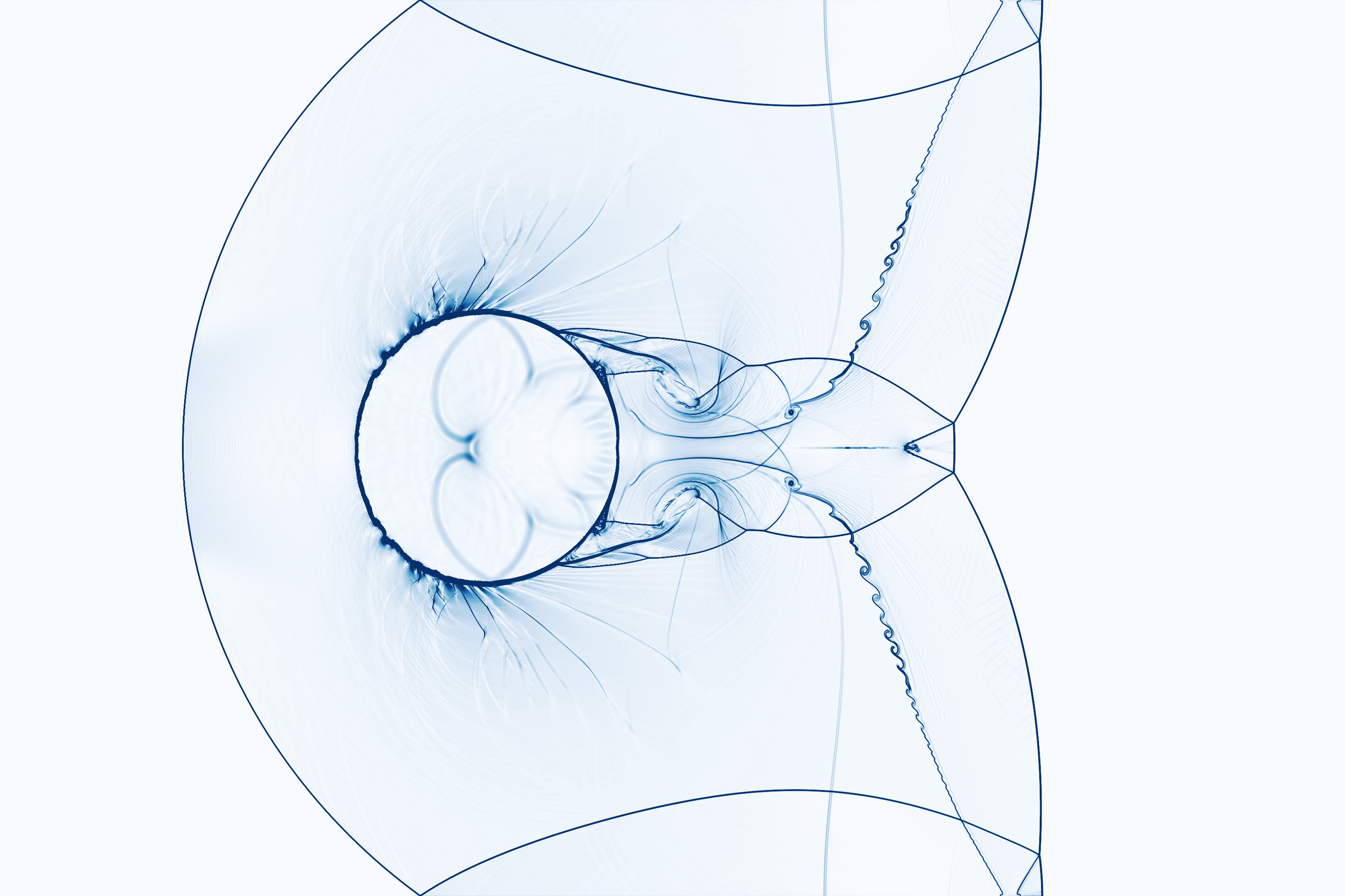}
\label{fig:wave_water}}
    \caption{Shock-water cylinder interaction, Example \ref{water-24}, on a 3072 $\times$ 2048 mesh, at t=67$\mu$s using various schemes.}
    \label{fig:w_adapt}
\end{figure} 
The numerical results using the direct reconstruction of primitive variables with the Wave-MP (Prim) scheme also reproduce the contact wave vortices, as shown in Figure \ref{fig:prim-water}. However, the primitive variable reconstruction introduces oscillations near the shock. In contrast, the adaptive primitive-characteristic reconstruction was free from oscillations near the shock and near the bubble. The complete characteristic reconstruction failed for this test case when using the MUSCL-THINC scheme, and neither the WENO nor MP schemes succeeded for both primitive and characteristic variables. Notably, computing the tangential velocities in either characteristic or primitive variable space produced vortical structures, highlighting the advantages of the proposed “multi-dimensional” upwinding approach.

For this test case, Wong et al. \cite{wong2021positivity} (see Figure 12 in their paper) and Zhang et al. \cite{zhang2024hybrid} (see Figure 9 in their paper) both performed numerical simulations using the same grid size and WENO schemes. However, neither simulation captured the vortical structures, likely due to the use of upwind schemes for all the variables. \textbf{Readers need to note that comparisons with other methods aim to highlight the strengths of the proposed algorithms rather than critique the work of others. Additionally, implementing and conducting simulations using every other published method is practically unfeasible.}
\begin{figure}[H]
%\begin{halfspacing}
\centering
\subfigure[\textcolor{black}{Wave-MP (Prim).}]{\includegraphics[width=0.54\textwidth]{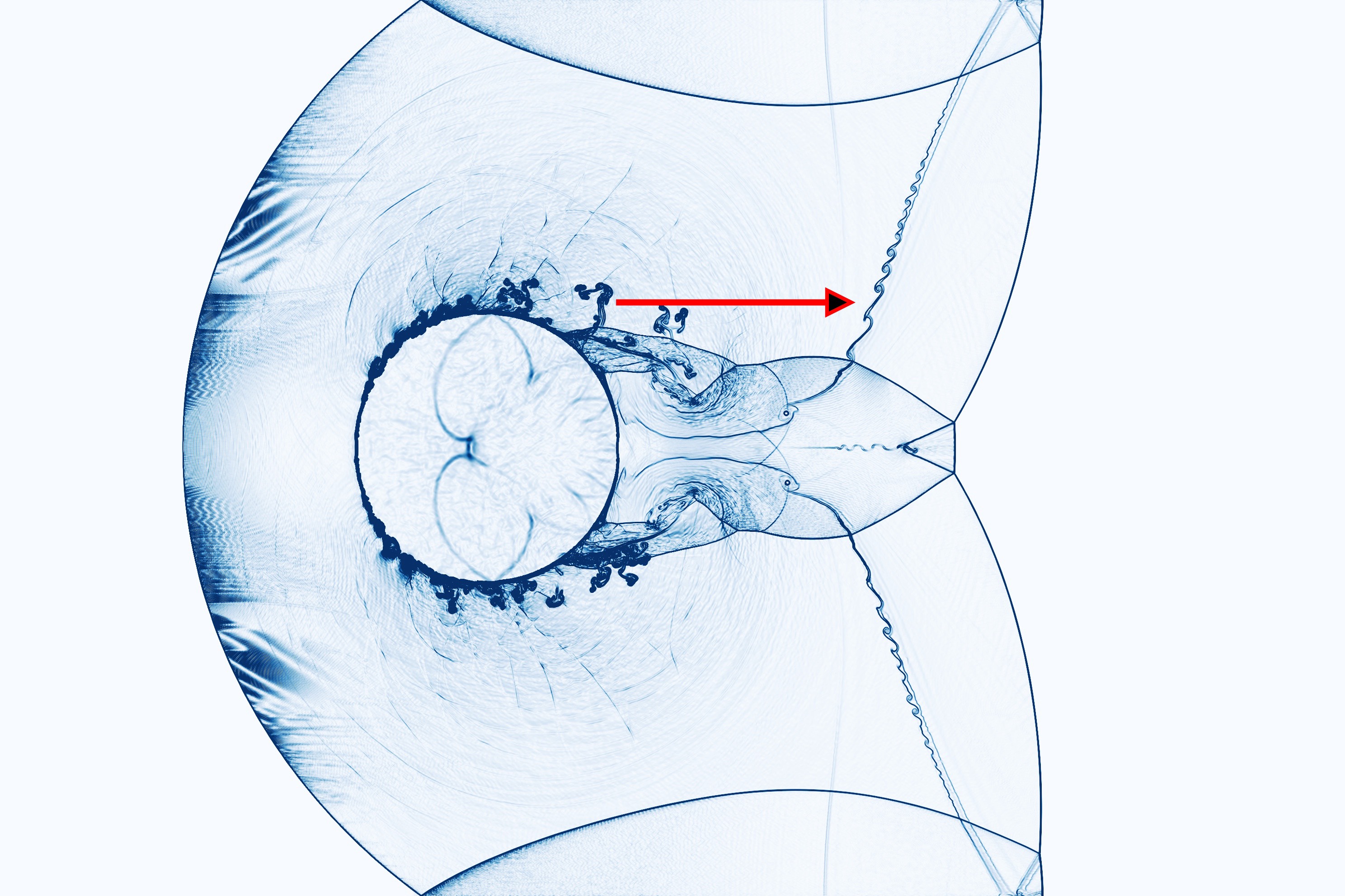}
\label{fig:prim-water}}
\subfigure[\textcolor{black}{Experiment.}]{\includegraphics[width=0.43\textwidth]{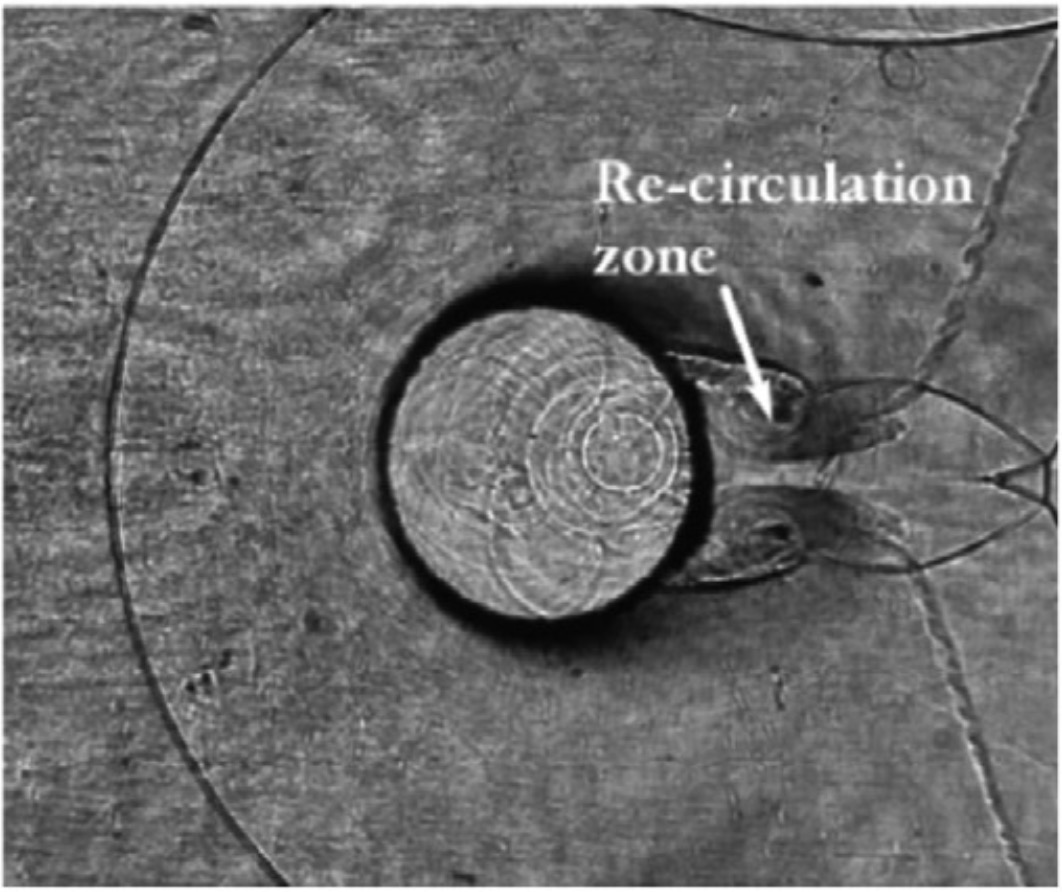}
\label{fig:exp}}
    \caption{Shock-water cylinder interaction, Example \ref{water-24}, on a 3072 $\times$ 2048 mesh, at t=67$\mu$s using primitive variable reconstruction approach, Wave-MP (Prim).}
    \label{fig:w_adapt_p}
\end{figure} 

\section{Conclusions}\label{clusions}

The paper introduces an algorithm grounded in the Euler equations' physical principles, leveraging their wave structure (in characteristic space) or the underlying physics of the physical variables to the extent possible. This approach effectively combines central, upwind, and interface-capturing schemes tailored to the specific characteristics of the flow to represent them better.

\begin{itemize}
	\item It has been demonstrated that employing an upwind scheme for acoustic waves while using a central scheme for vorticity waves mitigates the generation of spurious vortices in periodic shear layer cases. Specifically, vorticity waves are intentionally computed using a central reconstruction scheme. The monotonicity-preserving criterion \cite{suresh1997accurate}, serving as a shock or discontinuity detector, is sufficiently robust in preventing spurious vortices without introducing excessive numerical dissipation. \textbf{These results also indicate that the monotonicity-preserving approach \cite{suresh1997accurate} is more versatile as the MP criteria, \textcolor{red}{Equation (\ref{eqn:mp5Condition})}, for discontinuity detection can be used for both central and upwind schemes. This versatility is unlike that of the WENO schemes, which require evaluation of large and expensive smoothness indicators, similar observations are made about WENO scheme in \cite{li2021low}.} It has been shown that the TENO-THINC scheme \cite{takagi2022novel} will also benefit from applying the THINC scheme only for entropy waves, as shown for the periodic shear layer case. 
	\item The central scheme for vorticity wave computation performs better in reproducing vortical structures in various test cases. For instance, the proposed algorithm successfully captured vortical structures even on coarse grids for gas-gas and gas-liquid interactions. The algorithm accurately reproduced experimentally observed vortical structures in the shock-water-droplet interaction test case. In contrast, using an upwind scheme for vorticity wave computation resulted in excessive dissipation, suppressing vortices (readers can also see the results obtained for the inviscid Taylor Green vortex, shown in the Appendix, for further advantages of using a central scheme for vorticity waves).
	\item Similarly, computing entropy waves using a central scheme significantly improved the shock-entropy wave test case results. These findings also highlight that not all regions of the entropy wave correspond to contact discontinuities, suggesting that the THINC scheme should be applied selectively. Specifically, the THINC approach is appropriate for regions near material interfaces, mainly when using characteristic variable reconstruction. \textbf{These observations are one of the important contributions of the paper}.
	\item The algorithm employs an adaptive reconstruction strategy that switches between primitive and characteristic variables based on the flow conditions for multiphase flows. The stiffened gas parameter helps identify the liquid phase, where primitive variables are used for reconstruction. MUSCL/THINC schemes are applied in liquid regions for adaptive primitive-characteristic reconstruction, while the MP/THINC schemes are utilized in gas-phase regions. This adaptive strategy effectively balances accuracy and stability in multiphase flow simulations. Primitive variable reconstruction, while oscillatory near shocks, also reproduced the vortical structures noticed in the experimental results.
		\item The adaptive primitive-characteristic reconstruction algorithm is more robust than the complete primitive variable reconstruction algorithm, as the primitive variable algorithm failed for the air-cavity test case. The adaptive reconstruction algorithm is robust due to the use of the MUSCL/THINC scheme near the gas-liquid interface. In the adaptive primitive-characteristic reconstruction algorithm, the tangential velocities are still computed by the upwind MUSCL scheme in the regions of the gas-liquid interface. In contrast, in the primitive variable reconstruction, they are computed by a central scheme (sixth-order). It is one aspect that may be improved.
\end{itemize}

The wave (or physics) appropriate multi-dimensional algorithm(s) has benefited multiphase flow simulations. It can be further extended to other equations; further physics can be added, such as surface tension and phase change (which will add further physical constraints).

\section*{Appendix}\label{appen}
\renewcommand{\thesubsection}{\Alph{append}}
This appendix examines the performance of the proposed algorithms by applying them to the three-dimensional inviscid Taylor-Green vortex problem, a well-established benchmark in computational fluid dynamics. This problem, characterized by a low Mach number M=0.1, contains no discontinuities, making it an ideal case to evaluate the fidelity of the proposed methods. The THINC scheme should not affect (improve or otherwise) the entropy waves as there are no contact discontinuities, and using a central scheme for vorticity waves should improve the results.
The simulations are initialized in a periodic domain with dimensions $x,y,z \in [0,2\pi)$ and run up to time $t=10$ on a grid resolution of $64^3$. The specific heat ratio is set to $\gamma=5/3$, and the flow is effectively incompressible due to the dominance of the mean pressure. The initial conditions for this test case are as follows:\begin{equation}\label{itgv}
\begin{pmatrix}
\rho \\
u \\
v \\
w \\
p \\
\end{pmatrix}
=
\begin{pmatrix}
1 \\
\sin{x} \cos{y} \cos{z} \\
-\cos{x} \sin{y} \cos{z} \\
0 \\
100 + \frac{\left( \cos{(2z)} + 2 \right) \left( \cos{(2x)} + \cos{(2y)} \right) - 2}{16}
\end{pmatrix}.
\end{equation}

Figure \ref{fig_TGV} indicates that the contact discontinuity sensor) did not affect this test case as it should (also shown in \cite{sainadh2024consistent}). The results obtained by the MP5 and MP5-THINC scheme are one over the other for kinetic energy and enstrophy. When the vorticity waves, Wave- MP scheme, are computed using the central scheme (a non-dissipative scheme), the results are improved, shown with a green dashed line in Figure \ref{fig:TGV_KE}. Once again, it indicates that the proposed algorithm that treats each wave using different reconstructions relevant to physics improved the results.

In support of the current algorithm, results from the literature are analyzed. In Ref. \cite{takagi2022novel}, Takagi et al. performed the simulations using the TENO5 and TENO-THINC schemes. Figure \ref{fig:taku} is taken from the concerned paper, and the scheme denoted as \textit{Present} is the TENO-THINC scheme. The objective of the paper of Takagi et al. is, as written in the abstract, \textit{``based on a novel parameter-free discontinuity-detection criterion, a new shock-capturing framework is proposed by combining the standard TENO (targeted essentially non-oscillatory) scheme for smooth regions with the non-polynomial based THINC (tangent of hyperbola for INterface capturing) reconstruction for non-smooth discontinuities.''}

 Although this test case contains no discontinuities, the TENO-THINC scheme showed unexpected improvements in the results, which suggests that the TENO-based indicator is incorrectly identifying smooth flow regions as discontinuities; otherwise, both TENO5 and TENO-THINC should have had identical results. The proposed approach does not have these inconsistencies, as the central scheme is used to improve the results.

Figure \ref{fig:yon} presents the results obtained by Yang et al. \cite{yang2023novel} for the same test case. Notably, the TENO5 scheme outperformed the TENO6 scheme, suggesting that the adaptive upwind-central approach in TENO6 was less effective than the fully upwind TENO5 scheme. An adaptive upwind-central approach should perform better than a fully upwind scheme. While the current algorithm employs an adaptive upwind-central strategy, it achieves improved results. The Wave-MP scheme demonstrated improvements over the fully upwind MP5 scheme. MP criterion for discontinuity detection is not perfect either and can be improved.
\begin{figure}[H]
\centering
\subfigure[Kinetic energy]{\includegraphics[width=0.35\textwidth]{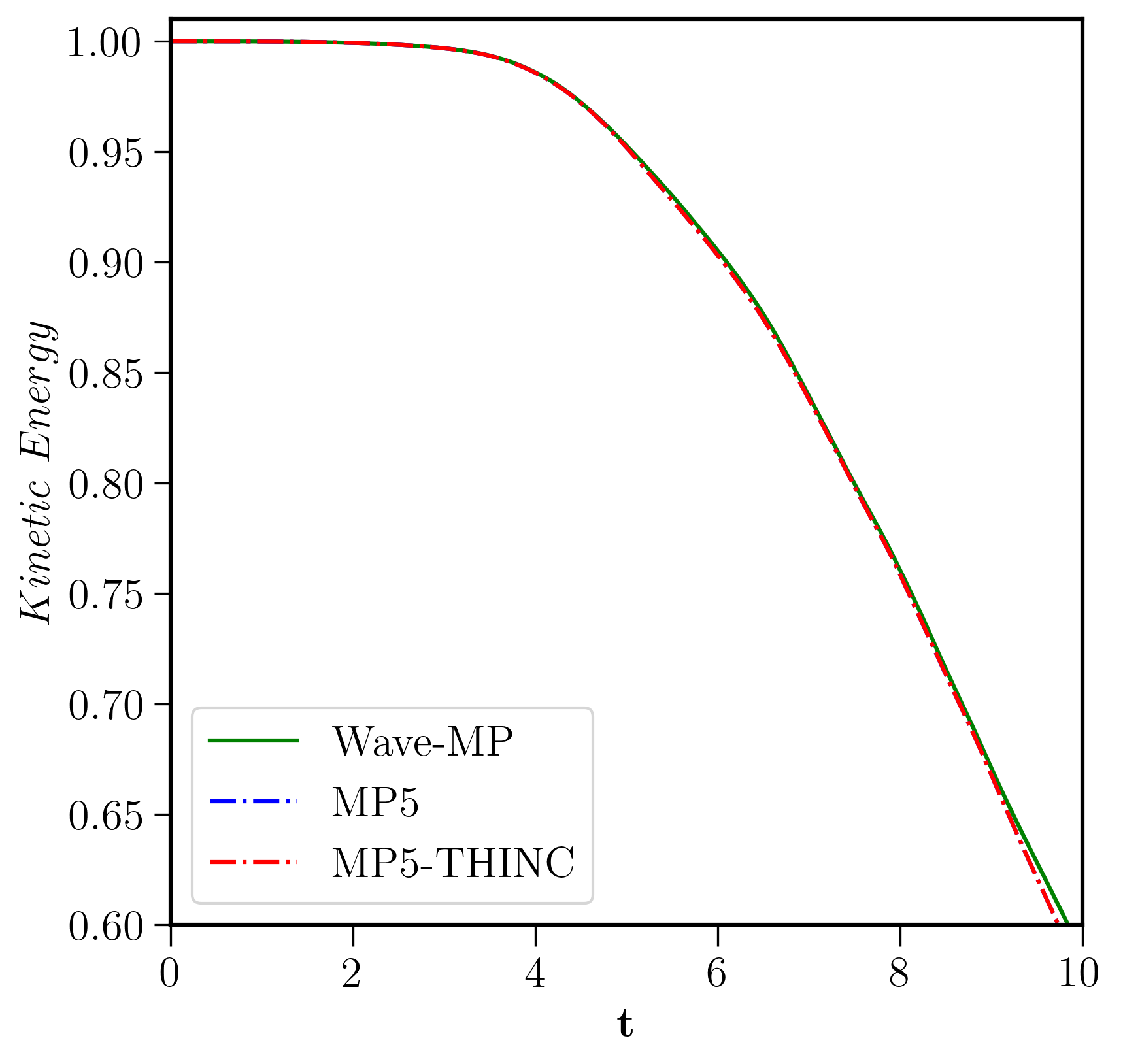}
\label{fig:TGV_KE}}
\subfigure[Enstrophy]{\includegraphics[width=0.35\textwidth]{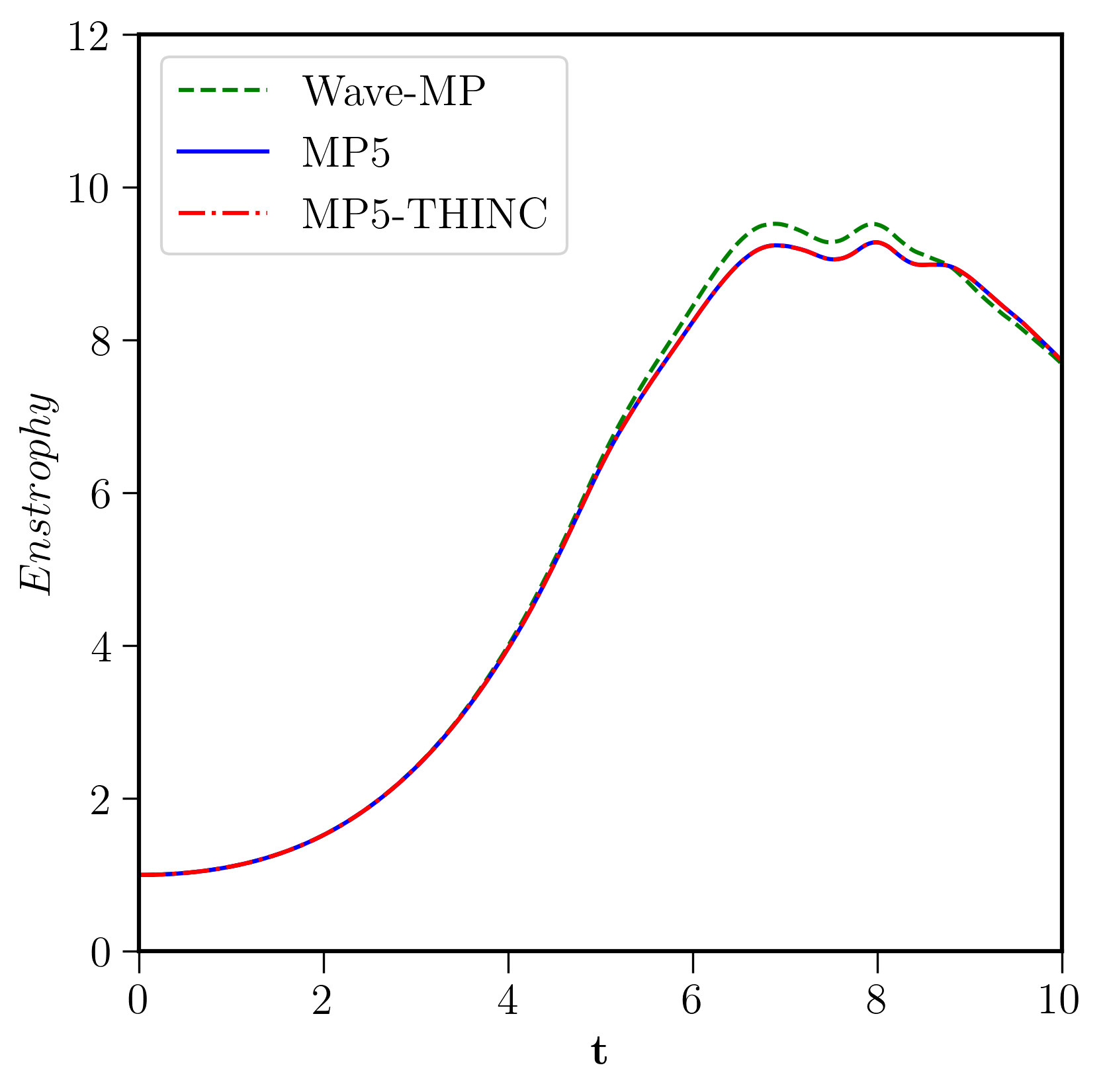}
\label{fig:TGV_ens}}
\caption{Normalised kinetic energy and enstrophy using various schemes on grid size of $64^3$. Dashed red line: MP5-THINC, dashed green line: Wave-MP and  blue line: MP5.}
\label{fig_TGV}
\end{figure}

\begin{figure}[H]
\centering
\subfigure[From \cite{takagi2022novel}]{\includegraphics[width=0.48\textwidth]{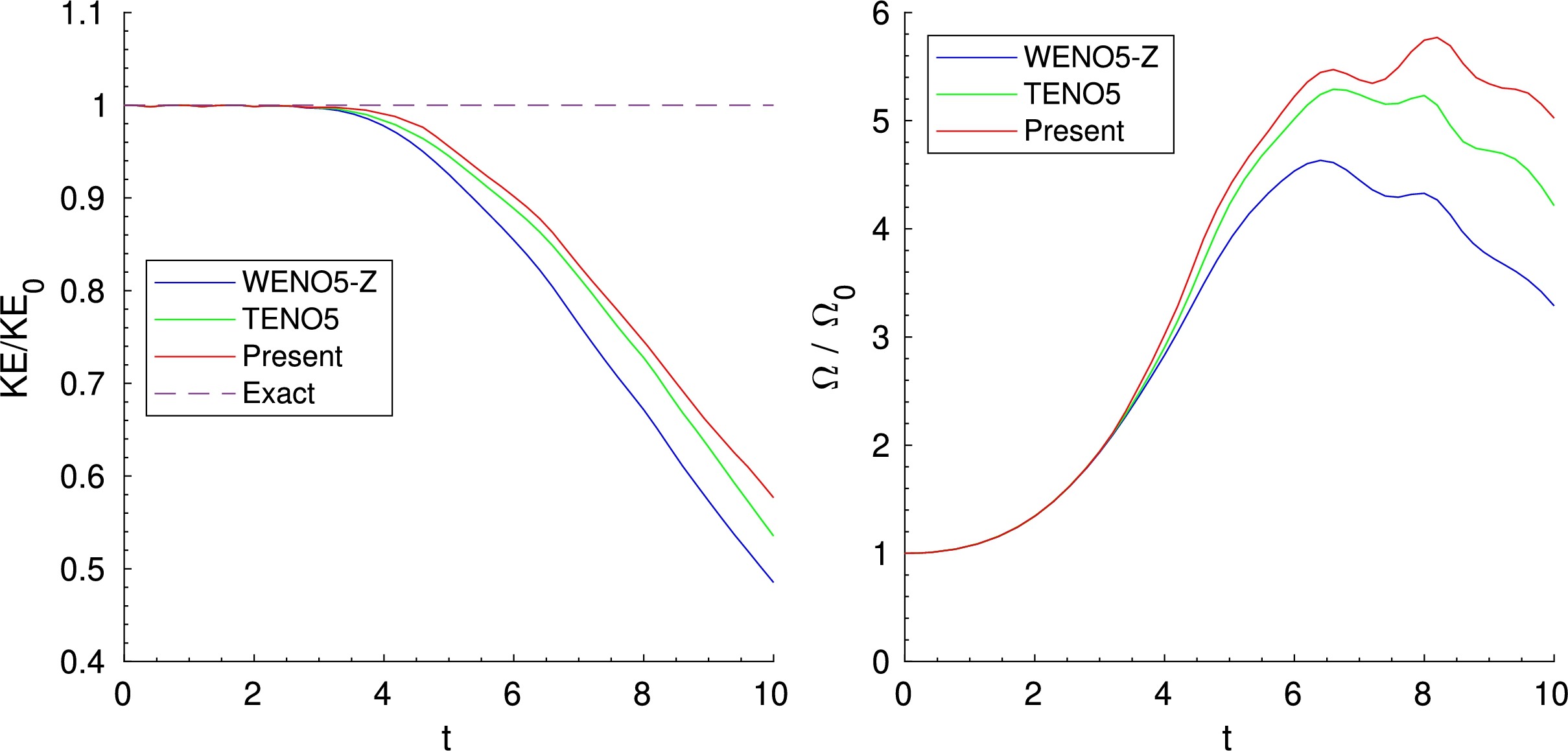}
\label{fig:taku}}
\subfigure[From\cite{yang2023novel}]{\includegraphics[width=0.48\textwidth]{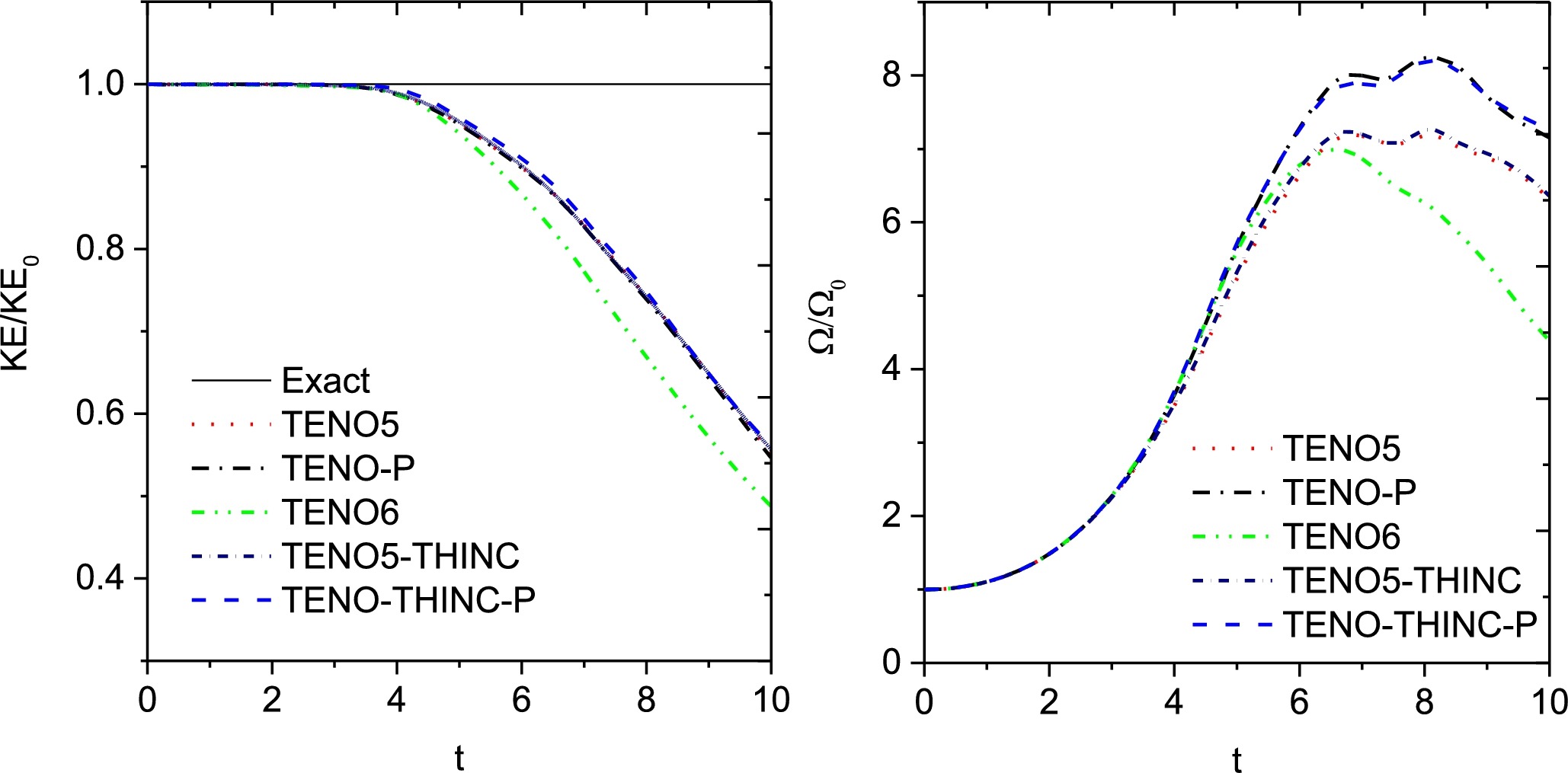}
\label{fig:yon}}
\caption{Results from the literature. Figure \ref{fig:taku} is reproduced from \cite{takagi2022novel} with permission from Elsevier BV 2024, License number 5941540100890. Figure \ref{fig:yon} is reproduced from \cite{yang2023novel} with permission from Elsevier BV 2024, License number 5941540221512. }
\label{fig_TGV_compare}
\end{figure}

\bibliographystyle{elsarticle-num}
\bibliography{contact_ref}

\end{document}